\documentclass[prd,twocolumn,eqsecnum]{revtex4}
\usepackage{bm}
\usepackage{amssymb}
\newcommand{\Adot}{\langle \dot{A} \rangle} 
\newcommand{\Mdot}{\langle \dot{M} \rangle} 
\newcommand{\Jdot}{\langle \dot{J} \rangle} 
\begin{document}
\title{Absorption of mass and angular momentum by a black hole:
Time-domain formalisms for gravitational perturbations, and the 
small-hole/slow-motion approximation} 
\author{Eric Poisson}
\affiliation{Department of Physics, University of Guelph, Guelph,
Ontario, Canada N1G 2W1}
\date{October 19, 2004 [Final version, as published 
in Physical Review D]} 
\begin{abstract}
The first objective of this work is to obtain practical
prescriptions to calculate the absorption of mass and angular momentum
by a black hole when external processes produce gravitational
radiation. These prescriptions are formulated in the time domain (in
contrast with the frequency-domain formalism of Teukolsky and Press)
within the framework of black-hole perturbation theory. Two such
prescriptions are presented. The first is based on the Teukolsky
equation and it applies to general (rotating) black holes. The second
is based on the Regge-Wheeler and Zerilli equations and it applies to
nonrotating black holes. The second objective of this work is to apply
the time-domain absorption formalisms to situations in which the black
hole is either small or slowly moving; the mass of the black hole is
then assumed to be much smaller than the radius of curvature of the
external spacetime in which the hole moves. In the context of this  
small-hole/slow-motion approximation, the equations of black-hole 
perturbation theory can be solved analytically, and explicit
expressions can be obtained for the absorption of mass and angular
momentum. The changes in the black-hole parameters can then be
understood in terms of an interaction between the tidal gravitational
fields supplied by the external universe and the hole's
tidally-induced mass and current quadrupole moments. For a nonrotating
black hole the quadrupole moments are proportional to the rate of
change of the tidal fields on the hole's world line. For a rotating
black hole they are proportional to the tidal fields themselves. When
placed in identical environments, a rotating black hole absorbs more
energy and angular momentum than a nonrotating black hole.   
\end{abstract}
\maketitle

\section{Introduction and summary}

\subsection{Goals and motivations}

The work described in this article is concerned with the absorption of
energy and angular momentum by a black hole when physical processes in
its exterior produce gravitational radiation. It is assumed throughout
that the rates of change of mass and angular momentum are sufficiently
low that they can be calculated within the framework of first-order
perturbation theory, in which the black hole differs only slightly
from a stationary and axisymmetric Kerr hole. 

The first goal of this work is to obtain practical prescriptions to
calculate the black-hole absorption, and to modernize the tools
fashioned in the early seventies by Teukolsky and Press 
\cite{teukolsky-press:74}. An
essential aspect of the new prescriptions is that they present the
absorption formulae in the time domain instead of the frequency
domain; they presuppose that in accordance with current trends, the
equations of black-hole perturbation theory have been solved as
partial differential equations in the time domain instead of ordinary
differential equations in the frequency domain. Two such prescriptions 
are presented here: the first is based on the Teukolsky equation 
\cite{teukolsky:73} and
it applies to general (rotating) black holes, while the second is
based on the Regge-Wheeler 
\cite{regge-wheeler:57} 
and Zerilli 
\cite{zerilli:70} 
equations and it applies to nonrotating black holes.      

That the Teukolsky equation can be separated in all of its variables 
is surely one of its most important properties. To a large extent, it
is this property that has permitted progress during the continuing
exploration of physical processes taking place in black-hole
spacetimes (see, for example, the book by Frolov and Novikov, 
Ref.~\cite{frolov-novikov:98}).
But it has to be acknowledged that the historical
importance of the separation property has diminished in recent years,
as a number of time-domain integrators of the Teukolsky equation have
been developed 
\cite{krivan-etal:96, krivan-etal:97, burko-khanna:03, 
      scheel-etal:04, martel:phd} 
and put to use in various applications. The numerical
task of solving the Teukolsky equation in the time domain is still
challenging: after decomposition into azimuthal modes one must solve
for a function of time and two spatial coordinates. But time-domain
methods appear now to be at least competitive with frequency-domain 
methods, with which one must solve for a number of radial and angular
functions, the number increasing as the spectrum of relevant
frequencies becomes wider. And it appears likely that the future 
will witness an increasing dominance of time-domain methods over
frequency-domain methods.  

While the superiority of time-domain methods is still to be proved in 
the case of the Teukolsky equation, it has clearly been established 
\cite{gundlach-etal:94a, gundlach-etal:94b, lousto-price:97, 
      martel-poisson:02, martel:04} 
in the context of the Regge-Wheeler 
\cite{regge-wheeler:57} 
and Zerilli 
\cite{zerilli:70} 
equations, which determine
the metric perturbations of a nonrotating black hole. In these cases  
the angular dependence of the perturbation variables can be completely
separated, and the integrator faces the relatively simple task of
solving for a function of two variables (time and a radial
coordinate). Simple, but powerful, numerical methods have been devised
\cite{lousto-price:97} 
for such problems, and these can even handle, without approximations, 
a singular source term contributed by a point particle. The
time-domain Regge-Wheeler and Zerilli equations are thus very easy to
integrate, and there is now little reason to go back to a
frequency-domain formulation.     

The new popularity of time-domain methods to solve the equations of 
black-hole perturbation theory calls for new prescriptions to
calculate the black-hole absorption of energy and angular
momentum. The only recipe currently available is the formalism of 
Teukolsky and Press
\cite{teukolsky-press:74}, 
which is based on the frequency-domain formulation of the Teukolsky
equation
\cite{teukolsky:73}. 
This formalism is not well adapted to time-domain calculations, and
in this work I provide the required translation of the
Teukolsky-Press recipe to the time domain. Another limitation of the
Teukolsky-Press formalism is that although it can be applied without
difficulty to a nonrotating black hole, this requires the use of the
Teukolsky equation instead of the more practical Regge-Wheeler and
Zerilli equations. Another objective of this work is therefore to
relate the absorption of mass and angular momentum by a Schwarzschild
black hole to the time-domain solutions to these equations.   

In effect, this work is about providing practical time-domain formulae 
for the fluxes of mass and angular momentum across a perturbed
black-hole horizon. For a nonrotating black hole these formulae are
based on the Regge-Wheeler and Zerilli equations, which govern the
behavior of the metric perturbations. For a rotating black hole the
formulae are based instead on the Teukolsky equation, which determines  
the perturbations of the Weyl curvature tensor.  

The second goal of this work is to apply the time-domain absorption
formalisms to physical situations in which the black hole can be
considered to be either small or slowly moving. In the context of this  
small-hole/slow-motion approximation (which I will describe in Sec.~I
E below), the equations of black-hole perturbation theory can be
solved analytically, and explicit expressions can be obtained for
the absorption of mass and angular momentum. While many results have
been obtained along those lines in the past
\cite{poisson-sasaki:95, death:96, tagoshi-etal:97, alvi:01}, 
they were all restricted to various special cases; the results
presented here consolidate and generalize these previous works.     

The absorption of mass and angular momentum by a black hole is
generally very small. In particular, the effect is likely to be too
small to be observed in a gravitational-wave signal that would be
measured by ground-based detectors such as LIGO, VIRGO, and GEO600. 
For example, Alvi
\cite{alvi:01} 
has calculated that for binary systems involving holes with masses
ranging from 5 to 50 solar masses, black-hole absorption is truly
negligible: It contributes only a small fraction of a wave cycle
during the signal's sweep through the detector's frequency band. For
this type of source the tools developed in this paper are not needed.      

In some circumstances, however, the black-hole absorption is a
significant effect that should not be neglected
\cite{price-whelan:01}.   
In particular, it is likely to be observed in gravitational-wave
signals that would be measured by a space-based detector such as
LISA. For example, Martel  
\cite{martel:04} 
has shown that during a close encounter between a massive black hole   
and a compact body (of a much smaller mass), up to approximately five 
percent of the total radiated energy is absorbed by the black hole,
the rest being transported out to infinity. Hughes   
\cite{hughes:01} 
has calculated that when the massive hole is rapidly rotating, the    
absorption has the effect of slowing down the inspiral of the orbiting  
body, thereby increasing the duration of the gravitational-wave
signal. For example, a $1\ M_\odot$ compact body on a slightly 
inclined, circular orbit around a $10^6\ M_\odot$ black hole of
near-maximum spin would spend approximately two years in the LISA
frequency band before its final plunge into the hole; Hughes shows
that the black-hole absorption contributes approximately 20 days
(and $10^4$ wave cycles) to these two years. For this kind of
situation the absorption is important, and the tools developed in this
paper will be useful.  

\subsection{Perturbative methods} 

A natural starting point for the calculation of black-hole absorption 
would be the definition of a dynamical mass $M(v)$ and angular
momentum $J(v)$ on a cross section $v=\mbox{constant}$ of an evolving
event horizon; here $v$ is a suitable advanced-time coordinate on the
horizon. Armed with such definitions, one would differentiate with
respect to $v$ to obtain $\dot{M}(v)$ and $\dot{J}(v)$, and seek to
express the right-hand sides in terms of standard perturbation
variables. Such an approach to black-hole absorption has recently been
pursued by a number of workers
\cite{ashtekar-krishnan:02, ashtekar-krishnan:03, booth-fairhurst:04, 
      hayward:04}, 
and the resulting (inequivalent)
formalisms can be formulated exactly in fully nonlinear general
relativity. These formalisms are based not on the event horizon, but 
instead on the hole's {\it trapping horizon}, a generally spacelike
hypersurface foliated by marginally trapped surfaces; and in the
Ashtekar-Krishnan formalism 
\cite{ashtekar-krishnan:02, ashtekar-krishnan:03} 
the definitions for $M(v)$ and $J(v)$ come
from the Hamiltonian formulation of general relativity. These
formalisms are interesting (and useful in the context of numerical
relativity) because they are fully general, and because they involve a
hypersurface (the trapping horizon) whose intersection with a given
Cauchy slice is easy to identify; the event horizon, on the other
hand, can be identified only once the future history of the spacetime
is completely known.  

The approach adopted here to calculate the black-hole absorption is 
not the one described in the preceding paragraph; it is based instead
on black-hole perturbation theory, and it assumes that the evolving
black hole is only slightly different from a stationary and
axisymmetric Kerr hole. Because the analysis is restricted to
first-order perturbation theory, it is possible to proceed without the 
specification of a mass function $M(v)$ and an angular-momentum
function $J(v)$, so long as only the {\it long-term} changes in mass
and angular momentum need to be calculated. In this long-term view one
imagines that the black hole starts in an initial stationary state
characterized by the parameters $(M,J)$, is perturbed for a time
$\Delta v$ by some external process, and then returns to another
stationary state characterized by the parameters $(M+\delta M,
J+\delta J)$. One then defines the {\it averaged rates of change} of
mass and angular momentum by $\Mdot = (\delta M)/(\Delta v)$ and 
$\Jdot = (\delta J)/(\Delta v)$, and one manipulates the equations of 
black-hole perturbation theory to calculate these quantities. This is
what I set out to do in this work. The perturbative techniques demand  
that $\delta M \ll M$, $\delta J \ll J$, and the long-term view
demands that $\Delta v \gg M$. While the price to pay is a
substantial loss of generality with respect to an exact formulation,
the perturbative-long-view approach adopted in this paper allows one 
to proceed without having to choose a specification of $M(v)$ and
$J(v)$, with the derived benefit that the final results are robust
with respect to a change of definitions. Another benefit is that the 
approach is based on the event horizon (the true boundary of the
black-hole region) instead of the trapping horizon; while locating the
event horizon in a nonlinear situation can be difficult, this is not a 
problem in the perturbative-long-view approach.  

The mathematical techniques required for the description of a
perturbed event horizon go back to the pioneering work of Hawking and
Hartle
\cite{hawking-hartle:72}, 
and these form the basis of this work (they also
formed the basis of the Teukolsky-Press prescription
\cite{teukolsky-press:74}). 
These techniques are reviewed in Secs.~II, III, and IV of the paper. I
begin in Sec.~II with a description of the unperturbed horizon of a
stationary Kerr black hole. In Sec.~III, I consider the dynamics of a
general evolving horizon, and in Sec.~IV, I specialize the discussion
to event horizons that are perturbed versions of the Kerr horizon. The
equations that govern the behavior of the horizon's null generators
are given in Sec.~III and IV in a form that closely resembles the
treatment provided by Price and Thorne
\cite{price-thorne:86}, 
and in Chapter VI of the book by Thorne, Price, and Macdonald 
\cite{thorne-etal:86}. 
Although these equations are well known,
I derive them {\it ab initio} in order to clearly identify the
simplifying assumptions that are incorporated along the way; in
particular, the averaging procedure involved in calculating $\Mdot$ 
and $\Jdot$ is explained fully in Secs.~IV B and C.    

The main ideas behind the Hawking-Hartle techniques 
\cite{hawking-hartle:72} 
are as follows. The intrinsic geometry of an evolving event horizon is  
described by a two-dimensional metric tensor $\gamma_{AB}$, which
depends on the advanced-time coordinate $v$ as well as angular
coordinates $\theta^A$ ($A = 2,3$). The metric is degenerate (and
explicitly two-dimensional) because the horizon is a null
hypersurface; $v$ is a parameter on the horizon's null generators, and
$\theta^A$ are generator labels that stay constant as the generators
move. The evolution of $\gamma_{AB}$ is determined by the behavior of 
the generators, which is described in terms of an expansion scalar
$\Theta$ and a shear tensor $\sigma_{AB}$. The evolution of the shear
is driven by the spacetime's Weyl curvature, and the evolution of the
expansion is driven by the square of the shear tensor. The evolution
equations can all be integrated (assuming that the Weyl curvature is
specified), and the solution for $\langle \Theta \rangle$ determines
$\Adot$, the (averaged) rate of change of the black hole's surface
area. The final step is to use the first law of black-hole mechanics
(see, for example, Chapter 12 of Ref.~\cite{wald:84}, or Chapter 5 of 
Ref.~\cite{poisson:04c})  
to relate this to $\Mdot$ and $\Jdot$. The end result is
Eqs.~(\ref{4.22})--(\ref{4.24}) in Sec.~IV C, which express
$\Mdot$, $\Jdot$, and $\Adot$ in terms of $\sigma_{AB}$ contracted
with derivatives of $\gamma_{AB}$, integrated over a cross section of
the event horizon. These equations are not new: they were first
presented by Thorne, Price, and Macdonald
\cite{thorne-etal:86}, 
but the derivation provided here is substantially different from
theirs. And while these equations are not themselves very practical,
they form an excellent starting point for the development of practical
formalisms.    

\subsection{Curvature formalism} 

The development of a time-domain formalism to calculate $\Mdot$,
$\Jdot$, and $\Adot$ in terms of standard curvature variables is
undertaken in Sec.~V. The end result of this reformulation of 
Eqs.~(\ref{4.22})--(\ref{4.24}) is the following prescription 
(Sec.~V D) to calculate the black-hole absorption. 

First, define a Teukolsky function $\Psi \equiv -\psi_0({\rm HH})$ as  
in Eq.~(\ref{5.6}), in terms of a null-tetrad decomposition of the
perturbed Weyl tensor. The label ``HH'' indicates that the Weyl tensor
is decomposed in the Hartle-Hawking null tetrad
\cite{hawking-hartle:72}, 
which is well behaved on the future horizon of the Kerr spacetime.  

Second, decompose the Teukolsky function in terms of azimuthal modes
proportional to $e^{i m \psi}$, 
\begin{equation} 
\Psi(v,r,\theta,\psi) = \sum_{m=-\infty}^\infty 
\Psi^m(v,r,\theta) e^{im\psi},
\label{1.1}
\end{equation}  
where $m$ is an integer. Because the Kerr spacetime is axially
symmetric, each mode $\Psi^m(v,r,\theta)$ evolves independently. Note
that the coordinates $(v,r,\theta,\psi)$ are ingoing Kerr coordinates
(Sec.~II A), and that they are well behaved on the event horizon.    

Third, integrate the Teukolsky equation 
\cite{teukolsky:73} 
for each relevant mode
$\Psi^m(v,r,\theta)$, and evaluate the result at $r = r_+ \equiv M + 
\sqrt{M^2 - a^2}$, the position of the unperturbed horizon; $a \equiv
J/M$ is the specific angular momentum of the Kerr black hole.     

Fourth, calculate the integrated curvatures 
\begin{equation} 
\Phi^m_+(v,\theta) = e^{\kappa v} \int_v^\infty e^{-(\kappa 
- im\Omega_{\rm H})v'} \Psi^m(v',r_+,\theta)\, dv' 
\label{1.2}
\end{equation}
and 
\begin{equation} 
\Phi^m_-(v,\theta) = \int_{-\infty}^v e^{im\Omega_{\rm H} v'}
\Psi^m(v',r_+,\theta)\, dv',    
\label{1.3}
\end{equation}
where $\kappa = (r_+-M)/(r_+^2+a^2)$ is the surface gravity of the
Kerr horizon, and $\Omega_{\rm H} = a/(r_+^2+a^2)$ its angular
velocity. Notice that $\Phi^m_+$ at advanced time $v$ depends on the
behavior of $\Psi^m$ at {\it later times}; this is a consequence of
the teleological nature of the event horizon.    

Fifth, and finally, insert the integrated curvatures and their
complex conjugates (indicated with an overbar) into the flux formulae   
\begin{eqnarray} 
\Mdot &=& \frac{r_+^2+a^2}{4\kappa} \sum_{m=-\infty}^\infty 
\biggl[ 2\kappa \int
\bigl\langle |\Phi_+^m|^2 \bigr\rangle \sin\theta\, d\theta 
\nonumber \\ & & \mbox{} 
- i m \Omega_{\rm H} \int \bigl\langle \bar{\Phi}_+^m \Phi_-^m 
- \Phi_+^m \bar{\Phi}_-^m \bigr\rangle \sin\theta\, d\theta \biggr],  
\qquad\quad \label{1.4} \\ 
\Jdot &=& -\frac{r_+^2+a^2}{4\kappa} \sum_{m=-\infty}^\infty (im)   
\nonumber \\ & & \mbox{} \times \int \bigl\langle
\bar{\Phi}_+^m \Phi_-^m - \Phi_+^m \bar{\Phi}_-^m \bigr\rangle
\sin\theta\, d\theta, 
\label{1.5}
\end{eqnarray}
and
\begin{equation} 
\frac{\kappa}{8\pi} \Adot = \frac{1}{2} (r_+^2 + a^2) 
\sum_{m=-\infty}^\infty \int  
\bigl\langle |\Phi_+^m|^2 \bigr\rangle \sin\theta\, d\theta. 
\label{1.6}
\end{equation} 
These equations reduce to those of Teukolsky and Press 
\cite{teukolsky-press:74} 
when $\Psi^m(v,r,\theta)$ is a pure mode of frequency $\omega$,
$\Psi^m \propto e^{-i\omega v}$; this is established in 
Sec.~V C. Equations  (\ref{1.4})--(\ref{1.6}) are therefore the
time-domain equivalent to the standard frequency-domain prescription.  

\subsection{Metric formalism} 

The curvature formalism of the preceding subsection applies to a
general rotating black hole, and the special case of a nonrotating
hole can be handled simply by setting $a = 0$. But in this case it is
often desirable to work with metric perturbations instead of curvature
perturbations, and it becomes useful to present the flux formulae in
terms of $\Psi^{lm}_{\rm RW}(v,r)$ and $\Psi^{lm}_{\rm ZM}(v,r)$, the 
standard Regge-Wheeler and Zerilli-Moncrief functions, instead of the
Teukolsky function $\Psi^m(v,r,\theta)$. Here the decomposition into  
modes involves spherical-harmonic functions of degree $l$ and
azimuthal number $m$.   

The development of a time-domain formalism to calculate $\Mdot$,
$\Jdot$, and $\Adot$ in terms of standard metric variables is
undertaken in Sec.~VII, after laying some important foundations in 
Sec.~VI. The end result of this reformulation of 
Eqs.~(\ref{4.22})--(\ref{4.24}) is the following prescription 
(Sec.~VII C) to calculate the absorption of mass and angular momentum
by a Schwarzschild black hole.    

First, integrate the Regge-Wheeler equation 
\cite{regge-wheeler:57} 
for $\Psi^{lm}_{\rm RW}(v,r)$, which describes the odd-parity sector
of the metric perturbations. This gauge-invariant function is defined  
in subsection 3 of the Appendix. 

Second, integrate the Zerilli equation 
\cite{zerilli:70} 
for $\Psi^{lm}_{\rm ZM}(v,r)$, which describes the even-parity sector
of the metric perturbations. This gauge-invariant function is defined
in subsection 4 of the Appendix.  

Third, and finally, evaluate the Regge-Wheeler and Zerilli-Moncrief
functions at $r = r_+ \equiv 2M$ and insert them into the flux
formulae   
\begin{eqnarray} 
\Mdot &=& \frac{1}{64\pi} \sum_{l=2}^\infty \sum_{m=-l}^l 
(l-1) l (l+1) (l+2) 
\nonumber \\ & & \mbox{} \times 
\Bigl\langle 4 \bigl| \Psi^{lm}_{\rm RW}(v,r_+) \bigr|^2 
+ \bigl| \dot{\Psi}^{lm}_{\rm ZM}(v,r_+) \bigr|^2 \Bigr\rangle 
\qquad
\label{1.7}
\end{eqnarray}
and 
\begin{eqnarray} 
\Jdot &=& \frac{1}{64\pi} \sum_{l=2}^\infty \sum_{m=-l}^l 
(l-1) l (l+1) (l+2) (im) 
\nonumber \\ & & \mbox{} \times 
\Bigl\langle 4 \Psi^{lm}_{\rm RW}(v,r_+) \int^v 
\bar{\Psi}^{lm}_{\rm RW}(v',r_+)\, dv' 
\nonumber \\ & & \mbox{} 
+ \dot{\Psi}^{lm}_{\rm ZM}(v,r_+)
\bar{\Psi}^{lm}_{\rm ZM}(v,r_+) \Bigr\rangle.
\label{1.8}
\end{eqnarray} 
Except for the substitution $(v \to u, r_+ \to \infty)$, these
formulae are identical to Eqs.~(\ref{A.26}) and (\ref{A.27}) which 
give the rates at which energy and angular momentum are transported to
future null infinity. Note that for a nonrotating black hole, the
first law of black-hole mechanics reduces to $(\kappa/8\pi) \Adot =
\Mdot$.

The flux formulae of Eqs.~(\ref{1.7}), (\ref{1.8}) were first
presented and used by Martel 
\cite{martel:04} 
in his numerical exploration of gravitational-wave processes
associated with the motion of a small-mass body in the field of a
Schwarzschild black hole. Although he arrived at the
correct results, the derivation of Eqs.~(\ref{1.7}) and (\ref{1.8})
presented by Martel is flawed, and the analysis presented in Sec.~VII
puts them on a firm footing. Martel's derivation incorporates both a
conceptual and a computational error, the latter compensating for the
former. Martel based his derivation of Eq.~(\ref{1.7}) and (\ref{1.8})
on Isaacson's effective stress-energy tensor for gravitational waves
\cite{isaacson:68a, isaacson:68b}, 
incorrectly assuming
that Isaacson's high-frequency description is always applicable near
the event horizon of a black hole. This story is related more fully in 
Sec.~VII C, and its proper telling requires the connection between
$\Mdot$, $\Jdot$ and the Isaacson stress-energy tensor established in
Sec.~VI C. The limitations of the high-frequency description become 
especially clear in view of this connection. 

\subsection{Small-hole/slow-motion approximation} 

A concrete evaluation of the flux formulae would typically require
the numerical integration of the Teukolsky equation, or the
Regge-Wheeler and Zerilli equations; an illustration is provided by
Martel's recent work
\cite{martel:04}. 
But in some circumstances it is possible to
solve these equations analytically, and to obtain approximate
expressions for $\Mdot$ and $\Jdot$. I carry out such calculations in  
Secs.~VIII and IX, in the context of a small-hole/slow-motion
approximation that I now describe. 

Consider a situation in which the black hole is immersed in an 
external universe whose radius of curvature ${\cal R}$ is such that
$M/{\cal R} \ll 1$. For example, suppose that the black hole is moving
on a circular orbit of radius $b$ in the gravitational field of
another body of mass $M_{\rm ext}$. Then ${\cal R}^{-1}$ is of the
order of the hole's angular velocity, and we have 
\[
\frac{M}{\cal R} \sim \frac{M}{M+M_{\rm ext}} V^3, \qquad
V = \sqrt{\frac{M + M_{\rm ext}}{b}}, 
\]
where $V$ is the hole's orbital velocity. One way to make this ratio
small is to let $M/M_{\rm ext} \ll 1$; then $M/{\cal R}$ will be small
irrespective of the magnitude of $V$. This is the {\it small-hole
approximation}, which allows the small black hole to move at
relativistic speeds in the strong gravitational field of the external
body. Another way is to let $V \ll 1$; then $M/{\cal R}$ will be small
for all mass ratios. This is the {\it slow-motion approximation},
which allows the slowly-moving black hole to have a mass comparable to
(or even much larger than) $M_{\rm ext}$. These two limiting
approximations are special cases of the fundamental requirement that 
$M/{\cal R}$ be small; I call this the small-hole/slow-motion
(SH/SM) approximation.    

When viewed on the large scale ${\cal R}$, the black hole occupies a   
very small region of the actual spacetime, and this region can be
idealized as a world line $\gamma$ in the external spacetime. Let
$u^\alpha$ be the (normalized) tangent vector to this world line, and 
call this the four-velocity of the black hole in the external
spacetime. Assume that the Ricci tensor of the external spacetime
vanishes on $\gamma$, so that the black hole's neighborhood will be
empty of matter. The curvature of the external spacetime in this
neighborhood is then described entirely by the Weyl tensor. This
can be decomposed into its electric and magnetic components
(see, for example, Ref.~\cite{thorne-hartle:85}), respectively   
\begin{equation}
{\cal E}_{\alpha\beta} = C_{\mu\alpha\nu\beta} u^\mu u^\nu 
\label{1.9}
\end{equation}
and
\begin{equation}
{\cal B}_{\alpha\beta} = \frac{1}{2} u^\mu 
\varepsilon_{\mu\alpha}^{\ \ \ \gamma\delta} 
C_{\gamma\delta\beta\nu} u^\nu,
\label{1.10}
\end{equation} 
where the Levi-Civita tensor $\varepsilon_{\mu\alpha\nu\beta}$ and 
the Weyl tensor $C_{\mu\alpha\nu\beta}$ are evaluated on the world 
line $\gamma$. The tensors ${\cal E}_{\alpha\beta}$ and 
${\cal B}_{\alpha\beta}$ are orthogonal to $u^\alpha$, and they are 
both symmetric and tracefree; they comprise all ten independent 
components of the Weyl tensor. These tensors represent the tidal 
gravitational fields that are supplied by the external universe, and 
these act on the black hole so as to produce a tidal distortion. This 
distortion, in turn, gives rise to a change of mass and angular
momentum that can be computed with the formalisms described in the
preceding subsections. 

In Sec.~VIII, I calculate $\Mdot$ and $\Jdot$ for a Schwarzschild 
black hole moving in an external universe, to leading order in a
SH/SM approximation. The results are 
\begin{equation} 
\Mdot = \frac{16M^6}{45} \bigl\langle \dot{\cal E}_{\alpha\beta} 
\dot{\cal E}^{\alpha\beta} + \dot{\cal B}_{\alpha\beta}
\dot{\cal B}^{\alpha\beta} \bigr\rangle 
\label{1.11}
\end{equation} 
and 
\begin{equation}
\Jdot = -\frac{32M^6}{45} u^\mu \varepsilon_{\mu\alpha\gamma\delta}
\bigl\langle \dot{\cal E}^\alpha_{\ \beta} {\cal E}^{\beta\gamma} 
+ \dot{\cal B}^\alpha_{\ \beta} {\cal B}^{\beta\gamma} \bigr\rangle
s^\delta,   
\label{1.12}
\end{equation}   
where $s^\alpha$ is a unit vector, orthogonal to $u^\alpha$, that
gives the direction of the vector $\langle \dot{J}^\alpha \rangle =
\Jdot s^\alpha$ (a more precise definition is found in Sec.~VIII F),
and  
$\dot{\cal E}_{\alpha\beta} \equiv {\cal E}_{\alpha\beta;\mu} u^\mu$, 
$\dot{\cal B}_{\alpha\beta} \equiv {\cal B}_{\alpha\beta;\mu} u^\mu$
are the proper-time derivative of the tidal gravitational fields. From
these expressions we infer that $\Mdot$ scales as $M^6/{\cal R}^6$,
while $\Jdot$ scales as $M^6/{\cal R}^5$. In Sec.~VIII G, I show that
the change in mass and angular momentum can be understood in terms of
a coupling between the tidal fields and the hole's induced mass and
current quadrupole moments, which are given by $M_{\alpha\beta} = 
\frac{32}{45} M^6 \dot{\cal E}_{\alpha\beta}$ and $J_{\alpha\beta} = 
\frac{8}{15} M^6 \dot{\cal B}_{\alpha\beta}$, respectively. As
illustrative examples, Eq.~(\ref{1.11}) and (\ref{1.12}) are evaluated
in two different limits in the case of circular binary motion: In
Sec.~VIII H, I calculate $\Mdot$ and $\Jdot$ for a slowly-moving
binary system consisting of bodies of comparable masses (one being
the black hole); and in Sec.~VIII I, I take the mass ratio to be small 
($M/M_{\rm ext} \ll 1$) but allow the black hole to move rapidly in
the strong gravitational field of the external body.     

In Sec.~IX, I calculate $\Mdot$ and $\Jdot$ for a Kerr black hole 
moving in an external universe. I again work to leading order in a
SH/SM approximation, but the statement of the approximation must now
be refined to $M/{\cal R} \ll \chi$, where $\chi \equiv a/M \equiv
J/M^2$ is the dimensionless rotational parameter of the black hole. 
The results, which were obtained previously by D'Eath
\cite{death:96}, 
are 
\begin{equation} 
\Mdot = O(M^5/{\cal R}^5) 
\label{1.13}
\end{equation} 
and 
\begin{eqnarray} 
\Jdot &=& -\frac{2}{45} M^5\chi \Bigl[ 8(1 + 3\chi^2) \langle E_1 
+ B_1 \rangle 
\nonumber \\ & & \mbox{} 
- 3(4 + 17\chi^2) \langle E_2 + B_2 \rangle 
\nonumber \\ & & \mbox{} 
+ 15\chi^2 \langle E_3 + B_3 \rangle \Bigr]. 
\label{1.14}
\end{eqnarray} 
where $E_1 = {\cal E}_{\alpha\beta} {\cal E}^{\alpha\beta}$,  
$E_2 = {\cal E}_{\alpha\beta} s^\beta {\cal E}^\alpha_{\ \gamma}
s^\gamma$, $E_3 = \bigl({\cal E}_{\alpha\beta} s^\alpha s^\beta
\bigr)^2$, and $B_1 = {\cal B}_{\alpha\beta} {\cal B}^{\alpha\beta}$,   
$B_2 = {\cal B}_{\alpha\beta} s^\beta {\cal B}^\alpha_{\ \gamma}
s^\gamma$, $B_3 = \bigl({\cal B}_{\alpha\beta} s^\alpha s^\beta
\bigr)^2$. The leading-order calculations carried out in Sec.~IX are
not sufficient to determine $\Mdot$, but they indicate that $\Jdot$
scales as $M^5/{\cal R}^4$. This result can also be understood in
terms of a coupling between the tidal fields and the hole's induced
mass and current quadrupole moments; here, as I show in Sec.~IX E, the   
relationship between $M_{\alpha\beta}$ and ${\cal E}_{\alpha\beta}$,
and the relationship between $J_{\alpha\beta}$ and 
${\cal B}_{\alpha\beta}$, do not involve a time derivative (as they do 
in the case of a Schwarzschild black hole). Three illustrative
applications of Eq.~(\ref{1.14}) are worked out: In Sec.~IX G, I
examine a Kerr black hole in circular motion in a slowly-moving binary
system; in Sec.~IX H, I consider instead the case of a small hole in
relativistic circular motion; and in Sec.~IX I, the Kerr black hole is
placed in a static tidal gravitational field. 

The main results of Sec.~VIII, Eqs.~(\ref{1.11}) and (\ref{1.12}),
hold to leading order in $M/{\cal R} \ll 1$, and they reveal that for
a Schwarzschild black hole, $\Mdot = O(M^6/{\cal R}^6)$ and $\Jdot =
O(M^6/{\cal R}^5)$. On the other hand, the main results of Sec.~IX,
Eqs.~(\ref{1.13}) and (\ref{1.14}), hold to leading order in 
$M/{\cal R} \ll \chi$, and they reveal that for a Kerr black hole,
$\Mdot = O(M^5/{\cal R}^5)$ and $\Jdot = O(M^5/{\cal R}^4)$. The
scalings are thus very different, and the condition $M/{\cal R} \ll
\chi$ implies that the Schwarzschild results cannot straightforwardly
be obtained from the Kerr results in a limit $\chi \to 0$. These
scalings indicate that when a rotating and a nonrotating black hole
are placed in identical environments, the rotating hole will absorb
larger quantities of energy and angular momentum. The agent
responsible for this enhanced absorption is evidently the hole's
rotation, and some insight into this matter is offered in Sec.~IX F. 

This concludes the summary of the work presented in this article.    

\subsection{Organization of the paper} 

The rest of the paper contains the derivations of the results
summarized previously, as well as additional results and details. 
I begin in Sec.~II with a description of the event horizon of an
unperturbed Kerr black hole, and I derive a number of results that
will be used in later sections of the paper. A description of a
general evolving horizon is presented in Sec.~III. This discussion is
specialized, in Sec.~IV, to the case of a perturbation of the Kerr
horizon, and I derive the flux formulae of
Eqs.~(\ref{4.22})--(\ref{4.24}). 

These formulae are translated into a practical curvature formalism in
Sec.~V; this was summarized in Sec.~I C. In Sec.~VI they are
translated into a metric formalism that applies to both rotating and
nonrotating black holes; in that section I introduce ``preferred'' and
``radiation'' gauges for the metric perturbations, and I establish a
connection between the fluxes and Isaacson's effective stress-energy
tensor for gravitational waves
\cite{isaacson:68a, isaacson:68b}. 
The metric formalism is specialized to a Schwarzschild black hole in
Sec.~VII; this was summarized in Sec.~I D. 

The small-hole/slow-motion approximation is worked out in the last two 
sections of the paper, first for a Schwarzschild black hole immersed
in an external universe (Sec.~VIII), and next for a Kerr black hole
(Sec.~IX). These results were summarized in Sec.~I E. 

The Appendix contains a brief summary of the theory of metric
perturbations of a Schwarzschild spacetime. The material contained in
this Appendix is well known, but it is convenient to record the main
results there because they are referred to on many occasions in the 
article's main body.  

Throughout this work I use geometrized units in which $c = G = 1$, and
I adopt the conventions of Misner, Thorne, and Wheeler \cite{MTW:73}. 

\section{Kinematics of the Kerr horizon} 

To prepare the way for the discussion of dynamical event horizons in
the next two sections, in this section I cover the kinematics of a 
stationary event horizon described by the Kerr metric. I shall
introduce a parametric description of the horizon's null generators,
and derive from this an intrinsic description of the horizon. Part of
this discussion will be devoted to the construction of null tetrads on
the horizon, and a description of the (well-known) algebraic structure
of the Weyl tensor. 

\subsection{Kerr metric} 

Throughout this work the Kerr metric will be written in terms of
ingoing Kerr coordinates $(v,r,\theta,\psi)$, so that its form will be
regular on the event horizon. It is given by (see, for example, Box
33.2 of Ref.~\cite{MTW:73}, or Sec.~5.3 of Ref.~\cite{poisson:04c})   
\begin{eqnarray}
ds^2 &=& -\frac{\Delta - a^2 \sin^2\theta}{\rho^2}\, dv^2 + 2\, dvdr  
- \frac{4 M a r \sin^2\theta}{\rho^2}\, dv d\psi  
\nonumber \\ & & \mbox{} 
- 2 a \sin^2\theta\, dr d\psi + \frac{\Sigma \sin^2\theta}{\rho^2}\,
d\psi^2 + \rho^2\, d\theta^2, 
\label{2.1}
\end{eqnarray}
where $M$ is the black-hole mass, $J \equiv Ma$ its angular momentum,   
$\rho^2 = r^2 + a^2 \cos^2\theta$, $\Delta = r^2 - 2Mr + a^2$, 
and $\Sigma = (r^2+a^2)^2 - a^2 \Delta \sin^2\theta$. The
transformation from the more usual Boyer-Lindquist coordinates
$(t_{\rm BL}, r_{\rm BL}, \theta_{\rm BL}, \phi_{\rm BL})$ is
given by $v = t_{\rm BL} + \int (r^2 + a^2) \Delta^{-1}\, dr$,
$r=r_{\rm BL}$, $\theta = \theta_{\rm BL}$, and $\psi = \phi_{\rm BL} 
+ a \int \Delta^{-1}\, dr$; the Kerr coordinates are sometimes denoted
$(\tilde{V},r,\theta,\tilde{\phi})$, as is done in
Ref.~\cite{MTW:73}. 
The event horizon is situated at the largest root of $\Delta$, at  
$r = r_+ \equiv M + \sqrt{M^2 - a^2}$.   

The Kerr spacetime admits the Killing vectors $t^\alpha = \partial
x^\alpha/\partial v$ and $\phi^\alpha = \partial x^\alpha/\partial
\psi$. The vector 
\begin{equation} 
k^\alpha = t^\alpha + \Omega_{\rm H} \phi^\alpha, 
\label{2.2}
\end{equation}
with 
\begin{equation}
\Omega_{\rm H} = \frac{a}{r_+^2 + a^2} = \frac{a}{2M r_+}, 
\label{2.3}
\end{equation}
is also a Killing vector, and it is null on the event horizon; it is  
tangent to the horizon's null generators. The quantity 
$\Omega_{\rm H}$ is the angular velocity of the black hole. 
 
\subsection{Parametric description of the horizon} 

We wish to introduce a system of coordinates $(v,\theta^A)$ on the 
horizon, adopting $v$ as a parameter on the generators, and $\theta^A$
($A = 2, 3$) as generator labels that stay constant as the generators 
move. Because $k^\alpha = (1,0,0,\Omega_{\rm H})$ in the spacetime
coordinates $(v,r,\theta,\psi)$, we have that $\theta$ is constant on
each generator, and it can therefore be chosen as one of the comoving
coordinates. On the other hand, $d\psi/dv = \Omega_{\rm H}$ and $\psi$
increases linearly as the generators wrap around the event
horizon; a suitable choice of comoving coordinate is therefore $\phi =
\psi - \Omega_{\rm H} v$, which stays constant. Our horizon
coordinates are therefore 
\begin{equation}
(v,\theta^A) = (v,\theta,\phi = \psi - \Omega_{\rm H} v). 
\label{2.4}
\end{equation} 
It is important not to confuse the horizon coordinate $\phi$ with the
Boyer-Lindquist coordinate $\phi_{\rm BL}$; these are not equal. 

The horizon generators can now be described by parametric equations of
the form $x^\alpha = z^\alpha(v,\theta^A)$, in which $z^\alpha$
gives the spacetime-coordinate positions of the generators in terms
of the intrinsic horizon coordinates. Explicitly, the parametric
description is $v=v$, $r=r_+$, $\theta = \theta$, and $\psi = \phi +
\Omega_{\rm H} v$. The vectors 
\begin{equation} 
k^\alpha = \frac{\partial z^\alpha}{\partial v}, \qquad 
e^\alpha_A = \frac{\partial z^\alpha}{\partial \theta^A} 
\label{2.5}
\end{equation}
are tangent to the horizon; $k^\alpha$ is tangent to each generator 
while $e^\alpha_A$ points in the directions transverse to the
generators. In the spacetime coordinates $(v,r,\theta,\psi)$ we have
$k^\alpha = (1,0,0,\Omega_{\rm H})$ as before, $e^\alpha_\theta =
(0,0,1,0)$, and $e^\alpha_\phi = (0,0,0,1)$. Because the coordinates
$\theta^A$ are comoving, the transverse vectors $e^\alpha_A$ are Lie
transported along the generators, and they therefore satisfy 
\begin{equation} 
e^\alpha_{A;\beta} k^\beta = k^\alpha_{\ ;\beta} e^\beta_A. 
\label{2.6}
\end{equation} 
They are also Lie transported along one another, so that
$e^\alpha_{A;\beta} e^\beta_B = e^\alpha_{B;\beta} e^\beta_A$. 

The basis vectors also satisfy 
\begin{equation}
k_\alpha k^\alpha = 0 = k_\alpha e^\alpha_A.
\label{2.7}
\end{equation} 
The only nonvanishing inner products are 
\begin{equation}
\gamma_{AB} = g_{\alpha\beta} e^\alpha_A e^\beta_B, 
\label{2.8}
\end{equation}
and these form the components of the induced metric on the horizon. To
see this, deduce from Eq.~(\ref{2.5}) that a displacement on the
horizon is described by $dx^\alpha = k^\alpha\, dv + e^\alpha_A\,
d\theta^A$ and calculate $ds^2 = g_{\alpha\beta} dx^\alpha dx^\beta$
for this displacement; use of Eqs.~(\ref{2.7}) and (\ref{2.8}) returns
$ds^2 = \gamma_{AB}\, d\theta^A d\theta^B$, with the interpretation 
that $\gamma_{AB}$ is indeed the induced metric. Notice that the 
horizon metric is degenerate, and explicitly two-dimensional in the
comoving coordinates. The nonvanishing components of the horizon
metric are $\gamma_{\theta\theta} = r_+^2 + a^2\cos^2\theta$,
$\gamma_{\phi\phi} = (r_+^2+a^2)^2
\sin^2\theta/(r_+^2+a^2\cos^2\theta)$, and 
\begin{equation} 
\sqrt{\gamma} = (r_+^2 + a^2)\sin\theta 
\label{2.9}
\end{equation}
is the square root of the metric determinant. 

The vector basis on the horizon can be completed with another null
vector $N^\alpha$ that satisfies 
\begin{equation} 
N_\alpha N^\alpha = 0 = N_\alpha e^\alpha_A, \qquad 
N_\alpha k^\alpha = -1. 
\label{2.10}
\end{equation}
These conditions determine the vector uniquely, and we find 
\begin{equation}
N_\alpha\, dx^\alpha = -dv + \frac{a^2 \sin^2\theta}{2(r_+^2+a^2)}\,
dr. 
\label{2.11}
\end{equation} 
The four basis vectors give us completeness relations for the inverse 
metric evaluated on the event horizon, 
\begin{equation}
g^{\alpha\beta} = -k^\alpha N^\beta - N^\alpha k^\beta + \gamma^{AB}
e^\alpha_A e^\beta_B, 
\label{2.12}
\end{equation}
where $\gamma^{AB}$ is the inverse of $\gamma_{AB}$. In the sequel we
will use the horizon metric and its inverse to lower and raise
upper-case Latin indices. We will also introduce a two-dimensional
connection $\Gamma^A_{BC}$ compatible with $\gamma_{AB}$, and denote
covariant differentiation in this connection with a vertical stroke;
for example, $\gamma_{AB | C} \equiv 0$.  

\subsection{Horizon connections} 

The tangential derivatives of the basis vectors are given by 
\begin{eqnarray} 
k^\alpha_{\ ;\beta} k^\beta &=& \kappa k^\alpha, 
\label{2.13} \\
k^\alpha_{\ ;\beta} e^\beta_A &=& \omega_A k^\alpha 
= e^\alpha_{A ;\beta} k^\beta, 
\label{2.14} \\
e^\alpha_{A ;\beta} e^\beta_B &=& p_{AB} k^\alpha + \Gamma^C_{AB} 
e^\alpha_C = e^\alpha_{B ;\beta} e^\beta_A, 
\label{2.15} 
\end{eqnarray} 
where $\kappa$, $\omega_A$, $p_{AB}$, and $\Gamma^C_{AB}$ are the
horizon connections. The surface gravity $\kappa = 
-N_\alpha k^\alpha_{\ ;\beta} k^\beta$ of a Kerr black hole is given
by 
\begin{equation} 
\kappa = \frac{r_+ - M}{r_+^2 + a^2} = \frac{\sqrt{M^2-a^2}}{2M r_+}, 
\label{2.16}
\end{equation} 
and Eq.~(\ref{2.13}) states that the vector $k^\alpha$ satisfies the
geodesic equation, but that the generator parameter $v$ is not an
affine parameter. Explicit expressions for $\omega_{A}$ and $p_{AB}$
will not be needed; the two-dimensional connection $\Gamma^C_{AB}$ can
easily be computed from $\gamma_{AB}$. 

The 2-vector $\phi^A = (0,1)$ is a Killing vector of the horizon's
intrinsic geometry, and it therefore satisfies Killing's equation, 
$\phi_{(A|B)} = 0$. This vector is related to the spacetime Killing
vector $\phi^\alpha$ by the relation $\phi^\alpha = \phi^A
e^\alpha_A$. The 2-tensor  
\begin{equation} 
c_{AB} \equiv -\phi_{A|B} = -\phi_{\alpha;\beta} e^\alpha_A e^\beta_B 
\label{2.17}
\end{equation}
will be needed in Sec.~VI of the paper. This tensor is antisymmetric
by virtue of Killing's equation; its only nonvanishing components are 
$c_{\theta\phi} =
(r_+^2+a^2)^3\sin\theta\cos\theta/(r_+^2+a^2\cos^2\theta)^2 = 
-c_{\phi\theta}$. 

The vector $k^\alpha$ introduced in Eq.~(\ref{2.5}) is defined on the 
horizon only, but Eq.~(\ref{2.2}) provides an extension away from the
horizon. The extended vector field is null on the horizon only, but it
is everywhere a Killing vector; it satisfies 
\begin{equation}
k_{\alpha;\beta} = -\kappa( k_\alpha N_\beta - N_\alpha k_\beta ) +
k_\alpha \omega_\beta - \omega_\alpha k_\beta 
\label{2.18}
\end{equation}
on the horizon, where $\omega^\alpha \equiv \omega^A e^\alpha_A$. 

\subsection{Null tetrads} 

The transverse vectors $e^\alpha_A$ can be combined into complex
vectors $e^\alpha = e^A e^\alpha_A$ that satisfy 
\begin{equation}
e_\alpha e^\alpha = 0 = \bar{e}_\alpha \bar{e}^\alpha, \qquad
e_\alpha \bar{e}^\alpha = 1, 
\label{2.19}  
\end{equation}
with an overbar indicating complex conjugation. In terms of the 
complex coefficients $e^A$, these relations read 
\begin{equation}
\gamma_{AB} e^A e^B = 0 = \gamma_{AB} \bar{e}^A \bar{e}^B, \qquad 
\gamma_{AB} e^A \bar{e}^B = 1, 
\label{2.20}
\end{equation}
and these produce the completeness relations $\gamma^{AB} = e^A
\bar{e}^B + \bar{e}^A e^B$. Substituting this into Eq.~(\ref{2.12})
yields 
\begin{equation}
g^{\alpha\beta} = -k^\alpha N^\beta - N^\alpha k^\beta + e^\alpha
\bar{e}^\beta + \bar{e}^\alpha e^\beta. 
\label{2.21}
\end{equation}
A particular (and traditional) choice of coefficients $e^A$ that
achieves these properties is 
\begin{equation}
e^\theta = \frac{1}{\sqrt{2}} \frac{r_+-i a \cos\theta}{r_+^2 
+ a^2\cos^2\theta}, \qquad 
e^\phi = \frac{i}{\sqrt{2}} \frac{r_+-i a \cos\theta}{(r_+^2 
+ a^2)\sin\theta}. 
\label{2.22}
\end{equation} 
Note that the inversion formula is $e^\alpha_A = \bar{e}_A
e^\alpha + e_A \bar{e}^\alpha$, where the upper-case Latin 
index was lowered with the horizon metric $\gamma_{AB}$. 

The basis $(k^\alpha,N^\alpha,e^\alpha,\bar{e}^\alpha)$ is a null
tetrad on the horizon, and it can be used to decompose various
tensors, as is customary in the Newman-Penrose formalism (see, for
example, the presentation of Ref.~\cite{chandrasekhar:83}). This 
tetrad, however, is not adapted to the algebraic structure of
$C_{\alpha\gamma\beta\delta}$, the Weyl tensor of the Kerr spacetime. 
With this tetrad we would find that the Weyl scalars $\psi_0$ and 
$\psi_1$ vanish, but that $\psi_3$ and $\psi_4$ do not. (These
quantities will be introduced below.) This is remedied by a null
rotation (a rotation of class I in the language of Chandrasekhar
\cite{chandrasekhar:83}) 
to a new tetrad $(k^\alpha, n^\alpha, m^\alpha, \bar{m}^\alpha)$ 
given by   
\begin{equation}  
n^\alpha = N^\alpha + |A|^2 k^\alpha + \bar{A} e^\alpha +
A \bar{e}^\alpha, \qquad 
m^\alpha = e^\alpha + A k^\alpha, 
\label{2.23}
\end{equation} 
where 
\begin{equation}
A = \frac{i}{\sqrt{2}} a\sin\theta \frac{r_+-ia\cos\theta}{r_+^2
+ a^2\cos^2\theta}. 
\label{2.24}
\end{equation} 
In the spacetime coordinates $(v,r,\theta,\psi)$, the components of
the new transverse vectors are  
\begin{equation}  
m^\alpha = \frac{1}{\sqrt{2}} \frac{r_+-ia\cos\theta}{r_+^2
+ a^2\cos^2\theta} \biggl( ia\sin\theta, 0, 1, \frac{i}{\sin\theta}
\biggr). 
\label{2.25}
\end{equation} 
Notice that this rotation leaves the vector $k^\alpha$ unchanged. 

The tetrad $(k^\alpha, n^\alpha, m^\alpha, \bar{m}^\alpha)$ introduced  
here is the {\it Hartle-Hawking null tetrad} of the Kerr spacetime 
\cite{hawking-hartle:72}
(as it was defined by Teukolsky 
\cite{teukolsky:73}
and Teukolsky and Press 
\cite{teukolsky-press:74}) 
restricted to the event horizon. The Hartle-Hawking tetrad is related
to the more standard {\it Kinnersley null tetrad} 
\cite{kinnersley:69} 
by a rescaling of the vectors $k^\alpha$ and $n^\alpha$ (a rotation of
class III in the language of Chandrasekhar
\cite{chandrasekhar:83}):  
\begin{equation} 
k^\alpha({\rm HH}) = \frac{\Delta}{2(r^2+a^2)} k^\alpha({\rm K})
\label{2.26}
\end{equation}
and 
\begin{equation} 
n^\alpha({\rm HH}) = \frac{2(r^2+a^2)}{\Delta} n^\alpha({\rm K})
\label{2.27}. 
\end{equation}
Notice that while the Hartle-Hawking tetrad is defined globally in the 
Kerr spacetime, the tetrad $(k^\alpha, n^\alpha, m^\alpha,
\bar{m}^\alpha)$ is defined on the horizon only; the extension of
$k^\alpha$ to $k^\alpha(\rm HH)$ is different from the extension of 
Eq.~(\ref{2.2}). Notice also that whereas the Hartle-Hawking tetrad is
well behaved on the event horizon, the Kinnersley tetrad is not.

In the tetrad $(k^\alpha, n^\alpha, m^\alpha, \bar{m}^\alpha)$ the
horizon Weyl scalars are defined by (see, for example,  
Ref.~\cite{chandrasekhar:83}) 
\begin{eqnarray} 
\psi_0 &=& -C_{\alpha\gamma\beta\delta} k^\alpha m^\gamma k^\beta
m^\delta, 
\label{2.28} \\ 
\psi_1 &=& -C_{\alpha\gamma\beta\delta} k^\alpha n^\gamma k^\beta
m^\delta, 
\label{2.29} \\ 
\psi_2 &=& -C_{\alpha\gamma\beta\delta} k^\alpha m^\gamma
\bar{m}^\beta n^\delta, 
\label{2.30} \\ 
\psi_3 &=& -C_{\alpha\gamma\beta\delta} k^\alpha n^\gamma
\bar{m}^\beta n^\delta, 
\label{2.31} \\ 
\psi_4 &=& -C_{\alpha\gamma\beta\delta} n^\alpha \bar{m}^\gamma
n^\beta \bar{m}^\delta, 
\label{2.32}
\end{eqnarray}
where $C_{\alpha\gamma\beta\delta}$ is the Weyl tensor of the Kerr
spacetime evaluated on the event horizon. We have that $\psi_0 =
\psi_1 = \psi_3 = \psi_4 = 0$, a property that is true globally with 
the Hartle-Hawking or Kinnersley tetrads, and 
\begin{equation} 
\psi_2 = \frac{M r_+ (r_+^2 - 3a^2\cos^2\theta)}{(r_+^2 
+ a^2\cos^2\theta)^3}  
+ \frac{iMa\cos\theta(3r_+^2 - a^2\cos^2\theta)}{(r_+^2 
+ a^2\cos^2\theta)^3}
\label{2.33}
\end{equation}
is the only nonvanishing horizon Weyl scalar. 

\subsection{Weyl identities}

For future reference I record here a number of identities satisfied 
by the Weyl tensor of the Kerr spacetime: 
\begin{eqnarray} 
C_{\alpha\gamma\beta\delta} k^\gamma e^\beta_A k^\delta &=& 0, 
\label{2.34} \\ 
C_{\alpha\gamma\beta\delta} \bigl( e^\alpha_A e^\beta_B + e^\alpha_B
e^\beta_A \bigr) k^\gamma &=& -2 \mbox{Re}(\psi_2) \gamma_{AB}
k_\delta, \qquad 
\label{2.35} \\ 
C_{\alpha\gamma\beta\delta} k^\gamma k^\delta &=& -2\mbox{Re}(\psi_2)
k_\alpha k_\beta.
\label{2.36}
\end{eqnarray} 
As a consequence of Eqs.~(\ref{2.34}), (\ref{2.35}), and (\ref{2.23})
we also obtain  
\begin{equation} 
C_{\alpha\gamma\beta\delta} k^\gamma m^\beta k^\delta = 0 
= C_{\alpha\gamma\beta\delta} m^\alpha m^\beta k^\gamma. 
\label{2.37}
\end{equation} 
These identities are formulated on the horizon only.   

\section{Dynamics of an evolving horizon} 

In this section I generalize the preceding discussion to a
nonstationary event horizon. The presentation is patterned after 
Price and Thorne 
\cite{price-thorne:86} 
and Chapter VI of the book by Thorne, Price, and Macdonald  
\cite{thorne-etal:86}. 

\subsection{Comoving coordinates and vector basis} 

As in Sec.~II we take $(v,\theta^A)$ as our system of intrinsic
coordinates on the horizon, with $v$ now promoted to an arbitrary
parameter on the null generators, and $\theta^A$ still denoting
constant generator labels. The horizon is still described by the
parametric equations $x^\alpha = z^\alpha(v,\theta^A)$, but the
coordinate positions of the dynamical horizon may be displaced with
respect to those of a stationary Kerr horizon.

The vectors $k^\alpha = \partial z^\alpha/\partial v$ and $e^\alpha_A
= \partial z^\alpha/\partial \theta^A$ form a partial basis on the
horizon; as before $k^\alpha$ is tangent to the generators,
$e^\alpha_A$ is transverse to them, and $k_\alpha k^\alpha = k_\alpha
e^\alpha_A = 0$. The nonvanishing inner products 
\begin{equation} 
\gamma_{AB}(v,\theta^A) = g_{\alpha\beta} e^\alpha_A e^\beta_B
\label{3.1}
\end{equation} 
give the components of the induced metric on the horizon. The basis is
completed by another null vector $N^\alpha$, which is orthogonal to
$e^\alpha_A$ and normalized by the condition $N_\alpha k^\alpha =
-1$. The completeness relations of Eq.~(\ref{2.12}) still apply. 

The vectors $k^\alpha$ and $e^\alpha_A$ are all Lie transported along
one another, so that $e^\alpha_{A;\beta} k^\beta = 
k^\alpha_{\ ; \beta} e^\beta_A$ and $e^\alpha_{A;\beta} e^\beta_B =  
e^\alpha_{B;\beta} e^\beta_A$.  

\subsection{Generator kinematics} 

The tangent vector $k^\alpha$ satisfies the geodesic equation in its 
generalized form 
\begin{equation} 
k^\alpha_{\ ;\beta} k^\beta = \kappa k^\alpha, 
\label{3.2}
\end{equation}
where $\kappa$ is the evolving surface gravity of the event horizon,
defined with respect to our choice of parameterization $v$. 

The transverse derivatives of the tangent vector can be decomposed as  
\begin{equation}
k^\alpha_{\ ;\beta} e^\beta_A = \omega_A k^\alpha + B_A^{\ B}
e^\alpha_B = e^\alpha_{A;\beta} k^\beta, 
\label{3.3}
\end{equation}
for some 2-vector $\omega_A(v,\theta^A)$ and 2-tensor
$B_{AB}(v,\theta^A)$; this generalizes Eq.~(\ref{2.14}). It is easy to
show that the right-hand side of Eq.~(\ref{3.3}) cannot include a term 
proportional to $N^\alpha$; this follows from the fact that $k^\alpha$
is a null vector field. And it can be established that $B_{AB}$ is a
symmetric tensor, because the congruence of null geodesics to which
$k^\alpha$ is tangent is necessarily hypersurface orthogonal (see, 
for example, Sec.~9.2 of Ref.~\cite{wald:84}, or Sec.~2.4 of
Ref.~\cite{poisson:04c}). The
tensor $B_{AB} = k_{\alpha;\beta} e^\alpha_A e^\beta_B$ can
be decomposed into its irreducible parts, 
\begin{equation} 
B_{AB} = \frac{1}{2} \Theta \gamma_{AB} + \sigma_{AB}, 
\label{3.4}
\end{equation}
thereby defining the {\it expansion scalar} $\Theta = \gamma^{AB}
B_{AB}$ and the {\it shear tensor} $\sigma_{AB} = B_{AB} - \frac{1}{2}
\Theta \gamma_{AB}$. Notice that the expansion is the trace of
$B_{AB}$, while the shear is the tracefree part of this tensor. 

\subsection{Generator dynamics} 

Evolution equations can easily be derived for $\gamma_{AB}$, $\Theta$,
and $\sigma_{AB}$, and these will form the basis of the discussion of
perturbed event horizons in the next section. 

Starting with the identity $\partial \gamma_{AB}/\partial v =
(g_{\alpha\beta} e^\alpha_A e^\alpha_B)_{;\gamma} k^\gamma$ and using
Eqs.~(\ref{3.3}) and (\ref{3.4}), we quickly arrive at an evolution
equation for the horizon metric, 
\begin{equation}
\frac{\partial \gamma_{AB}}{\partial v} = \Theta \gamma_{AB} +
2\sigma_{AB}. 
\label{3.5}
\end{equation}
From this it follows that $\partial \gamma^{AB} /\partial v = -\Theta
\gamma^{AB} - 2\sigma^{AB}$ and 
\begin{equation} 
\Theta = \frac{1}{\sqrt{\gamma}} \frac{\partial
\sqrt{\gamma}}{\partial v}, 
\label{3.6}
\end{equation}
where $\gamma$ is the metric determinant. 

To derive evolution equations for the expansion and shear we follow
the usual route that leads to Raychaudhuri's equation (see, 
for example, Sec.~9.2 of Ref.~\cite{wald:84}, or Sec.~2.4 of 
Ref.~\cite{poisson:04c}). Starting with 
the identity $\partial B_{AB}/\partial v =
(k_{\alpha;\beta} e^\alpha_A e^\alpha_B)_{;\gamma} k^\gamma$ and using
Eqs.~(\ref{3.2}), (\ref{3.3}), as well as Ricci's identity, we arrive
at $\partial B_{AB}/\partial v = \kappa B_{AB} + B_A^{\ C} B_{C B} -
R_{\alpha\gamma\beta\delta} e^\alpha_A k^\gamma e^\beta_B
k^\delta$. Taking the trace of this equation, using the fact that a
symmetric-tracefree tensor automatically satisfies $\sigma^A_{\ C}  
\sigma^C_{\ B} = \frac{1}{2} (\sigma_{CD} \sigma^{CD}) 
\delta^A_{\ B}$, produces Raychaudhuri's equation, 
\begin{equation} 
\frac{\partial \Theta}{\partial v} = \kappa \Theta - \frac{1}{2}
\Theta^2 - \sigma_{AB} \sigma^{AB} - 8\pi \rho, 
\label{3.7}
\end{equation}
where $\rho \equiv (R_{\alpha\beta}/8\pi) k^\alpha k^\beta =
T_{\alpha\beta} k^\alpha k^\beta$ after using the Einstein field 
equations. The tracefree part of the equation reduces to 
\begin{equation} 
\frac{\partial \sigma^A_{\ B}}{\partial v} = (\kappa - \Theta)
\sigma^A_{\ B} - C^A_{\ B}, 
\label{3.8}
\end{equation}
where
\begin{equation}
C_{AB} \equiv C_{\alpha\gamma\beta\delta} e^\alpha_A k^\gamma
e^\beta_B k^\delta 
\label{3.9}
\end{equation}
are tangential components of the Weyl tensor. In these equations
all upper-case Latin indices are lowered and raised with $\gamma_{AB}$
and $\gamma^{AB}$, respectively; the horizon metric evolves according
to Eq.~(\ref{3.5}). 

The area of any cross section $v = \mbox{constant}$ of the event
horizon is given by $A(v) = \oint \sqrt{\gamma}\, d^2\theta$. Assuming
that the number of generators stays constant as the horizon evolves 
(that is, assuming that no new generator joins the horizon at a
caustic), a change of area occurs when $\gamma$, the metric
determinant, varies with time. Equation (\ref{3.6}) yields 
\begin{equation} 
\frac{dA}{dv} = \oint \Theta\, dS, 
\label{3.10}
\end{equation} 
where $dS = \sqrt{\gamma}\, d^2\theta$ is an element of surface area
on the horizon cross sections.  

The equations derived in this section are all exact, and they apply to
an event horizon that evolves dynamically. [The assumption made in the
derivation of Eq.~(\ref{3.10}), that the number of generators must
be conserved during the horizon's evolution, represents a serious
restriction. In the perturbative context to be described in the next
paragraph, however, this limitation is lifted because the formation 
of a caustic is necessarily associated with a large, nonperturbative
value of $\Theta$.] The choice of parameter $v$ and generator labels
$\theta^A$ is completely arbitrary, and the quantities $\kappa$,
$\gamma_{AB}$, $\Theta$, and $\sigma_{AB}$ all refer to this
choice. For a stationary Kerr black hole, $v$ is chosen so that
$k^\alpha$ is given by Eq.~(\ref{2.2}), and the generator labels of
Eq.~(\ref{2.4}) are adopted. This means that $\kappa$ is given by
Eq.~(\ref{2.16}), $\gamma_{AB}$ by the expressions listed near
Eq.~(\ref{2.9}), and that $\Theta = 0 = \sigma_{AB}$, as can be seen
by comparing Eq.~(\ref{2.14}) with Eq.~(\ref{3.3}). Equations
(\ref{3.7}) and (\ref{3.8}) are then consistent if $\rho = 0$, which
follows because the Kerr metric is a vacuum solution to the Einstein
field equations, and $C_{AB} = 0$, which follows from Eq.~(\ref{2.35}). 

In the next section we will apply these equations to situations in
which the horizon is very close to being stationary, so that it can be
described as a slightly perturbed version of the Kerr horizon. The
horizon coordinates $(v,\theta^A)$ will then be chosen to be ``close
to'' the Kerr coordinates, and we will see that the ambiguities
associated with this choice never explicitly enter the
discussion; our final expressions will be gauge invariant. This
implies that the vectors $k^\alpha$, $N^\alpha$, $e^\alpha_A$ on the
perturbed horizon will be perturbed versions of the Kerr basis, and
that all derived quantities, such as $\kappa$, $\gamma_{AB}$,
$\Theta$, and $\sigma_{AB}$, will be perturbed versions of the
corresponding Kerr quantities.    

\section{Perturbed horizon} 

I now specialize the formalism of the preceding section to event
horizons that are slightly nonstationary, those that can be considered
to be perturbed versions of the Kerr horizon. I shall assume that the
perturbation is caused entirely by gravitational radiation, and no
matter will be allowed to cross the event horizon. The perturbation
formalism described here is adapted from Price and Thorne
\cite{price-thorne:86} 
and Chapter VI of the book by Thorne, Price, and
Macdonald
\cite{thorne-etal:86}; 
these methods go back to the pioneering work of Hawking and Hartle
\cite{hawking-hartle:72}.  

\subsection{Perturbation equations} 

Changing our notation with respect to the previous section, the
perturbed values for the horizon metric, surface gravity, expansion,
shear, and Weyl tensor will now be denoted $\hat{\gamma}_{AB}$,
$\hat{\kappa}$, $\hat{\Theta}$, $\hat{\sigma}_{AB}$, and
$\hat{C}_{AB}$, respectively; these quantities were all introduced in
Sec.~III. The unperturbed (Kerr) values will be denoted without a
decorating caret; for example $\Theta = 0$ is the background expansion
scalar, $\sigma_{AB} = 0$ the background shear tensor, and $C_{AB}
= 0$ the background Weyl tensor. The only nonvanishing background
quantities are the metric $\gamma_{AB}$ and surface gravity $\kappa$;
these were introduced in Sec.~II. 

The horizon perturbation is driven by the Weyl tensor $\hat{C}_{AB}$,
which we imagine to be a quantity of the first order in an expansion
parameter $\lambda$; we write this as 
\begin{equation}
\hat{C}^A_{\ B} = \lambda C^{\ A}_{1\ B} + O(\lambda^2). 
\label{4.1}
\end{equation} 
At the end of the calculation we will set $\lambda \equiv 1$ by 
absorbing it into the definition of the perturbations. Equation
(\ref{3.8}) indicates that the Weyl tensor drives a first-order
perturbation in the shear, and we have  
\begin{equation} 
\hat{\sigma}^A_{\ B} = \lambda \sigma^{\ A}_{1\ B} + O(\lambda^2). 
\label{4.2}
\end{equation} 
Equation (\ref{3.7}), on the other hand, shows that it is the square
of the shear tensor that is driving a perturbation in the expansion
(recall that we have set $\rho = 0$), and we must therefore have 
\begin{equation}
\hat{\Theta} = \lambda^2 \Theta_2 + O(\lambda^3). 
\label{4.3}
\end{equation} 
(A more careful treatment that incorporates a first-order term would
eventually lead to $\Theta_1 = 0$ and therefore back to this
assertion.) Finally, Eq.~(\ref{3.5}) shows that the shear produces a
first-order perturbation in the metric, 
\begin{equation} 
\hat{\gamma}_{AB} = \gamma_{AB} + \lambda \gamma^1_{AB} +
O(\lambda^2), 
\label{4.4}
\end{equation} 
and these considerations lead to the statement that the perturbed
surface gravity will differ from its Kerr value by a first-order
quantity: $\hat{\kappa} = \kappa + O(\lambda)$. 

Substituting the expansions of Eqs.~(\ref{4.1})--(\ref{4.4}) into
Eqs.~(\ref{3.5}), (\ref{3.7}), and (\ref{3.8}) gives us the following
set of perturbation equations: 
\begin{eqnarray} 
\frac{\partial \gamma^1_{AB}}{\partial v} &=& 2\sigma^1_{AB}, 
\label{4.5} \\ 
\frac{\partial \Theta_2}{\partial v} &=& \kappa \Theta_2 
-\sigma^1_{AB} \sigma^{AB}_1, 
\label{4.6} \\ 
\frac{\partial \sigma^1_{AB}}{\partial v} &=& \kappa \sigma^1_{AB} 
-C^1_{AB}.  
\label{4.7} 
\end{eqnarray} 
In these equations, $\kappa$ is the surface gravity of the Kerr black
hole, and the generator parameter $v$ can be identified with the
advanced-time coordinate of the Kerr spacetime (as introduced in
Sec.~II); the first-order deviations of these quantities with respect
to the Kerr values do not enter the perturbation equations. It should
be noted that upper-case Latin indices can now be manipulated with 
$\gamma_{AB}$, the background horizon metric, and its inverse. 

According to Eq.~(\ref{3.10}), a growth in the horizon area is driven
by the expansion scalar, and we therefore have $\dot{A} = \lambda^2
\dot{A}_2 + O(\lambda^3)$, with 
\begin{equation}
\dot{A}_2 \equiv \frac{d A_2}{dv} = \oint \Theta_2\, dS, 
\label{4.8}
\end{equation}
where $dS \equiv \sqrt{\gamma}\, d^2\theta$ is an element of surface
area on the cross sections of the {\it unperturbed} (Kerr) horizon.   

\subsection{Integration of the perturbation equations}  

We imagine a horizon that starts in an initial Kerr state, is
perturbed for some time by an external process, and then eventually
returns to another Kerr state. The expansion and shear vanish in the
initial state, and they must return to zero after the external process
has ended; this requires the imposition of {\it teleological
boundary conditions} (as opposed to retarded boundary conditions; see,
for example, Sec.~VI C 6 of Ref.~\cite{thorne-etal:86}) on 
the solutions to the perturbation equations. 

The teleological solution to the equation $(d/dv - \kappa) \psi =
-f(v)$ is $\psi(v) = \int_v^\infty e^{-\kappa(v'-v)} f(v')\, dv'$. It
shows that $\psi(v)$ depends on the future behavior of the driving
force, but that $\psi(v)$ goes to zero after the driving force is
switched off; the causal solution would depend only on the past
behavior of the driving force, but it would grow exponentially after
the force is switched off. If the driving force varies very slowly
over a time comparable to $1/\kappa$, then the teleological solution
reduces to the local expression $\psi(v) = \kappa^{-1} f(v)$, to a
fractional accuracy of order $(\kappa \tau)^{-1}$, where $\tau \sim
f/\dot{f}$ is the time scale over which the driving force varies. The
local expression can be simply obtained by noting that in this limit,
$d/dv \ll \kappa$ and the differential term can be neglected in the
differential equation; a more careful derivation starts with the
teleological solution and employs integration by parts. 

It is not permissible to neglect the differential term in 
Eq.~(\ref{4.7}), and one must write down a proper teleological
solution to this equation. To see this, suppose that the black hole is
a member of a binary system, and that it moves in the field of an
external body with an angular velocity $\Omega$. As seen in the
rotating frame of the generators, the Weyl tensor behaves as $C \sim
e^{-i\omega v}$, where $\omega \equiv \Omega - \Omega_{\rm H}$ is the
{\it relative} angular velocity between the external field and the
generators. Thus, unless the external field is nearly corotating with
the black hole, $\omega$ will be of order $\Omega_{\rm H}$, which is
itself of order $\kappa$, and the Weyl tensor will not vary slowly. 

The exact solution to Eq.~(\ref{4.7}) is 
\begin{equation} 
\sigma^1_{AB}(v,\theta^A) = \int_v^\infty e^{-\kappa(v'-v)}
C^1_{AB}(v',\theta^A)\, dv',  
\label{4.9}
\end{equation}     
and this can be substituted into the right-hand side of
Eq.~(\ref{4.6}). Here we shall allow ourselves some simplification. For
a Weyl tensor that behaves schematically as $C \sim e^{-i\omega v}$,
the shear tensor will go as $\sigma \sim e^{-i\omega v}/(\kappa 
+ i\omega)$, and it will vary as rapidly as $C$. The {\it square} of
the shear tensor, however, will contain a piece that oscillates at
twice the frequency $\omega$, and a piece that stays constant. The
force driving the expansion therefore contains both a slowly-varying
piece and a rapidly-varying piece. In a generic situation we expect
that $\sigma^1_{AB} \sigma_1^{AB}$ can always be decomposed into such
slowly-varying and rapidly-varying pieces, and we isolate the
slowly-varying component by averaging over a time scale that is long
compared with $\kappa^{-1}$: $(\sigma^1_{AB} \sigma_1^{AB})_{\rm slow}
= \langle \sigma^1_{AB} \sigma_1^{AB} \rangle$. If we now agree to
follow only the slow evolution of the expansion scalar, and to
ignore its rapid fluctuations around the slowly-evolving mean, then 
\begin{equation}
\langle \Theta_2 \rangle = \frac{1}{\kappa} \bigl\langle \sigma^1_{AB}
\sigma_1^{AB} \bigr\rangle 
\label{4.10}
\end{equation}
is an adequate approximate solution to Eq.~(\ref{4.6}). The meaning of
the averaging sign should be clear: The expansion scalar, and the
square of the shear tensor, are averaged over a time $\tau$ that is
long compared with $\kappa^{-1}$, the black-hole time scale. The time
scale $\tau$ is identified with a characteristic time associated with
the growth of the black-hole area, $\tau \sim \langle A
\rangle/\langle \dot{A} \rangle$. We note that it is a requirement of
the perturbative treatment that $\kappa \tau \gg 1$; the
simplification of Eq.~(\ref{4.10}) therefore represents no significant
loss of generality, other than a coarse-graining over short time
scales.        

Equation (\ref{4.9}) can also be inserted into Eq.~(\ref{4.5}) to
calculate the metric perturbation, which is given by 
\begin{equation} 
\gamma^1_{AB}(v,\theta^A) = 2 \int_{-\infty}^v
\sigma^1_{AB}(v',\theta^A)\, dv'.  
\label{4.11}
\end{equation} 
After altering the order of the integrations and performing one of the 
integrals, we obtain  
\begin{eqnarray} 
\gamma^1_{AB}(v,\theta^A) &=& \frac{2}{\kappa} \biggl[ 
\int_{-\infty}^v C^1_{AB}(v',\theta^A)\, dv' 
\nonumber \\ & & \mbox{} 
+ \int_v^\infty e^{-\kappa(v'-v)}
C^1_{AB}(v',\theta^A)\, dv' \biggr]; \qquad\quad
\label{4.12}
\end{eqnarray}  
this result is exact, and it does not involve a coarse-graining over
short time scales. 

The averaged rate of change of the horizon area can be calculated on
the basis of Eqs.~(\ref{4.8}) and (\ref{4.10}). The result is 
\begin{equation}
\langle \dot{A}_2 \rangle = \frac{1}{\kappa} \oint  
\bigl\langle \sigma^1_{AB} \sigma_1^{AB} \bigr\rangle\, dS, 
\label{4.13}
\end{equation} 
where an overdot indicates differentiation with respect to $v$. This
result can be expressed in terms of the Weyl tensor by means of 
Eq.~(\ref{4.9}). 

In the remainder of the paper we will denote $C^1_{AB}$ simply as
$C_{AB}$, $\sigma^1_{AB}$ as $\sigma_{AB}$, $\Theta_2$ as $\Theta$,
and $\dot{A}_2$ as $\dot{A}$; because these quantities all vanish for
a stationary horizon, this change of notation will not generate
ambiguities. But to avoid ambiguities we will continue to
denote the metric perturbation as $\gamma^1_{AB}$.  

\subsection{Fluxes of mass and angular momentum} 

We shall now derive expressions for the averaged rates of change of
the black-hole mass $M$ and angular momentum $J$, using
Eq.~(\ref{4.13}) as our main input.   

In the case of a horizon perturbed by a matter field it can be shown
(see, for example, Sec.~6.4.2 of Ref.~\cite{carter:79}) that these
rates are related by $(\kappa/8\pi) \dot{A} = \dot{M} - \Omega_{\rm H}
\dot{J}$, which is a statement of the first law of black-hole
mechanics (see, for example, Chapter 12 of Ref.~\cite{wald:84}, or
Chapter 5 of Ref.~\cite{poisson:04c}). In the present case of a
horizon perturbed by a purely gravitational perturbation, we shall
assume that this relation holds on the average, so that  
\begin{equation} 
\frac{\kappa}{8\pi} \Adot = \Mdot - \Omega_{\rm H} \Jdot. 
\label{4.14}
\end{equation}
The averaging introduced here is the same that was involved in
Eq.~(\ref{4.10}). If we imagine that the horizon evolves from an
initial Kerr state to a final Kerr state in a time $\Delta v \sim
\tau$, then a precise statement of Eq.~(\ref{4.14}) is $(\kappa/8\pi) 
(\delta A)/(\Delta v) = (\delta M)/(\Delta v) - \Omega_{\rm H} 
(\delta J)/(\Delta v)$, where $\delta A$, $\delta M$, and $\delta J$
are the total accumulated changes in the black-hole
parameters. Because these changes relate two stationary black-hole
states, we have here the usual statement of the first law divided by
the time interval $\Delta v$.  

In the case of a horizon perturbed by a matter field it can also be
shown (see, for example, Sec.~6.4.2 of Ref.~\cite{carter:79}) that if
the matter field is decomposed into modes proportional to 
$e^{-i\omega v} e^{i m \psi}$, where $\omega$ is a continuous
frequency and $m$ an integer, then each mode contribution to the
averaged rates is such that   
\begin{equation} 
\Mdot_{m,\omega} = \frac{\omega}{m} \Jdot_{m,\omega}. 
\label{4.15}
\end{equation} 
This mode decomposition is motivated by the symmetries of the 
background Kerr spacetime, and Eq.~(\ref{4.15}) states that a mode
labeled by $(m,\omega)$ carries across the horizon a quantity of
energy proportional to $\omega$ and a quantity of angular momentum  
proportional to $m$. This statement is easily understood on the basis 
of a quantum picture, but it holds for classical matter fields as
well. We shall assume that Eq.~(\ref{4.15}) is not restricted to
matter fields, but that it holds also for gravitational
perturbations. Such an assumption was made previously by 
Teukolsky and Press 
\cite{teukolsky-press:74} 
in their pioneering study of horizon fluxes.     

Equations (\ref{4.14}) and (\ref{4.15}) imply 
\begin{eqnarray} 
\Mdot_{m,\omega} = \frac{\omega}{k} \frac{\kappa}{8\pi}
\Adot_{m,\omega}, 
\label{4.16} \\
\Jdot_{m,\omega} = \frac{m}{k} \frac{\kappa}{8\pi}
\Adot_{m,\omega}, 
\label{4.17}
\end{eqnarray}
where 
\begin{equation} 
k \equiv \omega - m\Omega_{\rm H} 
\label{4.18}
\end{equation}
and $\Adot_{m,\omega}$ is the mode contribution to the averaged rate
of change of the horizon area. These equations will allow us to turn
Eq.~(\ref{4.13}) into useful expressions for $\Mdot$ and $\Jdot$. 

In the spacetime coordinates $(v,r,\theta,\psi)$, the mode
decomposition of the metric perturbation is given by 
\begin{equation}
\gamma^1_{AB} = \sum_m \int d\omega \gamma^{m,\omega}_{AB}(r,\theta)
e^{-i\omega v} e^{i m \psi}. 
\label{4.19}
\end{equation}
In the horizon coordinates $(v,\theta,\phi)$, where $\phi = \psi -
\Omega_{\rm H} v$, we have instead 
\begin{equation} 
\gamma^1_{AB} = \sum_m \int d\omega \gamma^{m,\omega}_{AB}(r_+,\theta) 
e^{-i k v} e^{i m \phi},  
\label{4.20}
\end{equation}
where we have set $r=r_+$, choosing the coordinate position of the
perturbed horizon to coincide with the position of the Kerr horizon. 
(That this choice can always be made is proved in Sec.~VI
A.) Substituting Eq.~(\ref{4.20}) into Eq.~(\ref{4.5}) we obtain  
\begin{equation} 
\sigma_{AB} = \frac{1}{2} \sum_m \int d\omega (-ik)
\gamma^{m,\omega}_{AB} e^{-i k v} e^{i m \phi},  
\label{4.21}
\end{equation}
the mode decomposition of the shear tensor. 

We now insert Eq.~(\ref{4.21}) into Eq.~(\ref{4.13}), but we do 
not decompose $\sigma^{AB}$ into modes. This gives  
\[
\frac{\kappa}{8\pi} \Adot = \sum_m \int d\omega \frac{-ik}{16\pi} 
\oint \Bigl\langle \sigma^{AB} \gamma_{AB}^{m,\omega} e^{-i k v} 
e^{i m \phi} \Bigr\rangle\, dS, 
\]
and from this we can read off each mode contribution to $(\kappa/8\pi)
\Adot$. According to Eqs.~(\ref{4.16}) and (\ref{4.17}), then, we have 
\[
\Mdot = \sum_m \int d\omega \frac{-i\omega}{16\pi} 
\oint \Bigl\langle \sigma^{AB} \gamma_{AB}^{m,\omega} e^{-i k v}  
e^{i m \phi} \Bigr\rangle\, dS
\]
and 
\[
\Jdot = \sum_m \int d\omega \frac{-im}{16\pi} 
\oint \Bigl\langle \sigma^{AB} \gamma_{AB}^{m,\omega} e^{-i k v}  
e^{i m \phi} \Bigr\rangle\, dS. 
\]
The metric perturbation can now be reconstructed from its mode
decomposition. It is easy to show that the factor of $-i\omega$ is
generated by applying the differential operator
$\partial/\partial v - \Omega_{\rm H} \partial/\partial \phi$ to  
$\gamma^1_{AB}(v,\theta,\phi)$; this can be written as a Lie
derivative in the direction of $t^\alpha = k^\alpha - \Omega_{\rm H}
\phi^\alpha$, the timelike Killing vector of the background Kerr
spacetime. Similarly, the factor of $i m$ is generated by acting with   
$\partial/\partial \phi$, which is a Lie derivative in the direction
of $\phi^\alpha$, the rotational Killing vector of the background Kerr
spacetime. 

The final expressions are 
\begin{eqnarray} 
\Mdot = \frac{1}{16\pi} \oint \bigl\langle \sigma^{AB} \pounds_t
\gamma^1_{AB} \bigr\rangle\, dS, 
\label{4.22} \\ 
\Jdot = -\frac{1}{16\pi} \oint \bigl\langle \sigma^{AB} \pounds_\phi 
\gamma^1_{AB} \bigr\rangle\, dS, 
\label{4.23}
\end{eqnarray}
and 
\begin{equation}
\frac{\kappa}{8\pi} \Adot = \frac{1}{16\pi} \oint \bigl\langle
\sigma^{AB} \pounds_k \gamma^1_{AB} \bigr\rangle\, dS,  
\label{4.24}
\end{equation} 
where we have used $\sigma_{AB} = \frac{1}{2} \pounds_k
\gamma^1_{AB}$. As we have seen, the Lie derivatives acting on the
metric perturbations refer to specific directions in the background
Kerr spacetime. In the horizon coordinates $(v,\theta,\phi)$, the Lie
derivatives take the explicit form 
\begin{equation} 
\pounds_k = \biggl(\frac{\partial}{\partial v}\biggr)_{\theta,\phi},
\qquad 
\pounds_\phi = \biggl(\frac{\partial}{\partial \phi}\biggr)_{v,\theta}, 
\label{4.25}
\end{equation}
and $\pounds_t = \pounds_k - \Omega_{\rm H} \pounds_\phi$. On the
other hand, in the spacetime coordinates $(v,r,\theta,\psi)$ they take
the form  
\begin{equation} 
\pounds_t = \biggl(\frac{\partial}{\partial v}\biggr)_{r,\theta,\psi},
\qquad 
\pounds_\phi = \biggl(\frac{\partial}{\partial \psi}\biggr)_{v,r,\theta},
\label{4.26}
\end{equation}
and $\pounds_k = \pounds_t + \Omega_{\rm H} \pounds_\phi$. We also
note that in Eqs.~(\ref{4.22})--(\ref{4.24}), the surface element 
$dS = \sqrt{\gamma} d^2\theta$ refers to the metric $\gamma_{AB}$ of
the unperturbed Kerr horizon.  

Equations (\ref{4.22})--(\ref{4.24}) are an excellent starting point
for the development of a practical formalism to calculate the horizon
fluxes, a topic we shall turn to in the next three sections. These
equations are not new: they appear in Sec.~VI C 11 of the book by
Thorne, Price, and Macdonald
\cite{thorne-etal:86}. 
The derivation presented here, however,
is substantially different from theirs, and it incorporates the
Teukolsky-Press assumption of Eq.~(\ref{4.15}); this assumption was
not part of the original derivation, and their route from
Eq.~(\ref{4.13}) to Eqs.~(\ref{4.22}), (\ref{4.23}) is not as direct.  
It should be clear that while the present derivation relies on a
mode decomposition of the metric perturbation, the final expressions
involve differential (as opposed to algebraic) operations and are 
independent of the decomposition.  

\subsection{Rigid rotation} 

The perturbed black hole is part of a system in rigid rotation when
the vector 
\begin{equation} 
\xi^\alpha = t^\alpha + \Omega \phi^\alpha 
\label{4.27}
\end{equation} 
is a Killing vector of both the background Kerr spacetime and the
perturbed spacetime; here $\Omega$ is a constant angular velocity. An 
example of a system in rigid rotation is when the black hole is a
member of a binary system, moving with a uniform angular velocity in
the field of the external body. The fact that $\xi^\alpha$ is a
Killing vector of the perturbed spacetime means that 
\begin{equation} 
\pounds_\xi \gamma^1_{AB} = 0. 
\label{4.28}
\end{equation}
If the metric perturbation is expressed in the spacetime coordinates
$(v,r,\theta,\psi)$, then Eq.~(\ref{4.28}) implies that its dependence
on $v$ and $\psi$ is through the combination $\psi - \Omega v$ only;
in a reference frame that is rotating at an angular velocity $\Omega$
with respect to the original inertial frame, the perturbation appears
to be stationary, and the system is indeed in rigid rotation.  

Equations (\ref{2.2}) and (\ref{4.27}) imply $k^\alpha = \xi^\alpha +
(\Omega_{\rm H} - \Omega) \phi^\alpha$ and $t^\alpha = \xi^\alpha -
\Omega \phi^\alpha$, and by virtue of Eq.~(\ref{4.28}) the Lie
derivatives of $\gamma^1_{AB}$ in the directions of $k^\alpha$ and
$t^\alpha$ can be expressed in terms of a derivative along
$\phi^\alpha$,  
\[
\pounds_k \gamma^1_{AB} = (\Omega_{\rm H} - \Omega) 
\pounds_\phi \gamma^1_{AB} 
\]
and 
\[
\pounds_t \gamma^1_{AB} = - \Omega \pounds_\phi \gamma^1_{AB}. 
\]
Substituting this into Eqs.~(\ref{4.22})--(\ref{4.24}), and using also 
$\sigma^{AB} = \frac{1}{2} (\Omega_{\rm H}-\Omega) \pounds_\phi
\gamma_1^{AB}$, we obtain 
\begin{eqnarray} 
\Mdot = \Omega(\Omega-\Omega_{\rm H}) {\cal K}, 
\label{4.29} \\ 
\Jdot = (\Omega-\Omega_{\rm H}) {\cal K}, 
\label{4.30} \\ 
\frac{\kappa}{8\pi} \Adot = (\Omega-\Omega_{\rm H})^2 {\cal K}, 
\label{4.31}
\end{eqnarray} 
where 
\begin{equation} 
{\cal K} = \frac{1}{32\pi} \oint \bigl\langle 
\bigl(\pounds_\phi \gamma_1^{AB} \bigr)  
\bigl(\pounds_\phi \gamma^1_{AB} \bigr) \bigr\rangle\, dS; 
\label{4.32}
\end{equation} 
we recall that $\gamma_1^{AB}$ is obtained from $\gamma^1_{AB}$ by
raising indices with $\gamma^{AB}$, the inverse background metric. 

Equation (\ref{4.30}) indicates that the black hole's angular momentum
will increase when $\Omega > \Omega_{\rm H}$, that is, when the
external rotation is faster than the rotation of the generators;
otherwise the angular momentum will decrease. Equation (\ref{4.29})
indicates that $\Mdot = \Omega \Jdot$ when the black hole is in rigid
rotation; the sign of $\Mdot$ is tied to the sign of $\Jdot$ and the
sign of the angular velocity (which is defined relative to
$\Omega_{\rm H}$). Finally, Eq.~(\ref{4.31}) shows that the black-hole
area will always increase, as is dictated by Hawking's area
theorem (see, for example, Sec.~12.2 of Ref.~\cite{wald:84}). These
equations also appear in Sec.~VII B 1 of the book by Thorne, Price,
and Macdonald \cite{thorne-etal:86}.   

\section{Curvature formalism} 

In this section I translate the flux formulae of
Eqs.~(\ref{4.22})--(\ref{4.24}) into a more practical language that 
involves {\it curvature variables}. The most important variable in
this formalism is the Weyl scalar $\psi_0$, which can be obtained by
solving Teukolsky's differential equation 
\cite{teukolsky:73}.   

\subsection{Relation between $C_{AB}$ and $\psi_0$} 

Our first task is to express $C_{AB}$, the Weyl tensor of the
perturbed horizon, in terms of the more practical curvature variable
$\psi_0$. The calculation is straightforward but somewhat lengthy; it 
requires a number of steps. 

The Weyl tensor of the perturbed spacetime is
$\hat{C}_{\alpha\gamma\beta\delta} = C_{\alpha\gamma\beta\delta} +
\lambda C^1_{\alpha\gamma\beta\delta} + O(\lambda^2)$, and the basis
vectors of the perturbed horizon are $\hat{k}^\alpha = k^\alpha +
\lambda k^\alpha_1 + O(\lambda^2)$ and $\hat{e}^\alpha_A = e^\alpha_A
+ \lambda e^\alpha_{1A} + O(\lambda^2)$. The Weyl tensor of the
perturbed horizon is then defined by 
\begin{equation} 
\hat{C}_{AB} = \hat{C}_{\alpha\gamma\beta\delta} \hat{e}^\alpha_A
\hat{k}^\gamma \hat{e}^\beta_B \hat{k}^\delta, 
\label{5.1}
\end{equation} 
which is the same equation as Eq.~(\ref{3.9}). By virtue of
Eq.~(\ref{2.35}) we have that $C_{AB} = 0$ for the Kerr horizon, and
$\hat{C}_{AB} = \lambda C^1_{AB} + O(\lambda^2)$. To comply with
preceding usage we shall now omit the label ``1'' on the Weyl tensor
and set $\lambda = 1$ after the expansion in powers of $\lambda$ has
been carried out.   

Direct evaluation of $\hat{C}_{AB}$ from the preceding information
gives  
\begin{eqnarray*} 
C_{AB} &=& C^1_{\alpha\gamma\beta\delta} e^\alpha_A k^\gamma e^\beta 
k^\delta + C_{\alpha\gamma\beta\delta} \bigl( e^\alpha_A k^\gamma
e^\beta_B k_1^\delta 
\\ \mbox{} & & 
+ e^\alpha_A k_1^\gamma e^\beta_B k^\delta 
+ e^\alpha_A k^\gamma e^\beta_{1B} k^\delta 
+ e^\alpha_{1A} k^\gamma e^\beta_B k^\delta \bigr). 
\end{eqnarray*}
We will now simplify this expression, and show that $C_{AB}$ can be
expressed solely in terms of $C^1_{\alpha\gamma\beta\delta}$, the
perturbation of the Weyl tensor.   

The main source of simplification comes from the algebraic properties
of the unperturbed Weyl tensor: The last two terms within the brackets
vanish by virtue of Eq.~(\ref{2.34}), and the first two can be
rewritten with the help of Eq.~(\ref{2.35}). This gives 
\[
C_{AB} = C^1_{\alpha\gamma\beta\delta} e^\alpha_A k^\gamma e^\beta_B
k^\delta - 2\mbox{Re}(\psi_2) \gamma_{AB} k_\alpha k_1^\alpha.
\]
Next we decompose $e^\alpha_A$ in the basis of complex vectors
$(e^\alpha,\bar{e}^\alpha)$ introduced in Sec.~II D; the relations are
$e^\alpha_A = \bar{e}_A e^\alpha + e_A \bar{e}^\alpha$ with the
coefficients $e_A = \gamma_{AB} e^B$ obtained from
Eq.~(\ref{2.22}). This yields  
\begin{eqnarray*} 
C^1_{\alpha\gamma\beta\delta} e^\alpha_A k^\gamma e^\beta_B k^\delta 
&=& \bar{e}_A \bar{e}_B C^1_{\alpha\gamma\beta\delta} e^\alpha
k^\gamma e^\beta k^\delta 
\\ & & \mbox{} 
+ e_A e_B C^1_{\alpha\gamma\beta\delta}
\bar{e}^\alpha k^\gamma \bar{e}^\beta k^\delta
\\ & & \mbox{} 
+ \bigl(\bar{e}_A e_B + e_A \bar{e}_B \bigr)
C^1_{\alpha\gamma\beta\delta} e^\alpha k^\gamma \bar{e}^\beta k^\delta 
\end{eqnarray*}
for the first term in the previous expression for $C_{AB}$. The factor
within brackets is $\gamma_{AB}$, and by virtue of the symmetries of
the Weyl tensor and the completeness relations of Eq.~(\ref{2.21}), we
also have $C^1_{\alpha\gamma\beta\delta} e^\alpha k^\gamma
\bar{e}^\beta k^\delta = \frac{1}{2} g^{\alpha\beta}
C^1_{\alpha\gamma\beta\delta} k^\gamma k^\delta$, where
$g^{\alpha\beta}$ is the inverse of the Kerr metric. But because the
perturbed Weyl tensor must be traceless in the perturbed metric, 
$0 = \hat{g}^{\alpha\beta} \hat{C}_{\alpha\gamma\beta\delta} =
(g^{\alpha\beta} - \lambda h^{\alpha\beta}) 
(C_{\alpha\gamma\beta\delta} + \lambda
C^1_{\alpha\gamma\beta\delta})$, where $\lambda h_{\alpha\beta} \equiv 
\hat{g}_{\alpha\beta} - g_{\alpha\beta}$ is the metric
perturbation. We therefore have $g^{\alpha\beta}
C^1_{\alpha\gamma\beta\delta} = h^{\alpha\beta} 
C_{\alpha\gamma\beta\delta}$, and gathering these results we obtain 
\begin{eqnarray*} 
C^1_{\alpha\gamma\beta\delta} e^\alpha_A k^\gamma e^\beta_B k^\delta 
&=& \bar{e}_A \bar{e}_B C^1_{\alpha\gamma\beta\delta} e^\alpha
k^\gamma e^\beta k^\delta 
\\ & & \mbox{} 
+ e_A e_B C^1_{\alpha\gamma\beta\delta}
\bar{e}^\alpha k^\gamma \bar{e}^\beta k^\delta
\\ & & \mbox{} 
+ \frac{1}{2} \gamma_{AB} h^{\alpha\beta} C_{\alpha\gamma\beta\delta}
k^\gamma k^\delta. 
\end{eqnarray*} 
We may simplify this further by using Eq.~(\ref{2.36}), and we now
have  
\begin{eqnarray*} 
C_{AB} &=& \bar{e}_A \bar{e}_B C^1_{\alpha\gamma\beta\delta} e^\alpha  
k^\gamma e^\beta k^\delta + e_A e_B C^1_{\alpha\gamma\beta\delta}
\bar{e}^\alpha k^\gamma \bar{e}^\beta k^\delta 
\\ & & \mbox{} 
- 2\mbox{Re}(\psi_2)
\gamma_{AB} \biggl( \frac{1}{2} h_{\alpha\beta} k^\alpha k^\beta +
k_\alpha k_1^\alpha \biggr).
\end{eqnarray*}
In the last step we recognize that the vector $\hat{k}^\alpha$ must be 
null in the metric $\hat{g}_{\alpha\beta}$, so that $0 =
(g_{\alpha\beta} + \lambda h_{\alpha\beta})(k^\alpha+\lambda
k_1^\alpha) (k^\beta + \lambda k_1^\beta) = \lambda(2 k_\alpha
k_1^\alpha + h_{\alpha\beta} k^\alpha k^\beta)$. We finally arrive at 
the expression
\begin{equation} 
C_{AB} = \bar{e}_A \bar{e}_B C^1_{\alpha\gamma\beta\delta} e^\alpha 
k^\gamma e^\beta k^\delta + e_A e_B C^1_{\alpha\gamma\beta\delta}
\bar{e}^\alpha k^\gamma \bar{e}^\beta k^\delta, 
\label{5.2}
\end{equation}
which involves only the perturbed Weyl tensor and the Kerr basis
vectors. 

The Weyl scalar of the perturbed spacetime is defined as in 
Eq.~(\ref{2.28}), 
\begin{equation}
- \hat{\psi}_0 = \hat{C}_{\alpha\gamma\beta\delta} \hat{k}^\alpha
\hat{m}^\gamma \hat{k}^\beta \hat{m}^\delta. 
\label{5.3}
\end{equation} 
Expansion in powers of $\lambda$ gives $\hat{\psi}_0 = \lambda
\psi^1_0 + O(\lambda^2)$. After dropping the label ``1'' and
setting $\lambda = 1$, we obtain 
\begin{eqnarray*} 
- \psi_0 &=& C^1_{\alpha\gamma\beta\delta} k^\alpha m^\gamma k^\beta 
m^\delta + 2 C_{\alpha\gamma\beta\delta} k^\alpha m^\gamma k^\beta
m^\delta_1 
\\ & & \mbox{} 
+ 2 C_{\alpha\gamma\beta\delta} k^\alpha m^\gamma k_1^\beta
m^\delta. 
\end{eqnarray*}
The last two terms vanish by virtue of Eq.~(\ref{2.37}), and we have
$-\psi_0 = C^1_{\alpha\gamma\beta\delta} k^\alpha m^\gamma k^\beta
m^\delta$. We now express $m^\alpha$ in terms $e^\alpha$ and
$k^\alpha$, as in Eq.~(\ref{2.23}). This yields 
\begin{equation} 
- \psi_0 = C^1_{\alpha\gamma\beta\delta} k^\alpha e^\gamma k^\beta
e^\delta
\label{5.4}
\end{equation}
as our final expression for the Weyl scalar.  

The relation between $C_{AB}$ and $\psi_0$ is obtained by substituting
Eq.~(\ref{5.4}) into Eq.~(\ref{5.2}). We write this as 
\begin{equation} 
C_{AB} = \bar{e}_A \bar{e}_B \Psi + e_A e_B \bar{\Psi}, 
\label{5.5}
\end{equation}
where 
\begin{equation} 
\Psi(v,\theta^A) \equiv -\psi_0(\mbox{HH}) \equiv -
\frac{\Delta^2}{4(r^2 + a^2)^2} \psi_0(\mbox{K}) 
\label{5.6}
\end{equation}
is our main curvature variable. We choose, for convenience, to absorb
a minus sign into the definition of $\Psi$. The first equality in
Eq.~(\ref{5.6}) states that apart from this minus sign, $\Psi$ is the
Weyl scalar $\psi_0$ as defined with the Hartle-Hawking tetrad
\cite{hawking-hartle:72, teukolsky:73},
evaluated on the horizon and expressed in terms of the horizon
coordinates. The second equality gives the relationship between $\Psi$
and the Weyl scalar as defined with the Kinnersley tetrad
\cite{kinnersley:69, teukolsky:73}, 
evaluated in the limit $r \to r_+$ in which $\psi_0(\mbox{K})$
diverges as $\Delta^{-2}$. 

The Weyl scalars in either choice of tetrad, and therefore $\Psi$, can
be obtained by solving the Teukolsky equation. Because $\Psi$ is a
scalar that vanishes in the Kerr spacetime, this quantity is gauge
invariant. These two properties make $\Psi(v,\theta^A)$ an especially
useful choice of variable to describe the horizon perturbation.    

\subsection{Fluxes}

We obtain the shear tensor by inserting the Weyl tensor of
Eq.~(\ref{5.5}) into Eq.~(\ref{4.9}). Because the transverse vectors
$e_A$ depend on $\theta^A$ only, we obtain 
\begin{equation}
\sigma_{AB}(v,\theta^A) = \bar{e}_A \bar{e}_B \Phi_+ 
+ e_A e_B \bar{\Phi}_+, 
\label{5.7}
\end{equation}
where 
\begin{equation} 
\Phi_+(v,\theta^A) = \int_v^\infty e^{-\kappa(v'-v)}
\Psi(v',\theta^A)\, dv' 
\label{5.8}
\end{equation} 
is the future integral of the Weyl scalar $\Psi$, weighted by the
exponential factor $e^{-\kappa(v'-v)}$ so that only the near future 
contributes significantly. 

The metric perturbation is obtained by substituting Eq.~(\ref{5.5})
into Eq.~(\ref{4.12}). Here we obtain 
\begin{equation}
\gamma^1_{AB}(v,\theta^A) = \frac{2}{\kappa} \Bigl( \bar{e}_A
\bar{e}_B \Phi + e_A e_B \bar{\Phi} \Bigr), 
\label{5.9}
\end{equation}
where 
\begin{equation} 
\Phi(v,\theta^A) = \Phi_-(v,\theta^A) + \Phi_+(v,\theta^A), 
\label{5.10}
\end{equation}
with 
\begin{equation}
\Phi_-(v,\theta^A) = \int_{-\infty}^v \Psi(v',\theta^A)\, dv' 
\label{5.11}
\end{equation} 
representing the past integral of the Weyl scalar (weighted
uniformly). 

Equations (\ref{5.7}) and (\ref{5.9}) can now be substituted into
Eqs.~(\ref{4.22})--(\ref{4.24}) to obtain expressions for the
fluxes. Because the basis vectors $e_A$ do not depend on $v$ nor
$\phi$, we have that $\pounds_k e_A = \pounds_\phi e_A = \pounds_t e_A
= 0$, and the derivative operators act only on $\Phi_\pm$. Using the
properties $e_A e^A = 0$ and $e_A \bar{e}^A = 1$ of the basis vectors
--- recall Eq.~(\ref{2.20}) --- we obtain, for example
\[
\Mdot = \frac{1}{8\pi \kappa} \oint \bigl\langle \Phi_+ \pounds_t
\bar{\Phi} + \bar{\Phi}_+ \pounds_t \Phi \bigr\rangle\, dS. 
\]
Using now Eq.~(\ref{2.9}), we arrive at the following expression for
the mass flux: 
\begin{equation} 
\Mdot = \frac{r_+^2+a^2}{8\pi \kappa} \oint \bigl\langle \Phi_+
\pounds_t \bar{\Phi} + \bar{\Phi}_+ \pounds_t \Phi \bigr\rangle\,
d\Omega, 
\label{5.12}
\end{equation} 
where $d\Omega = \sin\theta\, d\theta d\phi$ is an element of solid
angle on the unit two-sphere. We obtain similarly 
\begin{equation} 
\Jdot = -\frac{r_+^2+a^2}{8\pi \kappa} \oint \bigl\langle \Phi_+
\pounds_\phi \bar{\Phi} + \bar{\Phi}_+ \pounds_\phi \Phi
\bigr\rangle\, d\Omega, 
\label{5.13}
\end{equation} 
for the flux of angular momentum, and 
\begin{equation}
\frac{\kappa}{8\pi} \Adot = \frac{r_+^2+a^2}{4\pi} \oint \bigl\langle     
|\Phi_+|^2 \bigr\rangle\, d\Omega 
\label{5.14}
\end{equation} 
for the rate of increase of the horizon area. 

In Eqs.~(\ref{5.12})--(\ref{5.14}) it is understood that the
integrated-curvature fields $\Phi_\pm(v,\theta^A)$ are expressed in
terms of the horizon coordinates, and that the Lie-derivative
operators take the form given by Eq.~(\ref{4.25}). At a later stage we
will remove this remaining dependence on the horizon coordinates, and
express all quantities in terms of the original spacetime
coordinates. 

\subsection{Pure mode; comparison with Teukolsky and Press} 

Suppose that the Weyl scalar of Eq.~(\ref{5.6}) has the form 
\begin{equation}
\Psi(v,r_+,\theta,\psi) = \Psi^{m,\omega}(\theta) e^{-i\omega v} 
e^{i m \psi} 
\label{5.15}
\end{equation}
when expressed in terms of the spacetime coordinates; this solution to
the Teukolsky equation is then a pure mode of frequency $\omega$ and
azimuthal number $m$. In terms of the horizon coordinates we have 
\begin{equation}
\Psi(v,\theta^A) = \Psi^{m,\omega}(\theta) e^{-i k v} e^{i m \phi};
\label{5.16}
\end{equation}
we recall that $\phi = \psi - \Omega_{\rm H} v$, and $k = \omega - m 
\Omega_{\rm H}$ was first introduced in Eq.~(\ref{4.18}). We wish to
calculate the rates of change of mass, angular momentum, and area for
this pure mode, and to compare our results with those first obtained
by Teukolsky and Press \cite{teukolsky-press:74}.  

Substituting Eq.~(\ref{5.16}) into Eqs.~(\ref{5.8}), (\ref{5.11}), and
(\ref{5.10}) yields  
\begin{eqnarray} 
\Phi_+ &=& \frac{\Psi^{m,\omega}(\theta)}{\kappa + i k} e^{-i k v}  
e^{i m \phi}, 
\label{5.17} \\ 
\Phi_- &=& \frac{\Psi^{m,\omega}(\theta)}{- i k} e^{-i k v}  
e^{i m \phi}, 
\label{5.18} 
\end{eqnarray} 
and
\begin{equation}
\Phi = \frac{\kappa \Psi^{m,\omega}(\theta)}{-ik (\kappa + i k)} 
e^{-i k v} e^{i m \phi}. 
\label{5.19}
\end{equation}
It should be noted that while the integral defining $\Phi_+$ converges
properly for a pure mode, the integral defining $\Phi_-$ diverges in
the infinite past; this difficulty is remedied by inserting a
converging factor inside the integral, for example $e^{v'/v_1}$ with
$v_1 \gg k^{-1}$, to reflect the fact that the pure mode was turned on
in the finite (but remote) past.    

These expressions can now be substituted into
Eqs.~(\ref{5.12})--(\ref{5.14}). When acting on a pure mode,
$\pounds_t$ produces a factor of $-i\omega$, $\pounds_\phi$ a
factor of $im$, and $\pounds_k$ a factor of $-ik$. A simple
computation gives 
\begin{eqnarray} 
\Mdot &=& \frac{\omega(r_+^2+a^2)}{k(\kappa^2+k^2)} \frac{1}{4\pi}
\oint \bigl| \Psi^{m,\omega}(\theta) \bigr|^2\, d\Omega, 
\label{5.20} \\
\Jdot &=& \frac{m(r_+^2+a^2)}{k(\kappa^2+k^2)} \frac{1}{4\pi}
\oint \bigl| \Psi^{m,\omega}(\theta) \bigr|^2\, d\Omega, \qquad
\label{5.21} \\
\frac{\kappa}{8\pi} \Adot &=& \frac{r_+^2+a^2}{\kappa^2+k^2}
\frac{1}{4\pi} \oint \bigl| \Psi^{m,\omega}(\theta) \bigr|^2\,
d\Omega.  
\label{5.22}
\end{eqnarray} 
These relations are compatible with Eqs.~(\ref{4.29})--(\ref{4.31}) if
we define the angular velocity of the pure mode to be $\Omega =
\omega/m$; this follows from the fact that according to
Eq.~(\ref{5.15}), $\Psi$ depends on $\psi$ and $v$ only through the
combination $\psi - (\omega/m) v$, so that the perturbation rotates
rigidly with an angular velocity $\omega/m$. 

To compare our expressions with those of Teukolsky and Press 
\cite{teukolsky-press:74} 
we must first reconcile the different notations. In their Eq.~(4.40), 
Teukolsky and Press display the near-horizon behavior of
$\psi_0(\mbox{K})$ --- this is the Weyl scalar as defined with the 
Kinnersley tetrad --- in terms of Boyer-Lindquist coordinates  
$(t_{\rm BL}, r, \theta, \phi_{\rm BL})$. They have   
\[
\psi_0(\mbox{K}) \sim e^{-i\omega t_{\rm BL}} e^{im\phi_{\rm BH}}
e^{-i k r^*} \Delta^{-2} {}_2 S_{lm}(\theta) Y_{\rm hole},
\]
where $r^* = \int (r^2+a^2) \Delta^{-1}\, dr = v - t_{\rm BL}$, 
${}_2 S_{lm}(\theta)$ are the Teukolsky angular functions, and 
$Y_{\rm hole}$ is a normalization factor. It is easy to check that
after a transformation to the well-behaved Kerr coordinates
$(v,r,\theta,\psi)$, and after the rescaling of Eq.~(\ref{5.6}), this
expression becomes 
\[
\Psi = - \frac{e^{-i m \beta(r_+)} 
{}_2 S_{lm}(\theta) Y_{\rm hole}}{4 (r_+^2+a^2)^2}\, 
e^{-i\omega v} e^{im\psi},
\]
where $\beta(r) \equiv - a(r_+^2+a^2)^{-1} \int (r+r_+)(r-r_-)^{-1}\,
dr$ is a function that is well behaved at $r=r_+$. Here, $r_- \equiv M
- \sqrt{M^2 - a^2}$ denotes the position of the inner horizon. 

The preceding equation relates our definition of
$\Psi^{m,\omega}(\theta)$ in Eq.~(\ref{5.15}) with the quantities
introduced by Teukolsky and Press. Inserting this into
Eqs.~(\ref{5.20}) and (\ref{5.21}) returns  
\[
\biggl\langle \frac{d^2 M}{dv d\Omega} \biggr\rangle 
= \frac{\bigl| {}_2 S_{lm}(\theta) Y_{\rm hole} \bigr|^2}
{64\pi (r_+^2 + a^2)^3} \frac{\omega}{k(\kappa^2 + k^2)} 
\]
and 
\[
\biggl\langle \frac{d^2 J}{dv d\Omega} \biggr\rangle 
= \frac{\bigl| {}_2 S_{lm}(\theta) Y_{\rm hole} \bigr|^2}
{64\pi (r_+^2 + a^2)^3} \frac{m}{k(\kappa^2 + k^2)}. 
\]
These are precisely the results obtained by Teukolsky and Press 
\cite{teukolsky-press:74} 
and displayed in their Eq.~(4.44). Our formalism (which is based
partly on their work) is therefore consistent with theirs. 

\subsection{Decomposition into azimuthal modes}
 
Our aim in this subsection is to derive practical flux formulae that
are formulated in the time domain, in terms of fields expressed in the
spacetime coordinates $(v,r,\theta,\psi)$. We shall not, therefore,
follow Teukolsky and Press 
\cite{teukolsky-press:74}
and decompose $\Psi$ into modes
proportional to $e^{-i\omega v} e^{i m \psi}$, so as to work in the
frequency domain. But we will still decompose the Weyl scalar into 
{\it azimuthal modes} proportional to $e^{im\psi}$, and write  
\begin{equation} 
\Psi(v,r_+,\theta,\psi) = \sum_m \Psi^m(v,\theta) e^{im\psi}.  
\label{5.23}
\end{equation}  
This decomposition is motivated by the axial symmetry of the Kerr
spacetime, which implies that each mode labeled by $m$ will evolve
independently. Such a decomposition is therefore likely to be involved
in most attempts to integrate the Teukolsky equation numerically, in
the time domain. It will also allow us to remove the remaining 
dependence of our flux formulae on the horizon coordinates
$(v,\theta^A)$. It should be noted that the flux formulae of
Eqs.~(\ref{5.12})--(\ref{5.14}) do not require $\Psi$ to be decomposed
into modes; they are therefore ready to be used in situations where an
azimuthal decomposition is not attempted. But the implementation of
these formulae is delicate, because $\Phi_\pm$ must be evaluated by 
integrating $\Psi(v',r_+,\theta,\psi)$ along the horizon generators 
(integrating over $dv'$ keeping $\phi \equiv \psi - \Omega_{\rm H} v'$
constant). Our azimuthal decomposition will accomplish this
automatically. 

Substituting Eq.~(\ref{5.23}) into Eqs.~(\ref{5.8}) and (\ref{5.11})
gives us the azimuthal decomposition of the integrated curvatures,
which we express in terms of the horizon coordinate $\phi$ instead of
the spacetime angle $\psi$: 
\begin{equation} 
\Phi_\pm(v,\theta^A) = \sum_m \Phi^m_\pm(v,\theta) e^{im\phi}, 
\label{5.24}
\end{equation}
where 
\begin{equation} 
\Phi^m_+(v,\theta) = e^{\kappa v} \int_v^\infty e^{-(\kappa 
- im\Omega_{\rm H})v'} \Psi^m(v',\theta)\, dv' 
\label{5.25}
\end{equation}
and 
\begin{equation} 
\Phi^m_-(v,\theta) = \int_{-\infty}^v e^{im\Omega_{\rm H} v'}
\Psi^m(v',\theta)\, dv'.   
\label{5.26}
\end{equation}
Notice the presence of the oscillating factor 
$e^{im\Omega_{\rm H}v'}$ within the integrals; this comes from the
transformation between $\phi$ and $\psi$ and it reflects the fact that
the generators wrap around the horizon as $v'$ is integrated
forward. Notice also that Eqs.~(\ref{5.25})--(\ref{5.26}) are now
independent of $\phi$ or $\psi$, so that it is no longer necessary to
specify which is to remain constant during integration. 

Equation (\ref{5.24}) can now be substituted into
Eqs.~(\ref{5.12})--(\ref{5.14}), in which $\Phi = \Phi_+ +
\Phi_-$. In the horizon coordinates $(v,\theta,\phi)$ the operator
$\pounds_k$ is a partial derivative with respect to $v$,
$\pounds_\phi$ produces a factor of $i m$, and $\pounds_t =
\pounds_k - \Omega_{\rm H} \pounds_\phi$. Simple algebra and
integration over $d\phi$ give  
\begin{eqnarray} 
\Mdot &=& \frac{r_+^2+a^2}{4\kappa} \sum_m \biggl[ 2\kappa \int
\bigl\langle |\Phi_+^m|^2 \bigr\rangle \sin\theta\, d\theta 
\nonumber \\ & & \mbox{} 
- i m \Omega_{\rm H} \int \bigl\langle \bar{\Phi}_+^m \Phi_-^m 
- \Phi_+^m \bar{\Phi}_-^m \bigr\rangle \sin\theta\, d\theta \biggr],  
\qquad\quad \label{5.27} \\ 
\Jdot &=& -\frac{r_+^2+a^2}{4\kappa} \sum_m (im)  
\nonumber \\ & & \mbox{} \times \int \bigl\langle
\bar{\Phi}_+^m \Phi_-^m - \Phi_+^m \bar{\Phi}_-^m \bigr\rangle
\sin\theta\, d\theta, 
\label{5.28}
\end{eqnarray}
and
\begin{equation} 
\frac{\kappa}{8\pi} \Adot = \frac{1}{2} (r_+^2 + a^2) \sum_m \int  
\bigl\langle |\Phi_+^m|^2 \bigr\rangle \sin\theta\, d\theta. 
\label{5.29}
\end{equation} 
These are the final form of the flux formulae. Notice that these no
longer involve the angles $\phi$ and $\psi$, and that all fields are 
expressed in terms of $v$ and $\theta$, coordinates that are shared
by the spacetime and the horizon.

The steps required to compute $\Mdot$, $\Jdot$, and $\Adot$ are
therefore these (see also Sec.~I C): First, solve the Teukolsky
equation 
\cite{teukolsky:73} 
for the functions
$\Psi^m(v,\theta)$ defined by Eq.~(\ref{5.23}), for all relevant
values of $m$; recall from Eq.~(\ref{5.6}) that $\Psi$ is (minus) the  
Weyl scalar $\psi_0$ in the Hartle-Hawking tetrad, evaluated at
$r=r_+$. Second, compute the integrals of Eqs.~(\ref{5.25}) and
(\ref{5.26}) to obtain $\Phi_\pm^m(v,\theta)$. Third, and finally, 
substitute these values into the flux formulae of
Eqs.~(\ref{5.27})--(\ref{5.29}), integrate over $d\theta$, and sum
over $m$. 

\section{Metric formalism for general black holes} 

In this section I translate the flux formulae of
Eqs.~(\ref{4.22})--(\ref{4.24}) into a more practical language that 
involves {\it metric variables}. This translation is most useful in
the context of a Schwarzschild black hole, for which the theory of
metric perturbations is well developed; I shall consider this
specific case in the following section. In this section I keep the
discussion general, so that it applies to both rotating and
nonrotating black holes.

\subsection{Preferred gauge} 

We expand the metric of the perturbed black hole as 
\begin{equation}
\hat{g}_{\alpha\beta} = g_{\alpha\beta} + \lambda h_{\alpha\beta}, 
\label{6.1}
\end{equation}
where $g_{\alpha\beta}$ is the metric of the unperturbed spacetime
--- the Kerr metric --- and $\lambda h_{\alpha\beta}$ is the
perturbation. (As we did previously, we keep $\lambda$ for
book-keeping but we set it equal to unity at the end of the
calculation.) We wish first to impose a number of gauge conditions on 
$h_{\alpha\beta}$, which will simplify its relationship with the
quantity $\gamma^1_{AB}$ introduced in Sec.~IV A. 

Our preferred gauge is a ``horizon-locking gauge;'' it has the
property that the coordinate positions of the perturbed horizon are
the same as those of the unperturbed (Kerr) horizon. As we shall see
below, it is always possible to make this choice of gauge. In the
preferred gauge the parametric description of the horizon generators
is given by 
\begin{equation} 
\hat{z}^\alpha(v,\theta^A) = z^\alpha(v,\theta^A),  
\label{6.2}
\end{equation} 
with $z^\alpha(v,\theta^A)$ giving the parametric description of the
Kerr generators. This equality implies 
\begin{equation}
\hat{k}^\alpha = k^\alpha, \qquad 
\hat{e}^\alpha_A = e^\alpha_A, 
\label{6.3}
\end{equation}
so that the perturbation of the tangent vectors is identically zero: 
$k^\alpha_1 = 0 = e^\alpha_{1A}$ in the notation of Sec.~V A. 

The vector $k^\alpha$ must be null, and orthogonal to 
$e^\alpha_A$, in the metrics $g_{\alpha\beta}$ and 
$\hat{g}_{\alpha\beta}$. This observation gives rise to the three
gauge conditions 
\begin{equation}
h_{\alpha\beta} k^\alpha k^\beta = 0 = h_{\alpha\beta} k^\alpha
e^\beta_A \qquad \mbox{(preferred gauge)}. 
\label{6.4}
\end{equation}
These equations hold on the horizon only, and we shall not need to
extend them beyond the horizon. The preferred gauge is only partially
determined, and the space of transformations within the
preferred-gauge class is large.  

For some purposes it will be convenient to supplement the gauge
conditions of Eq.~(\ref{6.4}) with a fourth condition,
$h_{\alpha\beta} k^\alpha N^\beta = 0$, so that we have the four
conditions   
\begin{equation} 
h_{\alpha\beta} k^\beta = 0  \qquad \mbox{(radiation gauge)}. 
\label{6.5}
\end{equation} 
These conditions also hold on the horizon only, and again we have a
large space of gauge transformations within the radiation-gauge
class. The radiation gauge of Eq.~(\ref{6.5}) is similar to the one
introduced by Chrzanowski (his ingoing radiation gauge 
\cite{chrzanowski:75a}); 
but it is distinct because Chrzanowski imposes Eq.~(\ref{6.5}) as well
as $g^{\alpha\beta} h_{\alpha\beta} = 0$ globally in the Kerr  
spacetime [extending $k^\alpha$ away from the horizon as
$k^\alpha(\rm{HH})$, the first member of the Hartle-Hawking
tetrad]. The Chrzanowski radiation gauge is therefore much more
rigidly defined than (and a special case of) the radiation gauge of
Eq.~(\ref{6.5}).  

Equation (\ref{6.3}) states that three of the basis vectors on the
horizon are not changed by the perturbation. The fourth basis vector,
$\hat{N}^\alpha$, must be orthogonal to $e^\alpha_A$ in the perturbed
metric, and it must also satisfy $\hat{g}_{\alpha\beta} \hat{N}^\alpha 
k^\beta = -1$. It is easy to see that these requirements imply 
$\hat{N}_\alpha = N_\alpha - \frac{1}{2} \lambda (h_{\beta\gamma}
N^\beta N^\gamma) k_\alpha$. 

We have shown that the imposition of Eq.~(\ref{6.2}) implies the gauge 
conditions of Eq.~(\ref{6.4}). We now examine the reversed question: 
Does the imposition of the preferred-gauge conditions imply that the
coordinate description of the horizon is the same in the unperturbed
and perturbed spacetimes? We shall show that the answer is in the
affirmative. 

Suppose that on the contrary, the perturbed horizon is displaced with
respect to its unperturbed position. The parametric description of the
generators is then 
\begin{equation}
\hat{z}^\alpha(v,\theta^A) = z^\alpha(v,\theta^A) + \lambda
\xi^\alpha(v,\theta^A), 
\label{6.6}
\end{equation}
where the vector $\lambda \xi^\alpha$ points from a point identified
by $(v,\theta^A)$ on the unperturbed horizon to a point (carrying the 
same intrinsic coordinates) on the perturbed horizon. We shall show
below that if Eq.~(\ref{6.4}) holds, then $\xi^\alpha$ must be tangent
to the horizon; it can then be decomposed as $\xi^\alpha = a k^\alpha
+ a^A e^\alpha_A$ for some coefficients $a$ and $a^A$. If $\xi^\alpha$
is tangent to the horizon, then it maps a point $(v,\theta^A)$ on one
generator to another point $(v',\theta'_A)$ on another generator. (If
$\theta'_A = \theta_A$ the vector links two points on the same
generator; this happens when $a^A = 0$.) Because the mapping preserves
the coordinate labels, this amounts to performing a transformation
$(v',\theta'_A) \to (v,\theta_A)$ of the horizon's intrinsic
coordinates. This transformation can always be undone, and we conclude
that $\xi^\alpha$ can be made to vanish whenever it is tangent to the
horizon: $k_\alpha \xi^\alpha = 0 \Rightarrow \xi^\alpha = 0$. By
showing that $k_\alpha \xi^\alpha = 0$ follows from Eq.~(\ref{6.4}) we
therefore prove that Eq.~(\ref{6.4}) implies Eq.~(\ref{6.2}).  

According to Eq.~(\ref{6.6}) the perturbed basis vectors are
$\hat{k}^\alpha = k^\alpha + \lambda \xi^\alpha_{\ ,\beta} k^\beta$
and $\hat{e}^\alpha_A = e^\alpha_A + \lambda \xi^\alpha_{\ ,\beta}
e^\beta_A$. The perturbed metric at the new horizon position is
$\hat{g}_{\alpha\beta}(z+\lambda \xi) = g_{\alpha\beta}(z+\lambda \xi)
+ \lambda h_{\alpha\beta}(z) = g_{\alpha\beta} +
\lambda(g_{\alpha\beta,\gamma} \xi^\gamma + h_{\alpha\beta})$, with
all fields now evaluated at $z$, the position of the unperturbed
horizon. Let now $\hat{e}^\alpha_a$ stand for any one of the vectors
$\hat{k}^\alpha$, $\hat{e}^\alpha_A$, and $e^\alpha_a$ for the
corresponding unperturbed vector. The statement
$\hat{g}_{\alpha\beta}(z+\lambda \xi) \hat{k}^\alpha \hat{e}^\beta_a 
= 0$, when expanded in powers of $\lambda$, leads to  
\[
\bigl( \pounds_\xi g_{\alpha\beta} +
h_{\alpha\beta} \bigr) k^\alpha e^\beta_a = 0. 
\]
The gauge conditions $h_{\alpha\beta} k^\alpha e^\beta_a = 0$
therefore imply $k^\alpha e^\beta_a \pounds_\xi g_{\alpha\beta} = 0$, 
or 
\begin{equation} 
k^\alpha k^\beta \pounds_\xi g_{\alpha\beta} = 0 
= k^\alpha e^\beta_A \pounds_\xi g_{\alpha\beta}. 
\label{6.7}
\end{equation}
These equations come with an immediate interpretation: $\pounds_\xi
g_{\alpha\beta}$ is the change in $h_{\alpha\beta}$ produced by a
gauge transformation generated by the vector field $\xi^\alpha$;
Eq.~(\ref{6.7}) indicates that this transformation must preserve the 
gauge conditions of Eq.~(\ref{6.4}). 

The first of Eq.~(\ref{6.7}) can be expressed in the form $(k^\alpha
\xi_\alpha)_{;\beta} k^\beta - k^\alpha_{\ ;\beta} k^\beta \xi_\alpha
= 0$, or  
\begin{equation} 
\frac{\partial}{\partial v} (k_\alpha \xi^\alpha) = 
\kappa (k_\alpha \xi^\alpha)    
\label{6.8}
\end{equation}
after using Eq.~(\ref{3.2}). If we restrict ourselves to situations in
which the event horizon starts in a stationary state, then $\xi^\alpha
= 0$ initially, and Eq.~(\ref{6.8}) implies that $k_\alpha \xi^\alpha
= 0$ at all times. We therefore have the statement that $\xi^\alpha$
is tangent to the horizon, and the proof that Eq.~(\ref{6.4}) implies 
Eq.~(\ref{6.2}). 

The second of Eq.~(\ref{6.7}) can be expressed in the form 
\begin{equation} 
\frac{\partial}{\partial v} (e^\alpha_A \xi_\alpha) = 
- \frac{\partial}{\partial \theta^A} (k^\alpha \xi_\alpha) 
+ 2\omega_A (k^\alpha \xi_\alpha), 
\label{6.9}
\end{equation} 
where $\omega_A$ was introduced in Eq.~(\ref{2.14}). With $k^\alpha
\xi_\alpha = 0$ and $\xi^\alpha = 0$ initially, this equation states
that $e^\alpha_A \xi_\alpha = 0$ at all times. We therefore see that
$\xi^\alpha$ must be directed along $k^\alpha$, so that a point
$(v,\theta^A)$ is necessarily mapped to a point $(v',\theta^A)$ on the
same generator. This simply corresponds to a reparameterization $v \to
v'(v,\theta^A)$ of the generators, and we conclude that $\xi^\alpha$
can be set equal to zero whenever Eq.~(\ref{6.4}) is enforced.   
 
\subsection{Metric and Lie derivatives} 

In the preferred gauge defined by Eqs.~(\ref{6.2})--(\ref{6.4}), the  
perturbed horizon is in the same coordinate position as the Kerr
horizon, and the perturbed horizon metric is $\hat{\gamma}_{AB} =
\hat{g}_{\alpha\beta} \hat{e}^\alpha_A \hat{e}^\alpha_B = \gamma_{AB}
+ \lambda h_{\alpha\beta} e^\alpha_A e^\beta_B$, having used
Eq.~(\ref{6.3}). This means that the metric perturbation is given by 
\begin{equation}
\gamma^1_{AB} = h_{\alpha\beta} e^\alpha_A e^\beta_B. 
\label{6.10}
\end{equation}

The Lie derivative of $\gamma^1_{AB}$ in the direction of $k^\alpha$
is calculated as $\pounds_k \gamma^1_{AB} = (h_{\alpha\beta}
e^\alpha_A e^\beta_B)_{;\gamma} k^\gamma$. The covariant derivative
of $e^\alpha_A$ in the direction of $k^\alpha$ is evaluated in
Eq.~(\ref{2.14}), and we have $\pounds_k \gamma^1_{AB} =
h_{\alpha\beta;\gamma} e^\alpha_A e^\beta_B k^\gamma + (\omega_A 
e^\alpha_B + e^\alpha_A \omega_B) h_{\alpha\beta} k^\beta$. Using
Eq.~(\ref{6.4}), we arrive at 
\begin{equation} 
\pounds_k \gamma^1_{AB} = 2\sigma_{AB} = h_{\alpha\beta;\gamma}
e^\alpha_A e^\beta_B k^\gamma. 
\label{6.11}
\end{equation} 

The Lie derivative of $\gamma^1_{AB}$ in the direction of
$\phi^\alpha$ is calculated in a similar way. Here we need the
covariant derivative of $e^\alpha_A$ in the direction of
$\phi^\alpha$, which is evaluated as 
\[
e^\alpha_{A;\gamma} \phi^\gamma = e^\alpha_{A;\gamma} e^\gamma_C
\phi^C = (p_{AC} k^\alpha + \Gamma^B_{AC} e^\alpha_B) \phi^C, 
\]
in which we expressed $\phi^\gamma$ as $\phi^C e^\gamma_C$ --- refer
back to Sec.~II C --- and substituted Eq.~(\ref{2.15}). Gathering the
results and using Eq.~(\ref{6.10}) as well as the gauge conditions of
Eq.~(\ref{6.4}), we obtain $\pounds_\phi \gamma^1_{AB} =
h_{\alpha\beta;\gamma} e^\alpha_A e^\beta_B \phi^\gamma 
+ 2\gamma^1_{D(A} \Gamma^D_{B)C} \phi^C$. We now use the identity
$c_A^{\ D} \equiv \phi^D_{\ |A} = \Gamma^D_{AC} \phi^C$ (which follows
from the definition of intrinsic covariant differentiation and the
fact that $\phi^C$ does not depend on the horizon coordinates) to 
manipulate the terms involving the horizon connection. We arrive at  
\begin{equation} 
\pounds_\phi \gamma^1_{AB} =
h_{\alpha\beta;\gamma} e^\alpha_A e^\beta_B \phi^\gamma 
+ c_A^{\ C} \gamma^1_{CB} + c_B^{\ C} \gamma^1_{CA}. 
\label{6.12}
\end{equation} 
Recall that the antisymmetric two-tensor $c_{AB} \equiv - \phi_{A|B}$
was defined and evaluated in Sec.~II C; see Eq.~(\ref{2.17}).  

We now define a four-dimensional version of this tensor with the
relation 
\begin{equation} 
c^{\alpha\beta} = c^{AB} e^\alpha_A e^\beta_B, 
\label{6.13}
\end{equation}
where the indices on $c_{AB}$ are raised with $\gamma^{AB}$, the
inverse of the Kerr horizon metric; this relation is inverted by
$c_{AB} = c_{\alpha\beta} e^\alpha_A e^\beta_B$. With
Eqs.~(\ref{6.10}) and (\ref{6.13}) we have 
\[
c_A^{\ C} \gamma^1_{CB} = c_{\alpha\gamma} h_{\delta\beta} e^\alpha_A
e^\beta_B \bigl(\gamma^{CD} e^\gamma_C e^\delta_D \bigr). 
\]
Using now the completeness relations of Eq.~(\ref{2.12}) and the
properties $c_{\alpha\gamma} k^\gamma = c_{\alpha\gamma} N^\gamma =
0$, we obtain $c_A^{\ C} \gamma^1_{CB} = c_\alpha^{\ \gamma}
h_{\gamma\beta} e^\alpha_A e^\beta_B$ and Eq.~(\ref{6.12}) becomes 
\begin{equation} 
\pounds_\phi \gamma^1_{AB} =
h_{\alpha\beta;\gamma} e^\alpha_A e^\beta_B \phi^\gamma 
+ \bigl( c_\alpha^{\ \gamma} h_{\gamma\beta} 
+ c_\beta^{\ \gamma} h_{\gamma\alpha} \bigr) 
e^\alpha_A e^\beta_B. 
\label{6.14}
\end{equation} 

Finally, the Lie derivative of $\gamma^1_{AB}$ in the direction of
$t^\alpha$ is calculated as $\pounds_t \gamma^1_{AB} = \pounds_k 
\gamma^1_{AB} - \Omega_{\rm H} \pounds_\phi \gamma^1_{AB}$. From
Eqs.~(\ref{6.11}) and (\ref{6.14}) we obtain 
\begin{equation} 
\pounds_t \gamma^1_{AB} = 
h_{\alpha\beta;\gamma} e^\alpha_A e^\beta_B t^\gamma 
- \Omega_{\rm H} \bigl( c_\alpha^{\ \gamma} h_{\gamma\beta}  
+ c_\beta^{\ \gamma} h_{\gamma\alpha} \bigr) 
e^\alpha_A e^\beta_B. 
\label{6.15}
\end{equation} 
  
\subsection{Fluxes} 
 
It is a straightforward matter to substitute Eqs.~(\ref{6.11}),
(\ref{6.14}), and (\ref{6.15}) into the flux formulae of
Eqs.~(\ref{4.22})--(\ref{4.24}). We first obtain 
\begin{eqnarray*} 
\sigma^{AB} \pounds_k \gamma^1_{AB} &=& \frac{1}{2} F[k], \\ 
\sigma^{AB} \pounds_\phi \gamma^1_{AB} &=& \frac{1}{2} \bigl( F[\phi]
+ G \bigr), \\ 
\sigma^{AB} \pounds_t \gamma^1_{AB} &=& \frac{1}{2} \bigl( F[t]
- \Omega_{\rm H} G \bigr),
\end{eqnarray*} 
where 
\[
F[\xi] \equiv h_{\alpha\beta;\gamma} k^\gamma h_{\mu\nu;\lambda}
\xi^\lambda \bigl(e^{A\alpha} e^\mu_A \bigr) \bigl(e^{B\beta} e^\nu_B
\bigr) 
\]
and 
\[
G \equiv h_{\alpha\beta;\gamma} k^\gamma \bigl( c_\mu^{\ \lambda}
h_{\lambda\nu} + c_\nu^{\ \lambda} h_{\lambda\mu} \bigr) 
\bigl(e^{A\alpha} e^\mu_A \bigr) \bigl(e^{B\beta} e^\nu_B \bigr). 
\]

By virtue of Eq.~(\ref{2.12}), $e^{A\alpha} e^\mu_A = g^{\alpha\mu} +
k^\alpha N^\mu + N^\alpha k^\mu$, and this identity can be used to
simplify our expressions for $F[\xi]$ and $G$. To simplify things
further we also write  
\[
k^\beta h_{\alpha\beta;\gamma} \xi^\gamma = (h_{\alpha\beta}
k^\beta)_{;\gamma} \xi^\gamma - h_{\alpha\beta} k^\beta_{\ ;\gamma}
\xi^\gamma,
\]
and we note that according to Eqs.~(\ref{2.13}) and (\ref{2.14}), any
tangential derivative of the form $k^\beta_{\ ;\gamma} \xi^\gamma$ is
necessarily proportional to $k^\beta$. The preceding equation
therefore becomes    
\[
k^\beta h_{\alpha\beta;\gamma} \xi^\gamma = (h_{\alpha\beta}
k^\beta)_{;\gamma} \xi^\gamma - p[\xi] (h_{\alpha\beta} k^\beta), 
\]
where $p[\xi]$ is a proportionality factor that depends on the choice
of vector $\xi^\alpha$; for example, $p[k] = \kappa$ and $p[\phi] =
\omega_A \phi^A$. At this stage it is convenient to supplement the
three gauge conditions of Eqs.~(\ref{6.4}) with the fourth condition
implied by Eq.~(\ref{6.5}); adopting the radiation gauge allows us to
set $h_{\alpha\beta} k^\beta = 0$ on the horizon, and therefore to
discard all terms of the form $k^\beta h_{\alpha\beta;\gamma}
\xi^\gamma$. This greatly simplifies our expressions for $F[\xi]$ and
$G$. After carrying out these manipulations we obtain   
\[
F[\xi] = h_{\alpha\beta;\gamma} k^\gamma h^{\alpha\beta}_{\ \ ;\delta}
\xi^\delta 
\]
and 
\[
G = 2 h_{\alpha\beta;\gamma} k^\gamma c^{\alpha\delta} 
h_\delta^{\ \beta}.
\]

Substituting these results into Eqs.~(\ref{4.22})--(\ref{4.24}) we
arrive at 
\begin{eqnarray} 
\Mdot &=& \oint T_{\alpha\beta} k^\alpha t^\beta\, dS - \Omega_{\rm H}
\oint q\, dS, 
\label{6.16} \\ 
\Jdot &=& -\oint T_{\alpha\beta} k^\alpha \phi^\beta\, dS 
- \oint q\, dS,
\label{6.17} \\ 
\frac{\kappa}{8\pi} \Adot &=& \oint T_{\alpha\beta} k^\alpha k^\beta\,  
dS,  
\label{6.18}
\end{eqnarray} 
where 
\begin{equation} 
T_{\alpha\beta} \equiv \frac{1}{32\pi} \bigl\langle h_{\mu\nu;\alpha} 
h^{\mu\nu}_{\ \ ;\beta} \bigr\rangle 
\label{6.19}
\end{equation}
and 
\begin{equation} 
q \equiv \frac{1}{16\pi} \bigl\langle h_{\mu\nu;\alpha} k^\alpha
c^{\mu\lambda} h_\lambda^{\ \nu} \bigr\rangle. 
\label{6.20}
\end{equation} 
We recall that these results are formulated in the radiation gauge of
Eq.~(\ref{6.5}). And we mention that an alternative expression for $q$
is 
\begin{equation} 
q = -\frac{1}{16\pi} \bigl\langle \phi^\alpha_{\ ;\beta}
h^\beta_{\ \gamma} h^\gamma_{\ \alpha;\mu} k^\mu \bigr\rangle;  
\label{6.21}
\end{equation} 
this follows from substituting the (easily-derived) identity
$c_{\alpha\beta} = -\phi_{\alpha;\beta} - \phi_{\alpha;\mu} N^\mu
k_\beta + \phi_{\beta;\mu} N^\mu k_\alpha$ into Eq.~(\ref{6.20}) and
simplifying the result.

The integrals involving $T_{\alpha\beta}$ in
Eqs.~(\ref{6.16})--(\ref{6.18}) are formally identical to the flux
formulae that would be obtained for a horizon perturbed by a matter
field with stress-energy tensor $T_{\alpha\beta}$ (see, for example,
Sec.~6.4.2 of Ref.~\cite{carter:79}, or Sec.~4.3.6 of
Ref.~\cite{poisson:04c}). It is
therefore tempting to view Eq.~(\ref{6.19}) as a definition of an
effective stress-energy tensor for gravitational radiation crossing
the event horizon. While in general the integrals involving $q$ spoil
this interpretation, we see that there exists an approximate regime in 
which the interpretation is sound: this is the high-frequency regime
first investigated by Isaacson
\cite{isaacson:68a, isaacson:68b}. 
Schematically, $T \sim (\nabla h)^2$
while $q \sim h \nabla h$, and the additional derivative ensures that
$T_{\alpha\beta}$ dominates over $q$ in the high-frequency limit. And
indeed, our expression for $T_{\alpha\beta}$, as given by
Eq.~(\ref{6.19}), does agree with Isaacson's effective stress-energy
tensor. It should be noted that the time averaging involved in
Eq.~(\ref{6.19}) is different from the spacetime (Brill-Hartle
\cite{brill-hartle:64}) 
averaging used in Isaacson's construction; but it is plausible that
the two averaging procedures are reconciled after $T_{\alpha\beta}$ is 
integrated over $dS$. It should also be noted that while
Eq.~(\ref{6.19}) is formulated in the radiation gauge of
Eq.~(\ref{6.5}), Isaacson has shown that the expression is actually
gauge invariant in the high-frequency limit.  

The flux formulae of Eqs.~(\ref{6.16})--(\ref{6.18}) are not limited
to the high-frequency regime; they can applied in general situations, 
provided that the metric perturbation $h_{\alpha\beta}$ satisfies the
gauge conditions $h_{\alpha\beta} k^\beta = 0$ on the horizon. These
formulae could in principle be used in tandem with Chrzanowski's
metric reconstruction 
\cite{chrzanowski:75a, wald:78, lousto-whiting:02, ori:03} 
to calculate the absorption of mass and
angular momentum by a Kerr black hole. But to proceed like this would
be much more involved than to proceed directly with the curvature
formalism of Sec.~V. The flux formulae could also be used in the
context of a Schwarzschild black hole, but the formulation given here
is not optimal and I shall refine it in the following section. My
main purpose in this section was to introduce the preferred gauge
(which will be used also in Sec.~VII) and to establish the preceding 
connection with Isaacson's effective stress-energy tensor
\cite{isaacson:68a, isaacson:68b}.  

\section{Metric formalism for Schwarzschild black holes} 

In this section I fulfill the promise made in Sec.~VI, to translate
the flux formulae of Eqs.~(\ref{4.22})--(\ref{4.24}) into a more
practical language that involves the metric perturbations of a
Schwarzschild black hole. The key aspects of the theory of first-order
perturbations of this spacetime are summarized the Appendix. 

\subsection{Background spacetime} 

The Schwarzschild metric in Eddington-Finkelstein coordinates
$(v,r,\theta,\phi)$ is given by 
\begin{equation}
ds^2 = -f\, dv^2 + 2\, dvdr + r^2\, d\Omega^2, 
\label{7.1}
\end{equation}
where $f=1-2M/r$ and $d\Omega^2 = \Omega_{AB} d\theta^A d\theta^B =
d\theta^2 + \sin^2\theta\, d\phi^2$. The subset of coordinates
$(v,\theta,\phi)$ is used on the horizon; $v$ is a parameter on the
null generators, and $\theta^A = (\theta,\phi)$ are comoving
coordinates. In the spacetime coordinates $(v,r,\theta,\phi)$ the
basis vectors are $k^\alpha = (1,0,0,0)$, $N^\alpha = (0,-1,0,0)$,
$e^\alpha_\theta = (0,0,1,0)$, and $e^\alpha_\phi = (0,0,0,1)$. The
metric of the unperturbed horizon is $\gamma_{AB} = r_+^2
\Omega_{AB}$, where $r_+ = 2M$.  

\subsection{Metric perturbation} 

The metric perturbation $\lambda h_{\alpha\beta}$ (denoted $\delta 
g_{\alpha\beta}$ in the Appendix) is cast in the radiation gauge of 
Eq.~(\ref{6.5}). We therefore impose $h_{\alpha\beta} k^\beta = 0$ on
the perturbed horizon, which is still located at $r = r_+$ --- see the
discussion of Sec.~VI A. The gauge conditions imply that the
components $h_{v\alpha}$ of the metric perturbations all
vanish. According to Eq.~(\ref{6.10}) the perturbation of the horizon
metric is  
\begin{equation}
\gamma^1_{AB} = h_{\alpha\beta} e^\alpha_A e^\beta_B, 
\label{7.2}
\end{equation}
where $e^\alpha_A$ are the background basis vectors. 

The odd-parity sector of the perturbations is described by
Eqs.~(\ref{A.14}), (\ref{A.15}) and it involves the functions
$h_r^{lm}$ and $h_2^{lm}$ of the coordinates $(v,r)$. The
combinations of Eq.~(\ref{A.16}) are gauge invariant, and they are
used in Eq.~(\ref{A.18}) to form the Regge-Wheeler function 
$\Psi_{\rm RW}^{lm}(v,r)$ 
\cite{regge-wheeler:57}, 
which is also gauge invariant. A simple calculation shows that near
the horizon,  
\[
\Psi_{\rm RW}^{lm} = \frac{1}{2r_+} 
\frac{\partial h_2^{lm}}{\partial v} + O(f), 
\]
so that $h_2(v,r_+) = 2r_+ \int^v \Psi_{\rm RW}^{lm}(v',r_+)\,
dv'$. The odd-parity sector of Eq.~(\ref{7.2}) is therefore 
\begin{equation}
\gamma^{1,{\rm odd}}_{AB}(v,\theta^A) = 2 r_+ \sum_{lm}
X^{lm}_{AB}(\theta^A) \int^v \Psi_{\rm RW}^{lm}(v',r_+)\, dv', 
\label{7.3}
\end{equation}
where $X^{lm}_{AB}(\theta^A)$ are the odd-parity tensorial harmonics
introduced in Eq.~(\ref{A.8}). Here and below, the sum over $l$ is
restricted to $l \geq 2$, and the sum over $m$ extends from $-l$ to
$l$. 

The even-parity sector of the metric perturbations is described by
Eqs.~(\ref{A.20})--(\ref{A.22}) and it involves the functions
$h^{lm}_{rr}$, $j^{lm}_{r}$, $K^{lm}$, and $G^{lm}$ of the coordinates
$(v,r)$. The combinations of Eq.~(\ref{A.23}) are gauge invariant, and
they are used in Eq.~(\ref{A.24}) to form the Zerilli-Moncrief
function $\Psi_{\rm ZM}^{lm}(v,r)$ 
\cite{zerilli:70, moncrief:74, lousto-price:97}, which is also gauge
invariant. A simple calculation shows that near the horizon,  
\begin{eqnarray*} 
\Psi_{\rm ZM}^{lm} &=& -\frac{4 r_+^2}{l(l+1)(l^2+l+1)}
\frac{\partial}{\partial v} \biggl[ K^{lm} - \frac{1}{2} l(l+1) G^{lm}
\biggr] 
\\ & & \mbox{} 
+ \frac{2 r_+}{l(l+1)} K^{lm} + O(f). 
\end{eqnarray*}
On the other hand, an analysis of the linearized field equations near
the horizon shows that in the absence of sources, $K^{lm} =
\frac{1}{2} l(l+1) G^{lm} + O(f)$, so that $K^{lm}(v,r_+) =
\frac{1}{2} l(l+1) G^{lm}(v,r_+) = \frac{1}{2} l(l+1) r_+^{-1} 
\Psi^{lm}_{\rm ZM}(v,r_+)$. The even-parity sector of Eq.~(\ref{7.2})
is therefore  
\begin{equation}
\gamma^{1,{\rm even}}_{AB}(v,\theta^A) = r_+ \sum_{lm}
Z^{lm}_{AB}(\theta^A) \Psi_{\rm ZM}^{lm}(v,r_+), 
\label{7.4}
\end{equation}
where $Z^{lm}_{AB}(\theta^A)$ are the even-parity tensorial harmonics
introduced in Eq.~(\ref{A.7}).   

The complete perturbation of the horizon metric is given by the sum of
Eqs.~(\ref{7.3}) and (\ref{7.4}), 
\begin{equation}
\gamma^1_{AB} = r_+ \sum_{lm} \biggl[ 
2 X^{lm}_{AB} \int^v \Psi_{\rm RW}^{lm}(v')\, dv' 
+ Z^{lm}_{AB} \Psi_{\rm ZM}^{lm}(v) \biggr],  
\label{7.5}
\end{equation} 
where we have set $\Psi_{\rm RW}^{lm}(v') \equiv 
\Psi_{\rm RW}^{lm}(v',r_+)$ and $\Psi_{\rm ZM}^{lm}(v) \equiv  
\Psi_{\rm ZM}^{lm}(v,r_+)$. Notice that this tensor is tracefree:
$\Omega^{AB} \gamma^1_{AB} = 0$. The fact that $\gamma^1_{AB}$ is
related to the {\it integral} of the Regge-Wheeler function means that
this variable is rather ill-suited to describe ingoing gravitational 
radiation crossing the event horizon; a similar statement is made in
subsection 5 of the Appendix, about outgoing radiation at future null
infinity. 

The shear tensor is obtained by differentiating $\frac{1}{2}
\gamma^1_{AB}$ with respect to $v$, and 
\begin{equation} 
\sigma_{AB} = \frac{r_+}{2} \sum_{lm} \biggl[ 
2 X^{lm}_{AB} \Psi_{\rm RW}^{lm}(v) 
+ Z^{lm}_{AB} \dot{\Psi}_{\rm ZM}^{lm}(v) \biggr];  
\label{7.6}
\end{equation}  
this is also equal to $\frac{1}{2} \pounds_k \gamma^1_{AB}$ and 
$\frac{1}{2} \pounds_t \gamma^1_{AB}$. Because the spherical harmonics
are all proportional to $e^{im\phi}$, we also have 
\begin{eqnarray}
\pounds_\phi \gamma^1_{AB} &=& r_+ \sum_{lm} (im) \biggl[  
2 X^{lm}_{AB} \int^v \Psi_{\rm RW}^{lm}(v')\, dv' 
\nonumber \\ & & \mbox{} 
+ Z^{lm}_{AB} \Psi_{\rm ZM}^{lm}(v) \biggr]. 
\label{7.7}
\end{eqnarray} 

\subsection{Fluxes} 

It is a straightforward task to substitute the preceding results for
$\sigma_{AB}$, $\pounds_k \gamma_{AB}$, $\pounds_t \gamma_{AB}$, and
$\pounds_\phi \gamma_{AB}$ into the flux formulae of
Eqs.~(\ref{4.22})--(\ref{4.24}), and then to integrate over $dS =
r_+^2 \sin\theta\, d\theta d\phi$. The integrations can be carried out
explicitly with the help of the orthogonality relations of
Eqs.~(\ref{A.11}), (\ref{A.12}), and we arrive at 
\begin{eqnarray} 
\Mdot &=& \frac{1}{64\pi} \sum_{l m} (l-1) l (l+1)
(l+2) 
\nonumber \\ & & \mbox{} \times 
\Bigl\langle 4 \bigl| \Psi^{lm}_{\rm RW}(v) \bigr|^2 
+ \bigl| \dot{\Psi}^{lm}_{\rm ZM}(v) \bigr|^2 \Bigr\rangle 
\label{7.8}
\end{eqnarray}
and 
\begin{eqnarray} 
\Jdot &=& \frac{1}{64\pi} \sum_{l m} (l-1) l (l+1) (l+2) (im) 
\nonumber \\ & & \mbox{} \times 
\Bigl\langle 4 \Psi^{lm}_{\rm RW}(v) \int^v 
\bar{\Psi}^{lm}_{\rm RW}(v')\, dv' 
\nonumber \\ & & \mbox{} 
+ \dot{\Psi}^{lm}_{\rm ZM}(v)
\bar{\Psi}^{lm}_{\rm ZM}(v) \Bigr\rangle.
\label{7.9}
\end{eqnarray} 
Notice that except for the substitution $u \to v$, these formulae are
identical to Eqs.~(\ref{A.26}) and (\ref{A.27}), which give the rates
at which energy and angular momentum are transported to future null
infinity. Note also that for a nonrotating black hole, the first law
of black-hole mechanics reduces to $(\kappa/8\pi) \Adot =
\Mdot$. Finally, note that although it involves complex quantities,
the expression for $\Jdot$ is real; this property follows from the
identity $\Psi^{l,-m} = (-1)^m \bar{\Psi}^{lm}$, which is inherited
from the spherical harmonics, and which is satisfied by
both the Regge-Wheeler and Zerilli-Moncrief functions. 

Equations (\ref{7.8}) and (\ref{7.9}) give the final form of the flux
formulae for the case of a Schwarzschild black hole. The steps
required to compute $\Mdot$, $\Jdot$, and $\Adot$ are therefore these
(see also Sec.~I D): First, solve the Regge-Wheeler
\cite{regge-wheeler:57} and Zerilli \cite{zerilli:70} 
equations for the functions 
$\Psi^{lm}_{\rm RW}(v,r)$ and $\Psi^{lm}_{\rm ZM}(v,r)$ defined in the 
Appendix, for all relevant value of $l$ and $m$. Second, evaluate
the functions at $r=r_+$ and compute the integral of 
$\Psi^{lm}_{\rm RW}(v,r_+)$ and the derivative of 
$\Psi^{lm}_{\rm ZM}(v,r_+)$. Third, and finally, substitute these 
functions into the flux formulae of Eqs.~(\ref{7.8})--(\ref{7.9}) and 
sum over $l$ and $m$. 

These flux formulae were first presented and used by Martel 
\cite{martel:04} in his
exploration of gravitational-wave processes associated with the motion 
of a small-mass body around a Schwarzschild black hole. Although he
arrived at the correct results, the derivation of Eqs.~(\ref{7.8}) and
(\ref{7.9}) given by Martel is flawed --- it incorporates both a
conceptual and a computational error. The conceptual error is that
Martel based his derivation on Isaacson's effective stress-energy
tensor for gravitational waves
\cite{isaacson:68a, isaacson:68b}, 
incorrectly
assuming that the high-frequency description is always applicable near
the event horizon of a black hole (as it always is near future null
infinity). This assumption was motivated by the observation that for a
stationary observer just above the event horizon, any incoming
gravitational wave would appear highly blueshifted. While the
observation is of course valid, the observer-dependent blueshift does
not by itself produce a perturbation that satisfies the assumptions
underlying Isaacson's construction --- the static Schwarzschild
coordinates do not form a ``steady'' coordinate system near the
horizon. Martel's starting point was therefore Eqs.~(\ref{6.16}) and
(\ref{6.17}) without the integrations over $q$, and this should have
led him to the wrong formula for the flux of angular momentum (the $q$
integral does not contribute to $\Mdot$, because $\Omega_{\rm H} = 0$
for a nonrotating black hole). That he nevertheless obtained
Eq.~(\ref{7.9}) is due to a computational error that accidentally
compensated for the absence of the $q$ integral.  

\section{Small-hole/slow-motion approximation for a Schwarzschild
black hole} 

In this and the following section I describe an application of the
flux formulae obtained in Sec.~V and VII. I shall evaluate $\Mdot$
and $\Jdot$ in a small-hole/slow-motion (SH/SM) approximation in which 
the ratio $M/{\cal R}$, where $M$ is the black-hole mass and 
${\cal R}$ the radius of curvature of the spacetime in which the
black hole moves, is assumed to be small. I begin in this section
with a nonrotating black hole, and I will consider the case of a
rotating black hole in Sec.~IX.  

\subsection{The SH/SM approximation} 

We imagine a situation in which a black hole of mass $M$ is not
at rest and isolated, but moves in a spacetime that may contain a 
number of additional bodies. The radius of curvature of this
external spacetime is denoted ${\cal R}$, and although this may depend
on $M$ (if the geometry of the external spacetime is significantly 
influenced by the black hole), we assume that 
\begin{equation}
M/{\cal R} \ll 1 \quad \mbox{(small-hole/slow-motion
approximation)}. 
\label{8.1} 
\end{equation} 
More precisely, we assume that $M$ is much smaller than all of 
${\cal R}$, ${\cal L}$, and ${\cal T}$, where ${\cal L}$ is the scale
of inhomogeneity in the external universe, and ${\cal T}$ is the
time scale over which changes occur in the external universe. To
simplify the notation we take ${\cal R}$, ${\cal L}$, and ${\cal T}$
to be of the same order of magnitude. (These quantities, and many of
the concepts used throughout Secs.~VIII and IX, are defined precisely
in Thorne and Hartle
\cite{thorne-hartle:85}; 
the reader is referred to this paper for details.) 

Near the black hole the spacetime resembles closely the spacetime of
an isolated black hole: the gravitational field is strongly dominated
by the hole's contribution, and the influence of the external universe
is weak. But the hole is not truly isolated, and it is slightly
distorted by the tidal gravitational field supplied by the external
universe. As a result of this interaction, the hole's mass and angular
momentum change with time, and we wish here to calculate these
changes.   

When viewed on the large scale ${\cal R}$, the black hole occupies a  
very small region of the actual spacetime, and this region can be
idealized as a world line $\gamma$ in the external spacetime. Let
$u^\alpha$ be the (normalized) tangent vector to this world line, and
call this the four-velocity of the black hole in the external
spacetime. It can be shown that to a very good degree of accuracy, the
motion of the black hole is geodesic in this spacetime 
\cite{manasse:63, kates:80, death:96, mino-etal:97, quinn-wald:97,
      poisson:04b}.  
Let $e^\alpha_a$ (with the index $a$ running from 1 to 3) be a set of 
orthonormal vectors attached to $\gamma$; let these vectors be
orthogonal to $u^\alpha$ and choose them to be parallel transported on
the world line. The tetrad $(u^\alpha,e^\alpha_a)$ defines a reference
frame in a neighborhood of $\gamma$, and we shall call this frame the
{\it local asymptotic rest frame} of the black hole in the external
spacetime.     

We assume that the Ricci tensor of the external spacetime vanishes
in a neighborhood of $\gamma$, so that no matter will appear in the 
vicinity of the black hole. The curvature of the external spacetime  
in this neighborhood is therefore described entirely by the Weyl
tensor $C_{\alpha\gamma\beta\delta}$. The Weyl tensor evaluated on the
world line can be decomposed in the tetrad $(u^\alpha,e^\alpha_a)$; we 
write, for example, $C_{0c0d}(v) \equiv
C_{\alpha\gamma\beta\delta}(\gamma) u^\alpha e^\gamma_c u^\beta
e^\delta_d$ and $C_{0cbd}(v) \equiv
C_{\alpha\gamma\beta\delta}(\gamma) u^\alpha e^\gamma_c e^\beta_b
e^\delta_d$, with $v$ denoting proper time on $\gamma$ --- it will
later be identified with an advanced-time coordinate on the black-hole
horizon. It is easy to show that the frame tensors  
\cite{thorne-hartle:85} 
\begin{equation}
{\cal E}_{ab}(v) = C_{0a0b}(v), \qquad 
{\cal B}_{ab}(v) = \frac{1}{2} \varepsilon_a^{\ cd} C_{cdb0}(v),  
\label{8.2}   
\end{equation}
where $\varepsilon_{abc}$ is the permutation symbol (all frame indices
are lowered and raised with $\delta_{ab}$ and its inverse,
respectively), are symmetric and tracefree, and that their components
comprise all ten independent components of the Weyl tensor. These
frame tensors are the tidal gravitational fields supplied by the
external universe, and these are responsible for the tidal distortion
of the black hole.  

It will be convenient to promote the tidal fields ${\cal E}_{ab}$ and
${\cal B}_{ab}$ to four-dimensional spacetime tensors. We therefore
define 
\begin{equation}
{\cal E}_{\alpha\beta} = {\cal E}_{ab} e^a_\alpha e^b_\beta, \qquad
{\cal B}_{\alpha\beta} = {\cal B}_{ab} e^a_\alpha e^b_\beta,
\label{8.3}
\end{equation}
where $e^a_\alpha \equiv \delta^{ab} g_{\alpha\beta} e^\beta_b$. It is
not difficult to show that these tensors are also given by 
\begin{equation}
{\cal E}_{\alpha\beta} = C_{\mu\alpha\nu\beta} u^\mu u^\nu 
\label{8.4}
\end{equation}
and
\begin{equation}
{\cal B}_{\alpha\beta} = \frac{1}{2} u^\mu 
\varepsilon_{\mu\alpha}^{\ \ \ \gamma\delta} 
C_{\gamma\delta\beta\nu} u^\nu,
\label{8.5}
\end{equation} 
where the Levi-Civita tensor $\varepsilon_{\mu\alpha\nu\beta}$ and
the Weyl tensor $C_{\mu\alpha\nu\beta}$ are evaluated on the world
line $\gamma$.   

As an example of a SH/SM situation, consider a black hole of mass $M$
on a circular orbit of radius $b$ in the gravitational field of an
external body of mass $M_{\rm ext}$. The radius of curvature of the 
external spacetime at the position of the black hole is such that 
${\cal R}^{-2} \sim (M + M_{\rm ext})/b^3$, and we have 
\begin{equation}
\frac{M}{\cal R} \sim \frac{M}{M+M_{\rm ext}} V^3, \qquad
V = \sqrt{\frac{M + M_{\rm ext}}{b}}, 
\label{8.6}
\end{equation}
where $V$ is a measure of the hole's orbital velocity. There are many
ways by which $M/{\cal R}$ can be made small. One way is to let
$M/M_{\rm ext} \ll 1$; then $M/{\cal R}$ will be small irrespective of
the magnitude of $V$. This is the {\it small-hole approximation},
which allows the small black hole to move at relativistic speeds in
the strong gravitational field of the external body. Another way is to
let $V \ll 1$; then $M/{\cal R}$ will be small for all mass
ratios. This is the {\it slow-motion approximation}, which allows the
slowly-moving black hole to have a mass comparable to (or even much
larger than) $M_{\rm ext}$. These two limiting approximations are
special cases of the fundamental requirement that $M/{\cal R}$ be
small; we therefore call the approximation $M/{\cal R} \ll 1$ the
SH/SM approximation.   

\subsection{Metric of a tidally distorted black hole} 

My considerations thus far have been general, and they apply to 
rotating as well as nonrotating black holes. I now specialize to 
nonrotating black holes. 

The metric of a Schwarzschild black hole immersed in an external 
universe can be obtained by solving the Einstein field
equations. Because the tidal potentials scale as $(r/{\cal R})^2 
\ll 1$, where $r$ is a measure of distance from the black hole, it is 
sufficient to linearize the equations with respect to the
Schwarzschild solution; this is a standard application of black-hole
perturbation theory. Explicit forms for the metric were obtained by
Manasse \cite{manasse:63}, 
Alvi \cite{alvi:00}, 
Detweiler \cite{detweiler:01}, 
and Poisson \cite{poisson:04a}, 
and I summarize their results here. I follow the description of
Ref.~\cite{poisson:04a}, but I switch from
the retarded coordinates $(u,r,\theta^A)$ used there to a set of
advanced coordinates $(v,r,\theta^A)$ which are well behaved on the
event horizon; the expressions for the perturbed metric are identical,
except for the correspondence $du \to -dv$.    

The metric takes the form of an expansion in powers of $r/{\cal R}$, 
but it is correct to all orders in $M/r$. It is given by    
\begin{eqnarray} 
g_{vv} &=& -f(1+r^2 f {\cal E}^*) + O(r^3/{\cal R}^3), 
\label{8.7} \\ 
g_{vr} &=& 1, 
\label{8.8} \\ 
g_{vA} &=& -\frac{2}{3} r^3 f ({\cal E}^*_A + {\cal B}^*_A) +
O(r^4/{\cal R}^3), 
\label{8.9} \\ 
g_{AB} &=& r^2 \Omega_{AB} - \frac{1}{3} r^4 \biggl[
\biggl(1-\frac{2M^2}{r^2} \biggr) {\cal E}^*_{AB} + {\cal B}^*_{AB}
\biggr] 
\nonumber \\ & & \mbox{} 
+ O(r^5/{\cal R}^3), 
\label{8.10}
\end{eqnarray} 
where $f = 1-2M/r$. The irreducible tidal fields are defined by 
\begin{eqnarray} 
{\cal E}^* &=& \sum_{\sf m} {\cal E}_{\sf m} Y^{\sf m}, 
\label{8.11} \\ 
{\cal E}^*_A &=& \frac{1}{2} \sum_{\sf m} {\cal E}_{\sf m} 
Y^{\sf m}_{:A},   
\label{8.12} \\ 
{\cal E}^*_{AB} &=& \sum_{\sf m} {\cal E}_{\sf m} 
Z^{\sf m}_{AB}, 
\label{8.13} \\ 
{\cal B}^*_A &=& \frac{1}{2} \sum_{\sf m} {\cal B}_{\sf m} 
X^{\sf m}_{A},   
\label{8.14} \\ 
{\cal B}^*_{AB} &=& -\sum_{\sf m} {\cal B}_{\sf m} 
X^{\sf m}_{AB}, 
\label{8.15}
\end{eqnarray}
where $Y^{\sf m}$, $Y^{\sf m}_{:A}$, $Z^{\sf m}_{AB}$, $X^{\sf m}_A$,
and $X^{\sf A}_{AB}$ are the real spherical harmonics of degree $l=2$
that are introduced in subsection 2 of the Appendix, and 
\begin{eqnarray} 
{\cal E}_0 &=& {\cal E}_{33} = -({\cal E}_{11} + {\cal E}_{22}), 
\label{8.16} \\ 
{\cal E}_{1c} &=& 2 {\cal E}_{13}, 
\label{8.17} \\ 
{\cal E}_{1s} &=& 2 {\cal E}_{23}, 
\label{8.18} \\ 
{\cal E}_{2c} &=& \frac{1}{2} ({\cal E}_{11} - {\cal E}_{22}), 
\label{8.19} \\ 
{\cal E}_{2s} &=& {\cal E}_{12}, 
\label{8.20} 
\end{eqnarray} 
with corresponding relations defining ${\cal B}_{\sf m}$. (Notice that
a typographical error contained in Ref.~\cite{poisson:04a} is hereby 
corrected.) 

In the limit $M/r \to 0$ (keeping $r/{\cal R}$ fixed), the 
metric of Eqs.~(\ref{8.7})--(\ref{8.10}) becomes the metric of the 
external spacetime expressed as an expansion in powers of 
$r/{\cal R}$ about the timelike geodesic $\gamma$. In this limit the 
interpretation of $v$ as proper time on the world line becomes
precise, and ${\cal E}_{ab}(v)$, ${\cal B}_{ab}(v)$ are recognized
as frame components of the Weyl tensor evaluated on $\gamma$. In the
limit $r/{\cal R} \to 0$ (keeping $M/r$ fixed), the metric of
Eqs.~(\ref{8.7})--(\ref{8.10}) becomes the metric of an isolated
Schwarzschild black hole expressed in ingoing Eddington-Finkelstein
coordinates; there is no notion of a world line $\gamma$ in this
limit. For small values of $r/{\cal R}$ and arbitrary values of $M/r$,
the metric of Eqs.~(\ref{8.7})--(\ref{8.10}) describes a black hole 
distorted by the tidal gravitational fields supplied by the external
universe.  

The metric perturbation $h_{\alpha\beta}$ defined by
Eqs.~(\ref{8.7})--(\ref{8.10}) satisfies the conditions
$h_{\alpha\beta} k^\alpha k^\beta = h_{\alpha\beta} k^\alpha e^\beta_A
= 0$ at $r = r_+ = 2M$, because $h_{vv}$ and $h_{vA}$ are all
proportional to $f = 1-2M/r$. The metric perturbation therefore
satisfies the preferred gauge conditions of Eq.~(\ref{6.4}). It can
also be checked directly from the metric that the hypersurface $r=2M$
is null, and that its generators move with constant values of 
$\theta^A$. The parameter on the generators is $v$, and a short
calculation reveals that the surface gravity $\kappa$ is equal to its
Schwarzschild value $(4M)^{-1}$ up to a fractional correction of order 
$(M/{\cal R})^3$.  

\subsection{Odd-parity contribution to shear} 

Although the metric of the tidally distorted black hole is already
expressed in the preferred gauge, it is safer (and as it turns out,
necessary) to calculate the shear tensor $\sigma_{AB}$ by first
obtaining the gauge-invariant Regge-Wheeler 
\cite{regge-wheeler:57} 
and Zerilli-Moncrief
\cite{zerilli:70, moncrief:74, lousto-price:97}   
functions; the relation between these quantities is given by
Eq.~(\ref{7.6}). We begin here with the odd-parity piece of the shear
tensor. A description of this sector of the metric perturbations is
provided in subsection 3 of the Appendix.  

When $l=2$ the odd-parity perturbations can be expanded as 
\begin{equation}
h_{iA} = \sum_{\sf m} h_i^{\sf m}(v,r) X^{\sf m}_A(\theta^A) 
\label{8.21}
\end{equation}
and 
\begin{equation}
h_{AB} = \sum_{\sf m} h_2^{\sf m}(v,r) X^{\sf m}_{AB}(\theta^A).  
\label{8.22}
\end{equation}
The combinations 
\begin{equation} 
\tilde{h}^{\sf m}_i = h^{\sf m}_i + \frac{1}{2} h^{\sf m}_{2,i} -
\frac{1}{r} r_{,i} h^{\sf m}_2 
\label{8.23}
\end{equation}
are gauge invariant, and the Regge-Wheeler function is defined by 
\begin{equation} 
\Psi_{\rm RW}^{\sf m} = \frac{1}{r} r^{,i} \tilde{h}^{\sf m}_i. 
\label{8.24}
\end{equation} 

Comparison of Eqs.~(\ref{8.21}), (\ref{8.22}) with Eqs.~(\ref{8.9}),
(\ref{8.10}) using Eqs.~(\ref{8.14}), (\ref{8.15}) reveals that
$h_v^{\sf m} = -\frac{1}{3} r^3 f {\cal B}_{\sf m}$, $h_r^{\sf m} 
= 0$, and $h_2^{\sf m} = \frac{1}{3} r^4 {\cal B}_{\sf m}$. Equation
(\ref{8.23}) then gives $\tilde{h}_v^{\sf m} = -\frac{1}{3} r^3 f
{\cal B}_{\sf m}$ and $\tilde{h}_r^{\sf m} = \frac{1}{3} r^3 
{\cal B}_{\sf m}$, up to smaller terms proportional to 
$d{\cal B}_{\sf m}/dv \sim {\cal R}^{-3}$. From Eq.~(\ref{8.24}) we
obtain $\Psi_{\rm RW}^{\sf m} = 0$. This curious result leads to the
conclusion that the metric of Eqs.~(\ref{8.7})--(\ref{8.10}) is not
sufficiently accurate to calculate the Regge-Wheeler function, and
therefore the shear tensor.   
 
Fortunately, the Regge-Wheeler equation (\ref{A.19}) is sufficiently
simple that it can be solved directly. Assuming (as we shall verify
below) that derivatives of $\Psi_{\rm RW}$ with respect to $v$ can be
neglected compared with spatial derivatives, the Regge-Wheeler
equation for $l=2$ reduces to 
\begin{equation} 
\biggl[ r(r-2M) \frac{d^2}{dr^2} + 2 M \frac{d}{dr} - 6 \biggl(1 -
\frac{M}{r} \biggr) \biggr] \Psi_{\rm RW} = 0. 
\label{8.25}
\end{equation} 
The solution that is well behaved at the horizon is $\Psi_{\rm RW}
\propto r^3$, and to produce the correct metric perturbation we
write 
\begin{equation}
\Psi_{\rm RW}^{\sf m}(v,r) = -\frac{1}{12} r^3 \dot{B}_{\sf m}(v), 
\label{8.26}
\end{equation}
where the overdot indicates differentiation with respect to
$v$. While ${\cal B}_{\sf m}(v)$ scales as ${\cal R}^{-2}$, its time
derivative scales as ${\cal R}^{-3}$ and $v$-derivatives of the 
Regge-Wheeler function are indeed much smaller than its spatial
derivatives. 

To see that Eq.~(\ref{8.26}) is indeed the correct solution to the
Regge-Wheeler equation, we reconstruct the metric perturbation in the
preferred gauge and show that it agrees with the odd-parity sector of
Eqs.~(\ref{8.7})--(\ref{8.10}). We note first that according to
Eq.~(\ref{8.24}), $\Psi_{\rm RW} = r^{-1}(\tilde{h}_v + f
\tilde{h}_r)$, where we have removed the label $\sf m$ for
simplicity. On the other hand, the field equation (\ref{A.17}) implies
$\partial_v \tilde{h}_r + \partial_r (\tilde{h}_v + f \tilde{h}_r) 
= 0$. Solving these equations yields $\tilde{h}_v = -\frac{1}{3} r^3 f 
{\cal B}$ and $\tilde{h}_r = \frac{1}{3} r^3 {\cal B}$, up to
smaller terms involving $\dot{\cal B}$. The actual metric
perturbations are then recovered by using Eq.~(\ref{8.23}) along with
the gauge condition $h_r = 0$. The equation for $\tilde{h}_r$ gives
$(\frac{1}{2} \partial_r - r^{-1}) h_2 = \tilde{h}_r$, and this
differential equation has $h_2 = \frac{1}{3} r^4 {\cal B}$ as
solution; this agrees with our previous expression. Finally, the
equation for $\tilde{h}_v$ gives $h_v + \frac{1}{2} \partial_v h_2 =
\tilde{h}_v$, and we obtain $h_v = -\frac{1}{3} r^3 f {\cal B}$ up to
a smaller term involving $\dot{\cal B}$; this also agrees with our
previous expression. We conclude that the Regge-Wheeler function of
Eq.~(\ref{8.26}) is indeed compatible with the metric of
Eqs.~(\ref{8.7})--(\ref{8.10}).  

Substituting Eq.~(\ref{8.26}) into Eq.~(\ref{7.6}) we obtain
\begin{equation} 
\sigma^{\rm odd}_{AB}(v,\theta^A) = -\frac{1}{12} r_+^4 
\sum_{\sf m}\dot{\cal B}_{\sf m}(v) X^{\sf m}_{AB}(\theta^A).   
\label{8.27}
\end{equation} 
According to Eq.~(\ref{8.15}), this can also be written as
$\sigma^{\rm odd}_{AB} =  \frac{1}{12} r_+^4 \dot{\cal B}^*_{AB}$.   

\subsection{Even-parity contribution to shear}  

We turn next to the even-parity sector of the metric
perturbations; the reader is referred to the description given in 
subsection 4 of the Appendix.  

When $l=2$ the even-parity perturbations can be expanded as 
\begin{eqnarray}
h_{ij} &=& \sum_{\sf m} h_{ij}^{\sf m}(v,r) Y^{\sf m}(\theta^A), 
\label{8.28} \\ 
h_{iA} &=& \sum_{\sf m} j_{i}^{\sf m}(v,r) Y^{\sf m}_{:A}(\theta^A),  
\label{8.29} \\ 
h_{AB} &=& r^2 \sum_{\sf m} G^{\sf m}(v,r) Z^{\sf m}_{AB}(\theta^A),  
\label{8.30} 
\end{eqnarray} 
where we have incorporated our knowledge that $K^{\sf m} = 3 
G^{\sf m}$; this follows because the metric perturbation of
Eq.~(\ref{8.10}) is tracefree [refer also to the discussion that
precedes Eq.~(\ref{7.4})]. The combinations 
\begin{equation} 
\tilde{h}^{\sf m}_{ij} = h^{\sf m}_{ij} 
- 2 \varepsilon^{\sf m}_{(i.j)}, \qquad 
\tilde{K}^{\sf m} = 3 G^{\sf m} 
- \frac{2}{r} r^{,i} \varepsilon^{\sf m}_i, 
\label{8.31}
\end{equation} 
where $\varepsilon^{\sf m}_i = j^{\sf m}_i - \frac{1}{2} r^2 
G^{\sf m}_{,i}$, are gauge invariant, and the Zerilli-Moncrief
function 
\cite{zerilli:70, moncrief:74, lousto-price:97}   
is defined by 
\begin{equation} 
\Psi^{\sf m}_{\rm ZM} \equiv \frac{r}{3} \biggl[ \tilde{K}^{\sf m} 
+ \frac{2}{\Lambda} \bigl( r^{,i} r^{,j} \tilde{h}^{\sf m}_{ij} 
- r r^{,i} \tilde{K}^{\sf m}_{,i} \bigr) \biggr], 
\label{8.32}
\end{equation} 
where $\Lambda = 4 + 6M/r$.

Comparison of Eqs.~(\ref{8.28})--(\ref{8.30}) with
Eqs.~(\ref{8.7})--(\ref{8.10}) using Eqs.~(\ref{8.11})--(\ref{8.13})
reveals that $h_{vv}^{\sf m} = -r^2 f^2 {\cal E}_{\sf m}$,
$j_{v}^{\sf m} = -\frac{1}{3} r^3 f {\cal E}_{\sf m}$, 
$G^{\sf m} = -\frac{1}{3} r^2(1-2M^2/r^2) {\cal E}_{\sf m}$, 
and $K^{\sf m} = 3 G^{\sf m}$, with all other components vanishing. 
From this information the Zerilli-Moncrief function can be computed
straightforwardly (no need to solve the Zerilli equation directly),
and its value on the horizon is found to be 
\begin{equation}
\Psi^{\sf m}_{\rm ZM}(v,r_+) = -\frac{1}{6} r_+^3 
{\cal E}_{\sf m}(v).
\label{8.33}
\end{equation} 
Substituting this into Eq.~(\ref{7.6}) we obtain
\begin{equation} 
\sigma^{\rm even}_{AB}(v,\theta^A) = -\frac{1}{12} r_+^4 
\sum_{\sf m}\dot{\cal E}_{\sf m}(v) Z^{\sf m}_{AB}(\theta^A).   
\label{8.34}
\end{equation} 
According to Eq.~(\ref{8.13}), this can also be written as
$\sigma^{\rm even}_{AB} =  - \frac{1}{12} r_+^4 \dot{\cal E}^*_{AB}$.   

\subsection{Shear and Weyl tensor}   
 
The sum of Eqs.~(\ref{8.27}) and (\ref{8.34}) gives the complete shear
tensor, 
\begin{eqnarray}
\sigma_{AB}(v,\theta^A) &=& -\frac{1}{12} r_+^4 
\sum_{\sf m} \Bigl[ \dot{\cal E}_{\sf m}(v) 
Z^{\sf m}_{AB}(\theta^A) 
\nonumber \\ & & \mbox{} 
+ \dot{\cal B}_{\sf m}(v) X^{\sf m}_{AB}(\theta^A) \Bigr],  
\label{8.35}
\end{eqnarray}
and we observe that this 2-tensor is properly tracefree.  

We can use Eqs.~(\ref{8.35}) and Eq.~(\ref{4.7}) to calculate the Weyl
tensor $C_{AB}$ on the horizon. We may neglect time derivatives
and write $C_{AB} = \kappa \sigma_{AB}$, which gives 
\begin{equation}
C_{AB} = -\frac{1}{24} r_+^3 \sum_{\sf m} 
\Bigl[ \dot{\cal E}_{\sf m} Z^{\sf m}_{AB} 
+ \dot{\cal B}_{\sf m} X^{\sf m}_{AB} \Bigr]. 
\label{8.36}
\end{equation}
This expression shows that the (dimensionless) Weyl curvature on the
horizon is of order $(M/{\cal R})^3$, which is a factor $M/{\cal R}  
\ll 1$ smaller than the asymptotic value of the Weyl curvature (for 
$r \gg M$).   

For future reference we calculate the Weyl scalar $\Psi$ from
$C_{AB}$; this is defined by Eq.~(\ref{5.6}) and related to the
Weyl tensor in Eq.~(\ref{5.5}). From this equation and the properties
of the vectors $e^A$ we infer that $\Psi = C_{AB} e^A e^B$. With
Eqs.~(\ref{A.9}) and (\ref{A.10}) we can relate the tensorial
harmonics $Z^{\sf m}_{AB}$ and $X^{\sf m}_{AB}$ to the spin-weighted
spherical harmonics $\mbox{}_{\pm 2} Y^{\sf m}$. (Please note that
the vectors $\epsilon^A$ used in the Appendix are rescaled versions of
the vectors used here: $e^A = \epsilon^A/r_+$; in the notation of the
Appendix we have $\Psi = C_{AB} \epsilon^A \epsilon^B/r_+^2$.) Simple
algebra then gives  
\begin{equation}
\Psi(v,\theta^A) = -\frac{\sqrt{6}}{24} r_+ \sum_{\sf m} 
\Bigl[ \dot{\cal E}_{\sf m}(v) - i \dot{\cal B}_{\sf m}(v) \Bigr] 
\mbox{}_{2} Y^{\sf m}(\theta^A). 
\label{8.37}
\end{equation} 
This reveals that $\Psi = O(M/{\cal R}^3)$ on the horizon, while 
$\Psi = O(1/{\cal R}^2)$ asymptotically (for $r \gg M$). This shows
once more that the Weyl curvature on the horizon is suppressed by a
factor $M/{\cal R} \ll 1$ with respect to its asymptotic value. 

\subsection{Fluxes} 

The shear tensor of Eq.~(\ref{8.35}) is equal to $\frac{1}{2}
\pounds_k \gamma^1_{AB} = \frac{1}{2} \pounds_t \gamma^1_{AB}$, and
the metric perturbation $\gamma^1_{AB}(v,\theta^A)$ can be obtained by
direct integration with respect to $v$. Differentiation with respect
to $\phi$ then gives $\pounds_\phi \gamma^1_{AB}$; this can be worked
out by using the explicit forms for $Z^{\sf m}_{AB}$ and 
$X^{\sf m}_{AB}$ gathered from subsection 2 of the Appendix. Finally,
these results can be substituted into the flux formulae of
Eqs.~(\ref{4.22})--(\ref{4.24}), and integration over 
$dS = r_+^2\sin\theta\, d\theta d\phi$ is readily carried out using
the known angular dependence contained in the spherical harmonics. A
straightforward computation yields  
\begin{eqnarray*} 
\Mdot &=& \frac{8M^6}{45} \bigl\langle 3 \dot{\cal E}_0^2 
+ \dot{\cal E}_{1c}^2 + \dot{\cal E}_{1s}^2 + 4 \dot{\cal E}_{2c}^2  
+ 4 \dot{\cal E}_{2s}^2 
\\ & & \mbox{} 
+ 3 \dot{\cal B}_0^2 
+ \dot{\cal B}_{1c}^2 + \dot{\cal B}_{1s}^2 + 4 \dot{\cal B}_{2c}^2  
+ 4 \dot{\cal B}_{2s}^2 \bigr\rangle 
\end{eqnarray*} 
and 
\begin{eqnarray*} 
\Jdot &=& -\frac{8M^6}{45} \bigl\langle \dot{\cal E}_{1c} 
{\cal E}_{1s} - \dot{\cal E}_{1s} {\cal E}_{1c} 
+ 8 \dot{\cal E}_{2c} {\cal E}_{2s} - 8\dot{\cal E}_{2s} 
{\cal E}_{2c} 
\\ & & \mbox{} 
+ \dot{\cal B}_{1c} {\cal B}_{1s} 
- \dot{\cal B}_{1s} {\cal B}_{1c} + 8 \dot{\cal B}_{2c} {\cal B}_{2s} 
- 8\dot{\cal B}_{2s} {\cal B}_{2c} \bigr\rangle,  
\end{eqnarray*} 
where ${\cal E}_{\sf m}$ and ${\cal B}_{\sf m}$ are the harmonic
components of the tidal gravitational fields introduced in
Eqs.~(\ref{8.16})--(\ref{8.20}). 

These results can be expressed in terms of invariants formed from
${\cal E}_{ab}$ and ${\cal B}_{ab}$, the components of the tidal
fields in the local asymptotic rest frame of the moving black hole. We
also need the derivatives of these fields with respect to $v$ (denoted
with an overdot), and the unit vector $s^a \equiv (0,0,1)$ that points
in the direction of the third coordinate axis. (This direction is 
preferred because the angles $\theta$ and $\phi$ refer to it.) In
terms of these quantities, the previous expressions become   
\begin{equation} 
\Mdot = \frac{16M^6}{45} \bigl\langle \dot{\cal E}_{ab} 
\dot{\cal E}^{ab} + \dot{\cal B}_{ab} \dot{\cal B}^{ab} \bigr\rangle 
\label{8.38}
\end{equation} 
and 
\begin{equation}
\Jdot = -\frac{32M^6}{45} \varepsilon_{acd} \bigl\langle 
\dot{\cal E}^a_{\ b} {\cal E}^{bc} + \dot{\cal B}^a_{\ b} 
{\cal B}^{bc} \bigr\rangle s^d,  
\label{8.39}
\end{equation}   
where $\varepsilon_{acd}$ is the three-dimensional permutation
symbol. We also have $(\kappa/8\pi) \Adot = \Mdot$. We note that when
$\Jdot$ is expressed in the covariant form of Eq.~(\ref{8.39}), what
is actually meant by $\Jdot$ is the rate of change of the component of
the angular-momentum vector in the direction of $s^a$; in
three-dimensional vectorial language appropriate in the local
asymptotic rest frame, $\Jdot \equiv \langle \dot{J}^a \rangle s_a$,
where $J^a$ is the vectorial angular momentum. 

Alternatively, the flux formulae can be expressed in terms of the
spacetime tensors of Eq.~(\ref{8.4}) and (\ref{8.5}). The
translation is effected by Eq.~(\ref{8.3}) and the identity
$\varepsilon_{abc} = u^\mu \varepsilon_{\mu\alpha\beta\gamma}
e^\alpha_a e^\beta_b e^\gamma_c$, where
$\varepsilon_{\mu\alpha\beta\gamma}$ is the Levi-Civita tensor. 
We find
\begin{equation} 
\Mdot = \frac{16M^6}{45} \bigl\langle \dot{\cal E}_{\alpha\beta} 
\dot{\cal E}^{\alpha\beta} + \dot{\cal B}_{\alpha\beta}
\dot{\cal B}^{\alpha\beta} \bigr\rangle 
\label{8.40}
\end{equation} 
and 
\begin{equation}
\Jdot = -\frac{32M^6}{45} u^\mu \varepsilon_{\mu\alpha\gamma\delta}
\bigl\langle \dot{\cal E}^\alpha_{\ \beta} {\cal E}^{\beta\gamma} 
+ \dot{\cal B}^\alpha_{\ \beta} {\cal B}^{\beta\gamma} \bigr\rangle
s^\delta,   
\label{8.41}
\end{equation}   
where $s^\alpha = s^a e^\alpha_a$ is a unit spatial vector, and 
$\dot{\cal E}_{\alpha\beta} \equiv {\cal E}_{\alpha\beta;\mu} u^\mu$,
$\dot{\cal B}_{\alpha\beta} \equiv {\cal B}_{\alpha\beta;\mu} u^\mu$
are the proper-time derivative of the tidal gravitational fields. We
recall that in this SH/SM description, all vectors and tensors refer
to the spacetime of the external universe in which the black hole
moves. From Eqs.~(\ref{8.38})--(\ref{8.41}) we gather that $\Mdot$
scales as $M^6/{\cal R}^6$, while $\Jdot$ scales as 
$M^6/{\cal R}^5$. To the best of my knowledge,
Eqs.~(\ref{8.38})--(\ref{8.41}) have never appeared before in the
literature.  

\subsection{Comparison with Thorne, Hartle, and Zhang}  

The rate of change of angular momentum for a general body interacting 
with a tidal gravitational field was calculated, in the regime  
$M/{\cal R} \ll 1$, by Thorne and Hartle
\cite{thorne-hartle:85}; 
they obtained the expression 
\begin{equation} 
\langle \dot{J}^a \rangle = -\varepsilon^a_{\ bc} \Bigl\langle 
M^b_{\ d} {\cal E}^{dc} + \frac{4}{3} J^b_{\ d} {\cal B}^{dc}
\Bigr\rangle, 
\label{8.42}
\end{equation} 
where $M_{ab}$ is the body's mass quadrupole moment, while $J_{ab}$ is 
its current quadrupole moment (both defined in terms of the structure
of the gravitational field outside the arbitrary body). Zhang 
\cite{zhang:85}, 
on the other hand, calculated the rate at which the body changes its
mass; he obtained an expression equivalent to  
\begin{equation} 
\Mdot = \frac{1}{2} \Bigl\langle M_{ab} \dot{\cal E}^{ab} 
+ \frac{4}{3} J_{ab} \dot{\cal B}^{ab} \Bigr\rangle.
\label{8.43}
\end{equation} 
We wish to show that our previous results are compatible with these
expressions.  

An isolated Schwarzschild black hole is spherically symmetric, and its
intrinsic quadrupole moments vanish: $M_{ab} = J_{ab} = 0$. But a
black hole immersed in an external universe is tidally distorted and
therefore acquires nonvanishing moments. It is easy to see that
Eqs.~(\ref{8.38}), (\ref{8.39}) are compatible with the general
results of Eqs.~(\ref{8.42}), (\ref{8.43}) if the tidally-induced
quadrupole moments of a nonrotating black hole are given by 
\begin{equation}
M_{ab} = \frac{32 M^6}{45} \dot{\cal E}_{ab} 
\label{8.44}
\end{equation}
and 
\begin{equation}
J_{ab} = \frac{8 M^6}{15} \dot{\cal B}_{ab}.  
\label{8.45}
\end{equation}
These scale as $M^3 (M/{\cal R})^3$, and they both involve the 
{\it rates of change} of the tidal gravitational fields. This is a
rather surprising result, as one would expect the quadrupole
deformation of a tidally-distorted body to be proportional to the
tidal gravitational field itself, instead of its time derivative. But
the time derivative is present, and its origin can be traced back to
Eq.~(\ref{8.37}): the Weyl curvature at the horizon is proportional to
the time derivative of the asymptotic curvature. Since it is the
horizon curvature that produces the black-hole distortion, this
explains why a time derivative enters Eqs.~(\ref{8.44}) and
(\ref{8.45}).      

\subsection{Black hole in a circular binary: Slow-motion
approximation} 

If we specialize to a slow-motion situation, the tidal
gravitational fields of the external universe can be approximated by  
\begin{equation} 
{\cal E}_{ab} \simeq \Phi_{,ab}, \qquad
{\cal B}_{ab} \simeq 0, 
\label{8.46}
\end{equation}
where $\Phi$ is a Newtonian potential. For concreteness we take the 
Newtonian field to be produced by an external body of mass 
$M_{\rm ext}$ located at $\bm{r}_{\rm ext}(t)$ relative to the
system's center of mass. Then $\Phi(\bm{x}) = -M_{\rm ext}/|\bm{x} - 
\bm{r}_{\rm ext}|$, and we exclude the contribution $-M/|\bm{x} -
\bm{r}|$ from the black hole because this does not produce a tidal
field at the position $\bm{r}(t)$ of the black hole. Also for
concreteness we take the orbit to be circular, and we let $b \equiv
|\bm{r} - \bm{r}_{\rm ext}|$ be the constant relative separation
between the two bodies. The orbital angular velocity is  
\begin{equation}
\Omega = \sqrt{\frac{M+M_{\rm ext}}{b^3}},  
\label{8.47} 
\end{equation}
and the relative position vector is $\bm{r} - \bm{r}_{\rm ext}  
\equiv \bm{\rho} = b(\cos\Omega t, \sin\Omega t,0) \equiv b  
\bm{\hat{\rho}}(t)$. The relative velocity vector is $\bm{V} 
= b\Omega (-\sin\Omega t,\cos\Omega t, 0) \equiv b \Omega 
\bm{\hat{\phi}}(t)$. For simplicity we align the spin vector in the 
direction of the orbital angular momentum: $\bm{s} = (0,0,1)$.  

Using this information we calculate ${\cal E}_{ab}(t) = 
(M_{\rm ext}/b^3) (\delta_{ab} - 3 \hat{\rho}_a \hat{\rho}_b)$ and  
\[
\dot{\cal E}_{ab} = -\frac{3 M_{\rm ext} \Omega}{b^3} \bigl(
\hat{\rho}_a \hat{\phi}_b + \hat{\phi}_a \hat{\rho}_b \bigr).  
\]
Substituting this into Eq.~(\ref{8.38}) gives 
\begin{equation} 
\Mdot = \frac{32}{5} \eta^2 \biggl( \frac{M}{M+M_{\rm ext}} \biggr)^4
V^{18}, 
\label{8.48}
\end{equation}
where $\eta = M M_{\rm ext}/(M+M_{\rm ext})^2$ is a dimensionless
reduced-mass parameter and 
\begin{equation}
V = \sqrt{ \frac{M + M_{\rm ext}}{b} } \ll 1 
\label{8.49}
\end{equation}
is the relative orbital velocity. The rate of change of the hole's
angular momentum can be obtained directly from this and the
rigid-rotation relation $\Jdot = \Omega^{-1} \Mdot$; this gives 
\begin{equation} 
\Jdot = \frac{32}{5} \eta^2 \biggl( \frac{M}{M+M_{\rm ext}} \biggr)^4
(M + M_{\rm ext}) V^{15}. 
\label{8.50}
\end{equation}
These results agree (in a limit of no black-hole rotation) with
earlier expressions obtained by Alvi 
\cite{alvi:01}. 
In the regime $M_{\rm ext} \ll M$ they also agree with earlier results 
derived by Poisson and Sasaki
\cite{poisson-sasaki:95}.   

\subsection{Black hole in a circular binary: Small-hole
approximation}

We now allow the black hole to move rapidly in the strong
gravitational field of another Schwarzschild hole of mass 
$M_{\rm ext}$; to comply with the SH/SM condition $M/{\cal R} \ll 1$
we now impose $M/M_{\rm ext} \ll 1$, as was discussed in Sec.~VIII A. 
Once more we choose the orbit to be circular. In the standard 
Schwarzschild coordinates $(t,r,\theta,\phi)$ used in the background
spacetime of the large black hole, the orbital radius is $b$ and the
four-velocity of the small hole is $u^\alpha = \gamma(1,0,0,\Omega)$,
where $\gamma = (1-3M_{\rm ext}/b)^{-1/2}$ is a normalization factor
and 
\begin{equation}
\Omega = \sqrt{ \frac{M_{\rm ext}}{b^3} } 
\label{8.51}
\end{equation}
is the angular velocity. We again align the spin vector in the
direction of the orbital angular momentum, so that $s^\alpha 
= (0,0,-1/b,0)$. Calculation of ${\cal E}_{\alpha\beta}$,  
${\cal B}_{\alpha\beta}$ using Eqs.~(\ref{8.4}), (\ref{8.5}), and
substitution into Eqs.~(\ref{8.40}), (\ref{8.41}) gives 
\begin{equation} 
\Mdot = \frac{32}{5} \biggl( \frac{M}{M_{\rm ext}} \biggr)^6 V^{18} 
\frac{ (1-V^2)(1-2V^2) }{ (1-3V^2)^2 } 
\label{8.52}
\end{equation}
and 
\begin{equation} 
\Jdot = \frac{32}{5} \biggl( \frac{M}{M_{\rm ext}} \biggr)^6 
M_{\rm ext} V^{15} \frac{ (1-V^2)(1-2V^2) }{ (1-3V^2)^2 },  
\label{8.53}
\end{equation}
where $V = \sqrt{M_{\rm ext}/b} \leq 6^{-1/2}$ is a measure of the
hole's orbital velocity. These results agree with those of the
preceding subsection in a common domain of validity defined by
$M \ll M_{\rm ext}$ and $V \ll 1$. To the best of my knowledge, the 
results of Eqs.~(\ref{8.52}) and (\ref{8.53}), complete with all-order  
relativistic corrections, have never appeared before in the
literature.  

\section{Small-hole/slow-motion approximation for a Kerr black hole}  

In this section I apply the SH/SM approximation introduced in Sec.~V 
III A to the flux formulae derived in Sec.~V D,
Eqs.~(\ref{5.23})--(\ref{5.29}). I will proceed much as in Sec.~VIII,
except that I will deal with curvature perturbations --- and the
Teukolsky equation \cite{teukolsky:73} --- instead of metric
perturbations.  

\subsection{Flux formulae in the SH/SM approximation} 

We begin by isolating, in Eqs.~(\ref{5.27}) and (\ref{5.28}), the
terms for which $m = 0$:  
\begin{eqnarray} 
\Mdot &=& \frac{1}{2} (r_+^2+a^2) \int
\bigl\langle |\Phi_+^0|^2 \bigr\rangle \sin\theta\, d\theta
\nonumber \\ & & \mbox{} 
+ \frac{r_+^2+a^2}{4\kappa} \sum_{m\neq 0} \biggl[ 2\kappa \int 
\bigl\langle |\Phi_+^m|^2 \bigr\rangle \sin\theta\, d\theta 
\nonumber \\ & & \mbox{} 
- i m \Omega_H \int \bigl\langle \bar{\Phi}_+^m \Phi_-^m - \Phi_+^m
\bar{\Phi}_-^m \bigr\rangle \sin\theta\, d\theta \biggr], \qquad
\label{9.1} \\ 
\Jdot &=& -\frac{r_+^2+a^2}{4\kappa} \sum_{m\neq 0} (im) 
\nonumber \\ & & \mbox{} \times 
\int \bigl\langle \bar{\Phi}_+^m \Phi_-^m - \Phi_+^m \bar{\Phi}_-^m 
\bigr\rangle \sin\theta\, d\theta. 
\label{9.2}
\end{eqnarray}
We recall the definitions 
\begin{eqnarray} 
\Phi^m_+(v,\theta) &=& e^{\kappa v} \int_v^\infty e^{-(\kappa 
- im\Omega_{\rm H})v'} \Psi^m(v',\theta)\, dv', \qquad 
\label{9.3} \\ 
\Phi^m_-(v,\theta) &=& \int_{-\infty}^v e^{im\Omega_{\rm H} v'}  
\Psi^m(v',\theta)\, dv', 
\label{9.4}
\end{eqnarray} 
where
\begin{equation} 
\Psi(v,r_+,\theta,\psi) = \sum_{m} \Psi^m(v,\theta) e^{im\psi} 
\label{9.5}
\end{equation} 
is the horizon Weyl scalar introduced in Eq.~(\ref{5.6}).  

To see how we may specialize Eqs.~(\ref{9.3}) and (\ref{9.4}) to the 
SH/SM approximation, consider first an integral of the form $F_-(v)
= \int_{-\infty}^v e^{i\omega v'} f(v')\, dv'$, and suppose that 
$f(v')$ varies on a time scale $\tau$ that is large compared with
$\omega^{-1}$. (We also suppose that $f$ switches off sufficiently
rapidly in the infinite past so that the integral converges.) Then
$F_-(v)$ can be evaluated by successive integration by parts, each
integration generating a relative correction of order $\epsilon \equiv 
(\omega\tau)^{-1} \ll 1$. To leading order, $F_-(v) = -i\omega^{-1}
f(v) e^{i\omega v} [1 + O(i\epsilon)]$. Consider next an integral of
the form $F_+(v) = \int_v^\infty e^{-\lambda v'} f(v')\, dv'$, where
the real part of $\lambda$ is assumed to be positive; here we suppose
that $\epsilon' \equiv (\lambda\tau)^{-1} \ll 1$. Integration by parts
in this case leads to $F_+(v) = \lambda^{-1} f(v) e^{-\lambda v} [1 +
O(\epsilon')]$. These simple manipulations allow us, within the stated
conditions, to approximate the integrals by local expressions. This is
the technique we shall employ to evaluate
Eqs.~(\ref{9.3}) and (\ref{9.4}).       

In this way we obtain 
\[
\Phi_+^m = \frac{\Psi^m(v,\theta) e^{im\Omega_{\rm H} v}}{\kappa 
- im\Omega_{\rm H}} \biggl[ 1 + O\biggl(
\frac{1}{(\kappa-im\Omega_{\rm H})\tau} \biggr) \biggr] 
\]
and 
\[
\Phi_-^m = \frac{\Psi^m(v,\theta) 
e^{im\Omega_{\rm H} v}}{im\Omega_{\rm H}} 
\biggl[ 1 + O\biggl(\frac{1}{im\Omega_{\rm H}\tau} \biggr) \biggr], 
\]
where $\tau$ is the time scale associated with changes in
$\Psi^m(v,\theta)$. To see how the conditions $\kappa \tau \gg 1$ and
$\Omega_{\rm H} \tau \gg 1$ relate to the SH/SM approximation, we
first recall that changes in $\Psi(v,r_+,\theta,\psi)$ are governed by 
processes taking place in the external universe, so that $\tau \sim
{\cal R}$. We also express $\kappa$ and $\Omega_{\rm H}$ in terms of
the black-hole mass $M$ and its dimensionless rotational parameter
$\chi \equiv a/M \equiv J/M^2$: 
\begin{equation} 
\kappa = \frac{\sqrt{1-\chi^2}}{2M(1+\sqrt{1-\chi^2})}, \qquad 
\Omega_{\rm H} = \frac{\chi}{2M(1 + \sqrt{1-\chi^2})}; 
\label{9.6} 
\end{equation}
we recall that $\chi$ is limited to the interval $0 \leq \chi \leq 1$
and that $r_+ = M(1 + \sqrt{1-\chi^2})$. In orders of magnitude we
have $\kappa \sim 1/M$ and $\Omega_{\rm H} \sim \chi/M$, and to
achieve $\kappa \tau \gg 1$ and $\Omega_{\rm H} \tau \gg 1$ we need
$M/{\cal R} \ll 1$ and $M/{\cal R} \ll \chi$, respectively. The
stronger condition is  
\begin{equation} 
M/{\cal R} \ll \chi, 
\label{9.7}
\end{equation}
and we take this to be the precise statement of the
small-hole/slow-motion condition when we deal with rotating black
holes. Notice that by virtue of Eq.~(\ref{9.7}), the no-rotation
limit $\chi \to 0$ will be inaccessible in our analysis; this case was  
treated separately in Sec.~VIII. We shall write our previous results
as 
\begin{eqnarray} 
\Phi_+^m &=& \frac{\Psi^m(v,\theta) e^{im\Omega_{\rm H} v}}{\kappa 
- im\Omega_{\rm H}} \Bigl[ 1 + O(M/{\cal R}) \Bigr], 
\label{9.8} \\ 
\Phi_-^m &=& \frac{\Psi^m(v,\theta) 
e^{im\Omega_{\rm H} v}}{im\Omega_{\rm H}} 
\Bigl[ 1 + O(M/{\cal R}) \Bigr], 
\label{9.9}
\end{eqnarray}
with the understanding that the error terms are really of order
$M/(\chi {\cal R})$, and therefore small by virtue of
Eq.~(\ref{9.7}). For the remainder of this section we assume that  
$\chi$ is of order unity, and we allow ourselves to lose sight
of this distinction. 

Substituting Eqs.~(\ref{9.8}) and (\ref{9.9}) into Eq.~(\ref{9.1})
reveals that each $m\neq 0$ term vanishes to leading order in 
$M/{\cal R}$; what remains is 
\begin{equation} 
\Mdot = \frac{r_+^2+a^2}{2\kappa^2} \int \bigl\langle
|\Psi^0(v,\theta)|^2 \bigr\rangle \sin\theta\, d\theta 
+ O(M^5/{\cal R}^5). 
\label{9.10}
\end{equation} 
The scaling of the error term follows from the facts that each 
contribution to a $m \neq 0$ term is of order $(M/{\cal R})^4$, but
that the cancellation suppresses this by a factor of (at least)  
$M/{\cal R}$. An {\it a priori} estimate of the surviving term in 
Eq.~(\ref{9.10}) indicates that it is of order $(M/{\cal R})^4$, but
we shall see that it is in fact of order $(M/{\cal R})^6$. 
Inserting Eqs.~(\ref{9.8}) and (\ref{9.9}) into Eq.~(\ref{9.2})
gives 
\begin{eqnarray}
\Jdot &=& -\frac{r_+^2+a^2}{2\Omega_{\rm H}} \sum_{m\neq 0} 
(\kappa^2 + m^2 \Omega_{\rm H}^2)^{-1} 
\nonumber \\ & & \mbox{} \times 
\int \bigl\langle
|\Psi^m(v,\theta)|^2 \bigr\rangle \sin\theta\, d\theta, 
\label{9.11}
\end{eqnarray}
and this is of order $M^5/{\cal R}^4$. 

\subsection{Weyl scalar: asymptotic values} 

To proceed further we must compute the functions $\Psi^m(v,\theta)$
that enter into the simplified flux formulae of Eqs.~(\ref{9.10}) and
(\ref{9.11}). This requires solving the Teukolsky equation for the 
Weyl scalar $\psi_0(v,r,\theta,\psi)$, with appropriate boundary
data provided by the conditions in the external universe. In
this subsection I specify these boundary conditions; in the next
subsection (Sec.~IX C) I tackle the integration of the Teukolsky
equation and construct $\Psi^m(v,\theta)$ from the solution. My
presentation in these two subsections will not stray very far from
what is contained in Sec.~3.5 of the book by D'Eath
\cite{death:96}; 
and my end results will be equivalent to his.   

The function $\psi_0(v,r,\theta,\psi)$ we shall work with is 
\begin{equation} 
\psi_0 = -C^1_{\alpha\gamma\beta\delta} k^\alpha m^\gamma k^\beta
m^\delta, 
\label{9.12}
\end{equation}
where $C^1_{\alpha\gamma\beta\delta}$ is the perturbation of the Weyl
tensor, while $k^\alpha \equiv k^\alpha({\rm K})$ and $m^\alpha \equiv  
m^\alpha({\rm K})$ are members of {\it Kinnersley's} null tetrad
\cite{kinnersley:69, teukolsky:73}. 
The relation between $\psi_0 \equiv \psi_0({\rm K})$ and
$\Psi(v,r_+,\theta,\psi)$ is given by Eq.~(\ref{5.6}). 

We wish to calculate $\psi_0$ in the asymptotic regime $r \gg r_+$,
assuming that $r$ is still much smaller than ${\cal R}$, the radius of
curvature of the external spacetime. The asymptotic values will be 
constructed from ${\cal E}_{ab}(v)$ and ${\cal B}_{ab}(v)$, the tidal
gravitational fields introduced in Sec.~VIII A --- Eq.~(\ref{8.2}); 
recall that lower-case Latin indices refer to the hole's local
asymptotic rest frame, and that the hole's angular-momentum vector is 
directed along the third coordinate axis.   

In the asymptotic regime $r \gg r_+$ the coordinates
$(v,r,\theta,\psi)$ are easily related to a set of Lorentzian
coordinates $(t,x,y,z)$ that are adapted to the frame
$(u^\alpha,e^\alpha_a)$; the relations are $t = v - r$, $x 
= r\sin\theta\cos\psi$, $y = r\sin\theta\sin\psi$, and $z 
= r\cos\theta$. In this regime the null vector $k^\alpha$ can be  
decomposed as $k^\alpha \sim u^\alpha + r^\alpha$, where $u^\alpha$ is  
the hole's velocity vector in the external spacetime, and $r^\alpha$
is a spacelike vector that points radially outward. In the asymptotic
Lorentzian coordinates $(t,x,y,z)$ we have $u^\alpha = (1,0,0,0)$ and
$r^\alpha = (0,\sin\theta\cos\psi,\sin\theta\sin\psi,\cos\theta)$. In
the limit we also have $m^\alpha \sim 2^{-1/2}(0,\cos\theta\cos\psi 
- i\sin\psi, \cos\theta\sin\psi + i\cos\psi,-\sin\theta)$.

We have seen in Sec.~VIII A that in the vicinity of the black hole
(the region $r \ll {\cal R}$, which includes the asymptotic region 
$r \gg r_+$), the Weyl tensor of the external spacetime can be
decomposed into the symmetric, tracefree fields ${\cal E}_{ab}$ and
${\cal B}_{ab}$. In the asymptotic coordinates $(t,x,y,z)$ the
decomposition is given by $C_{tatb} = {\cal E}_{ab}$, $C_{tabc} 
= -\varepsilon_{bcd} {\cal B}^d_{\ a}$, and $C_{acbd} = \delta_{ab} 
{\cal E}_{cd} + \delta_{cd} {\cal E}_{ab} - \delta_{ad} {\cal E}_{bc} 
- \delta_{bc} {\cal E}_{ad}$. In these relations the Weyl tensor, and
the tidal gravitational fields ${\cal E}_{ab}$ and ${\cal B}_{ab}$,
are evaluated on the black hole's world line in the external
spacetime; they are functions of $t$ (or $v$) only. 

According to Eq.~(\ref{9.12}), the asymptotic value of $\psi_0$ is 
$-C_{\alpha\gamma\beta\delta}(u^\alpha+r^\alpha) m^\gamma
(u^\beta+r^\beta) m^\delta$, with $C_{\alpha\gamma\beta\delta}$
denoting the Weyl tensor of the external spacetime evaluated on the 
hole's world line. Using the information provided in the preceding
two paragraphs, we obtain the explicit expression
\begin{equation} 
\psi_0 \sim - 2 {\cal E}_{ab} m^a m^b + 2 r^a m^b \varepsilon_{abc}
{\cal B}^c_{\ d} m^d, 
\label{9.13}
\end{equation}
where $r^a$ and $m^a$ are the spatial components of the vectors
$r^\alpha$ and $m^\alpha$, respectively. The angular dependence
contained in these vectors is encoded in spin-weighted spherical
harmonics of degree $l=2$ (see subsection 2 of the Appendix for a
definition). It is convenient to introduce a set given by 
\begin{eqnarray}
\mbox{}_2 Y_2^0(\theta,\psi) &=& -\frac{3}{2} \sin^2\theta,  
\label{9.14} \\ 
\mbox{}_2 Y_2^{\pm 1}(\theta,\psi) &=& -\sin\theta(\cos\theta \mp 1)
e^{\pm i\psi},   
\label{9.15} \\ 
\mbox{}_2 Y_2^{\pm 2}(\theta,\psi) &=& \frac{1}{4} (1 \mp 2\cos\theta
+ \cos^2\theta ) e^{\pm 2i\psi}. \qquad  
\label{9.16} 
\end{eqnarray} 
This set is not normalized; we have instead $\int |\mbox{}_2
Y_2^0|^2\, d\Omega = 24\pi/5$, $\int |\mbox{}_2 Y_2^{\pm 1}|^2\, 
d\Omega = 16\pi/5$, and $\int |\mbox{}_2 Y_2^{\pm 2}|^2\,
d\Omega = 4\pi/5$. If we also introduce 
\begin{eqnarray} 
\alpha_0 &=& {\cal E}_{11} + {\cal E}_{22}, 
\label{9.17} \\
\alpha_{\pm 1} &=& {\cal E}_{13} \mp i {\cal E}_{23}, 
\label{9.18} \\
\alpha_{\pm 2} &=& {\cal E}_{11} - {\cal E}_{22} 
\mp 2 i {\cal E}_{12}
\label{9.19} 
\end{eqnarray}    
and
\begin{eqnarray} 
\beta_0 &=& {\cal B}_{11} + {\cal B}_{22}, 
\label{9.20} \\
\beta_{\pm 1} &=& {\cal B}_{13} \mp i {\cal B}_{23}, 
\label{9.21} \\
\beta_{\pm 2} &=& {\cal B}_{11} - {\cal B}_{22} 
\mp 2 i {\cal B}_{12},
\label{9.22} 
\end{eqnarray}    
then it is straightforward to show that Eq.~(\ref{9.13}) is equivalent
to  
\begin{equation} 
\psi_0 \sim - \sum_{m} \bigl[ \alpha_m(v) + i \beta_m(v) \bigr] 
\mbox{}_2 Y_2^m(\theta,\psi). 
\label{9.23}
\end{equation} 
This is the asymptotic value of the Weyl scalar
$\psi_0(v,r,\theta,\psi)$ in the regime $r_+ \ll r \ll {\cal R}$, 
expressed in terms of the tidal gravitational fields 
${\cal E}_{ab}(v)$ and ${\cal B}_{ab}(v)$. 

\subsection{Teukolsky equation} 

To relate $\Psi^m(v,\theta)$ to the asymptotic value of $\psi_0$
obtained in Eq.~(\ref{9.23}) it is necessary to solve the Teukolsky
equation 
\cite{teukolsky:73} 
for $s = 2$ and $l = 2$. Because the $v$-dependence of the
solution enters through the tidal gravitational fields 
${\cal E}_{ab}(v)$ and ${\cal B}_{ab}(v)$, and because this dependence
is slow (time scale of order ${\cal R}$), it is actually sufficient to 
integrate the {\it time-independent} Teukolsky equation. We therefore
write 
\begin{equation}
\psi_0(v,r,\theta,\psi) = - \sum_m \bigl[ \alpha_m(v) + i \beta_m(v)
\bigr] R_m(r) \mbox{}_2 Y_2^m(\theta,\psi), 
\label{9.24}
\end{equation}
where $R_m(r)$ is a radial function normalized so that $R_m(r \gg r_+)
\sim 1$; this function must be a solution to Eq.~(2.10) of Teukolsky
and Press \cite{teukolsky-press:74}, in which we set $\omega = 0$.      
 
The explicit form of the radial equation is 
\begin{eqnarray} 
\biggl\{ x(1+x) \frac{d^2}{dx^2} + \Bigl[ 3(2x+1) + 2 i m \gamma
\Bigr] \frac{d}{dx} \mbox{} & & \nonumber \\  
+ 4 i m \gamma \frac{2x+1}{x(1+x)} \biggr\} R_m(x)  
&=& 0, \qquad\quad
\label{9.25}
\end{eqnarray}
where 
\begin{equation}
x = \frac{r - r_+}{r_+ - r_-} 
\label{9.26}
\end{equation}
is a new independent variable, and
\begin{equation}
\gamma = \frac{a}{r_+ - r_-}; 
\label{9.27}
\end{equation} 
we recall that $r_\pm = M \pm \sqrt{M^2-a^2}$. The relevant solution
to Eq.~(\ref{9.25}) is 
\begin{equation}
R_m(r) = A_m x^{-2} (1+x)^{-2} F(-4,1;-1+2i m \gamma; -x), 
\label{9.28}
\end{equation}
in which the hypergeometric function is actually an ordinary
polynomial of degree 4 in the variable $-x$. Equation (\ref{9.28}) is 
essentially Eq.~(5) from Ref.~\cite{alvi:01}, and the superficial
difference is attributed to the fact that Alvi works in
Boyer-Lindquist coordinates instead of our Kerr coordinates. This is
also Eq.~(3.7) in Chapter VI of Teukolsky's Ph.D.\ dissertation
\cite{teukolsky:phd}, and Eq.~(\ref{9.28}) is equivalent to
Eq.~(3.5.7) of Ref.~\cite{death:96}. The constant $A_m$ must
be chosen so that the radial function approaches unity when $x \to
\infty$; a simple calculation shows that it must be given by 
\begin{equation}
A_m = -\frac{i}{6} m \gamma (1 + i m \gamma)(1 + 4 m^2 \gamma^2) 
\label{9.29}
\end{equation}
when $m \neq 0$. 

The case $m = 0$ must be considered separately. It is formally
obtained by setting $\gamma = 0$ in Eq.~(\ref{9.25}), and
Eq.~(\ref{9.27}) shows that this amounts to letting $a = 0$. For
$m=0$, therefore, Eq.~(\ref{9.25}) reduces to Teukolsky's radial 
equation in Schwarzschild spacetime; the independent variable is now 
given by $x = r/r_+ - 1$. The relation between $\Psi^m(v,\theta)$ and
the asymptotic value of $\psi_0$ was already worked out, for
Schwarzschild spacetime, in Sec.~VIII E --- Eq.~(\ref{8.37}). There it
was revealed that it is of the schematic form $\Psi^m \sim r_+
\dot{\psi}_0(v,r\gg r_+,\theta,\psi)$, and that it involves a 
derivative of the asymptotic field with respect to $v$. This relation 
is very different from what was anticipated in Eq.~(\ref{9.24}), and
therefore different from what is known to be true for $m \neq 0$.   
These considerations imply that for $m=0$ and $a \neq 0$, the relation
between $\Psi^m(v,\theta)$ and the asymptotic value of $\psi_0$ comes
with an additional factor of $M/{\cal R}$ relative to terms
with $m \neq 0$. We conclude that it is appropriate to neglect the
$m=0$ term in Eq.~(\ref{9.24}), which becomes 
\begin{eqnarray}
\psi_0 &=& - \sum_{m \neq 0} A_m 
\bigl[ \alpha_m(v) + i \beta_m(v) \bigr] 
x^{-2} (1+x)^{-2} 
\nonumber \\ & & \mbox{} \times 
F(-4,1;-1+2i m \gamma; -x)\,  
\mbox{}_2 Y_2^m(\theta,\psi), \qquad\quad
\label{9.30}
\end{eqnarray}
where $x = (r-r_+)/(r_+-r_-)$, $\gamma = a/(r_+-r_-)$, $A_m$ is given
by Eq.~(\ref{9.29}), and $\alpha_m(v)$, $\beta_m(v)$ are listed in
Eqs.~(\ref{9.17})--(\ref{9.22}).  

The functions $\Psi^m(v,\theta)$ are obtained by substituting
Eq.~(\ref{9.30}) into Eq.~(\ref{5.6}) and taking the limit $r \to
r_+$, or $x \to 0$; we recall that $\psi_0(v,r,\theta,\psi)$ is the
Weyl scalar constructed with the Kinnersley tetrad, and that
$\Psi(v,r_+,\theta,\psi)$ is decomposed as in Eq.~(\ref{9.5}). Simple
algebra, using $a = \chi M$, $r_\pm = M(1 \pm \sqrt{1-\chi^2})$, and
$\gamma = \frac{1}{2} \chi (1-\chi^2)^{-1/2}$, yields 
\begin{eqnarray}
\Psi^m(v,\theta) &=& 
- \frac{i m \chi (1-\chi^2)^{3/2}}{12(1+\sqrt{1-\chi^2})^2} 
(1 + i m \gamma)(1 + 4m^2 \gamma^2) 
\nonumber \\ & & \mbox{} \times 
\bigl[ \alpha_m(v) + i \beta_m(v) \bigr] 
\mbox{}_2 Y_2^m(\theta,0). 
\label{9.31}
\end{eqnarray}
This result holds when $m \neq 0$, and it reveals that 
$\Psi^{m \neq 0}$ is of order ${\cal R}^{-2}$; as we have seen, when
$m=0$ we have instead the Schwarzschild result $\Psi^0 = 
O(M/{\cal R}^3)$.  

\subsection{Fluxes} 
 
We now insert Eq.~(\ref{9.31}) into the approximate flux formulae
of Eqs.~(\ref{9.10}) and (\ref{9.11}). The fact that $\Psi^0 
= O(M/{\cal R}^3)$ implies that the error term of Eq.~(\ref{9.10}) is
in fact dominant, and we obtain 
\begin{equation} 
\Mdot = O(M^5/{\cal R}^5). 
\label{9.32}
\end{equation} 
This result indicates that to calculate $\Mdot$ requires information 
that is not accessible to the leading-order analysis carried out 
here. To go beyond this leading-order calculation should be feasible,
but this lies beyond the scope of this work. 

A more definite result can be obtained for $\Jdot$. Substitution of
Eq.~(\ref{9.31}) into Eq.~(\ref{9.11}) and integration over $\theta$
--- recall the explicit forms of the spin-weighted spherical harmonics 
specified by Eqs.~(\ref{9.14})--(\ref{9.16}) --- returns
\begin{eqnarray*}
\Jdot &=& -\frac{M^5\chi}{45} \sum_{m \neq 0} 
\bigl[1 + (m^2-1) \chi^2 \bigr]  
\bigl[4 + (m^2-4) \chi^2 \bigr]  
\\ & & \mbox{} \times 
\bigl\langle | \alpha_m(v) + i \beta_m(v) |^2 \bigr\rangle 
\end{eqnarray*}
after simplification; notice that a factor of $m^2$ is canceled by
the integral $\int |\mbox{}_2 Y_2^{m \neq 0}(\theta,0)|^2 \sin\theta\,
d\theta = 8/(5m^2)$. This becomes 
\begin{eqnarray*} 
\Jdot &=& -\frac{2}{45} M^5\chi \Bigl[ 
(4-3\chi^2) \bigl\langle{\cal E}_{13}^2 + {\cal E}_{23}^2 +   
{\cal B}_{13}^2 + {\cal B}_{23}^2 \bigr\rangle 
\\ & & \mbox{} 
+ 4(1 + 3\chi^2) \bigl\langle ({\cal E}_{11} - {\cal E}_{22})^2  
+ 4 {\cal E}_{12}^2 
\\ & & \mbox{} 
+ ({\cal B}_{11} - {\cal B}_{22})^2 
+ 4 {\cal B}_{12}^2 \bigr\rangle \Bigr] 
\end{eqnarray*}
after using Eqs.~(\ref{9.17})--(\ref{9.22}). 

At this stage we introduce the invariants 
\begin{eqnarray} 
E_1 &=& {\cal E}_{ab} {\cal E}^{ab} =    
{\cal E}_{\alpha\beta} {\cal E}^{\alpha\beta}, 
\label{9.33} \\ 
E_2 &=& {\cal E}_{ab} s^b {\cal E}^a_{\ c} s^c =    
{\cal E}_{\alpha\beta} s^\beta {\cal E}^\alpha_{\ \gamma} s^\gamma,  
\label{9.34} \\ 
E_3 &=& \bigl({\cal E}_{ab} s^a s^b \bigr)^2 = 
\bigl({\cal E}_{\alpha\beta} s^\alpha s^\beta \bigr)^2,
\label{9.35}
\end{eqnarray} 
and 
\begin{eqnarray} 
B_1 &=& {\cal B}_{ab} {\cal B}^{ab} =    
{\cal B}_{\alpha\beta} {\cal B}^{\alpha\beta}, 
\label{9.36} \\ 
B_2 &=& {\cal B}_{ab} s^b {\cal B}^a_{\ c} s^c =    
{\cal B}_{\alpha\beta} s^\beta {\cal B}^\alpha_{\ \gamma} s^\gamma,  
\label{9.37} \\ 
B_3 &=& \bigl({\cal B}_{ab} s^a s^b \bigr)^2 = 
\bigl({\cal B}_{\alpha\beta} s^\alpha s^\beta \bigr)^2, 
\label{9.38}
\end{eqnarray} 
where the unit vector $s^a$ gives the direction of the black hole's
spin in the local asymptotic rest frame, and $s^\alpha = s^a
e^\alpha_a$ is the corresponding spacetime vector. We therefore have 
$J^a = J s^a$, $J = \chi M^2$, and $\Jdot = \langle \dot{J}^a \rangle
s_a$. In terms of these invariants we have, for example, ${\cal
E}_{13}^2 + {\cal E}_{23}^2 = E_2 - E_3$ and $({\cal E}_{11} 
- {\cal E}_{22})^2 + 4 {\cal E}_{12}^2 = 2E_1 - 4E_2 + E_3$. Our final
expression for the rate of change of angular momentum is 
\begin{eqnarray} 
\Jdot &=& -\frac{2}{45} M^5\chi \Bigl[ 8(1 + 3\chi^2) \langle E_1 
+ B_1 \rangle 
\nonumber \\ & & \mbox{} 
- 3(4 + 17\chi^2) \langle E_2 + B_2 \rangle 
\nonumber \\ & & \mbox{} 
+ 15\chi^2 \langle E_3 + B_3 \rangle \Bigr]. 
\label{9.39}
\end{eqnarray} 
This result reveals that $\Jdot = O(M^5/{\cal R}^4)$. 

The first law of black-hole mechanics implies that the rate of change
of the horizon area is given by $(\kappa/8\pi) \Adot = -\Omega_{\rm H}
\Jdot + O(M^5/{\cal R}^5)$, with a leading term scaling as 
$M^4/{\cal R}^4$. The ratio $\Omega_{\rm H}/\kappa$ can be expressed
in terms of $M$ and $\chi \equiv a/M$, and we obtain 
\begin{eqnarray} 
\Adot &=& \frac{16\pi}{45} \frac{M^5\chi^2}{\sqrt{1-\chi^2}} 
\Bigl[ 8(1 + 3\chi^2) \langle E_1 + B_1 \rangle 
\nonumber \\ & & \mbox{} 
- 3(4 + 17\chi^2) \langle E_2 + B_2 \rangle 
\nonumber \\ & & \mbox{} 
+ 15\chi^2 \langle E_3 + B_3 \rangle \Bigr];
\label{9.40}
\end{eqnarray} 
this scales as $M^5/{\cal R}^4$. This result is equivalent to
Eq.~(3.5.39) of the book by D'Eath \cite{death:96}.  

\subsection{Comparison with Thorne and Hartle} 

The rate of change of angular momentum for a general body interacting
with a tidal gravitational field was calculated by Thorne and Hartle
\cite{thorne-hartle:85} 
and their result displayed in Eq.~(\ref{8.42}). We wish to compare
this general expression with our result for $\Jdot$ displayed in 
Eq.~(\ref{9.39}); recall that $\Jdot = \langle \dot{J}^a \rangle s_a$,
with $s^a$ giving the direction of the angular-momentum vector.  

The quadrupole moments of a Kerr black hole immersed in an external
universe include an {\it intrinsic} component that would be present
even if the black hole were isolated, and an {\it induced} component
that comes from the hole's tidal distortion. We write  
\begin{eqnarray}
M_{ab} &=& M^{\rm intrinsic}_{ab} + M^{\rm induced}_{ab},  
\label{9.41} \\ 
J_{ab} &=& J^{\rm intrinsic}_{ab} + J^{\rm induced}_{ab},  
\label{9.42}
\end{eqnarray} 
and we know that \cite{thorne-hartle:85} 
\begin{equation} 
M^{\rm intrinsic}_{ab} = \frac{1}{3} M^3 \chi^2 (\delta_{ab} 
- 3 s_a s_b), \qquad 
J^{\rm intrinsic}_{ab} = 0.  
\label{9.43}
\end{equation} 
We recall that $M_{ab}$ is the hole's mass quadrupole moment, while
$J_{ab}$ is its current quadrupole moment; both tensors are symmetric
and tracefree. We wish to see if we can determine 
$M^{\rm induced}_{ab}$, $J^{\rm induced}_{ab}$ and establish
compatibility between Eq.~(\ref{8.42}) and (\ref{9.39}).  

It is easy to show, by substituting Eq.~(\ref{9.43}) into
Eq.~(\ref{8.42}), that the coupling between the {\it intrinsic}
moments and the tidal gravitational fields does not affect the
magnitude of the angular-momentum vector; the only effect is to
produce a precession of $J^a$ described by $\dot{J}^a =
\varepsilon^a_{\ bc} \Omega_{\rm P}^b J^c$, where $\Omega_{\rm P}^a
\equiv -M \chi {\cal E}^a_{\ b} s^b$ is the precessional angular
velocity. We conclude that only the {\it induced} moments will
contribute to $\Jdot$, and we now seek to determine them.   

To ease the comparison between Eq.~(\ref{8.42}) and (\ref{9.39}) we
set $s^a = (0,0,1)$ and compute $\Jdot \equiv \langle \dot{J}^3
\rangle$, which we compare with the result displayed immediately
before Eq.~(\ref{9.33}). This reveals that $M^{\rm induced}_{ab}$ is 
partially determined by the relations $M_{11} - M_{22} \propto
16(1+3\chi^2) {\cal E}_{12}$, $M_{12} \propto -4(1+3\chi^2) 
({\cal E}_{11} - {\cal E}_{22})$, $M_{13} \propto (4-3\chi^2) 
{\cal E}_{23}$, and $M_{23} \propto -(4-3\chi^2) {\cal E}_{23}$, where 
the (unique) constant of proportionality is equal to
$\frac{2}{45} M^5\chi$. Analogous relations link $\frac{4}{3} 
J^{\rm induced}_{ab}$ to ${\cal B}_{ab}$. These relations determine
$M^{\rm induced}_{ab}$ (and $J^{\rm induced}_{ab}$) up to a term
proportional to $\delta_{ab} - 3 s_a s_b$; the coefficient must be a
scalar formed from ${\cal E}_{ab}$ (or ${\cal B}_{ab}$),
$\delta_{ab}$, $s_a$, and $\varepsilon_{abc}$, and the only possible
candidate is an arbitrary function of $\chi$ multiplying
${\cal E}_{ab} s^a s^b$ (or ${\cal B}_{ab} s^a s^b$). 

We therefore arrive at 
\begin{eqnarray} 
M^{\rm induced}_{ab} &=& \frac{2}{45} M^5 \chi \Bigl[ \lambda(\chi) 
(\delta_{ab} - 3 s_a s_b) {\cal E}_{cd} s^c s^d 
\nonumber \\ & & \mbox{} 
+ 8(1+3\chi^2) {\cal E}^c_{\ (a} \varepsilon_{b)cd} s^d 
\nonumber \\ & & \mbox{} 
+ 30\chi^2 s_{(a} \varepsilon_{b) cd} s^c {\cal E}^d_{\ e} s^e \Bigr]  
\label{9.44}
\end{eqnarray}
for the mass quadrupole moment, and 
\begin{eqnarray} 
J^{\rm induced}_{ab} &=& \frac{1}{30} M^5 \chi \Bigl[
\mu(\chi) (\delta_{ab} - 3 s_a s_b) {\cal B}_{cd} s^c s^d 
\nonumber \\ & & \mbox{} 
+ 8(1+3\chi^2) {\cal B}^c_{\ (a} \varepsilon_{b)cd} s^d 
\nonumber \\ & & \mbox{} 
+ 30\chi^2 s_{(a} \varepsilon_{b) cd} s^c {\cal B}^d_{\ e} s^e \Bigr]  
\label{9.45}
\end{eqnarray}
for the current quadrupole moment, where $\lambda(\chi)$ and
$\mu(\chi)$ are unknown functions of the hole's rotational
parameter. We conclude that our results are indeed compatible with the
general results of Thorne and Hartle \cite{thorne-hartle:85}. 
We observe that the relationships between the induced moments  
and the tidal gravitational fields have the schematic form    
$M^{\rm induced} \sim M^5 {\cal E}$, $J^{\rm induced} \sim M^5 
{\cal B}$ (but with a complicated tensorial structure), and that 
the moments scale as $M^3 (M/{\cal R})^2$. These results follow   
expectation, and they are markedly different from those of 
Sec.~VIII G; recall that for a nonrotating black hole the
relationships involve an additional factor of $M$ and a time
derivative.  

We emphasize that the induced moments are only partially  
determined: the functions $\lambda(\chi)$ and $\mu(\chi)$ cannot be 
determined by the comparison with Thorne and Hartle, because the terms
to which they belong in $M_{ab}$ and $J_{ab}$ do not affect the
magnitude of the angular-momentum vector. They produce instead a
small fractional correction of order $(M/{\cal R})^2$ to 
$\Omega_{\rm P}^a$, the precessional angular velocity.   

\subsection{Comparison between Kerr and Schwarzschild results}  
 
The main results of Sec.~VIII, Eqs.~(\ref{8.38}) and (\ref{8.39}), or 
equivalently Eqs.~(\ref{8.40}) and (\ref{8.41}), hold to leading order
in $M/{\cal R} \ll 1$, and they reveal that for a Schwarzschild black
hole, $\Mdot = O(M^6/{\cal R}^6)$ and $\Jdot = O(M^6/{\cal R}^5)$. On
the other hand, the main results of this section, Eqs.~(\ref{9.32})
and (\ref{9.39}), hold to leading order in $M/{\cal R} \ll \chi$, and
they reveal that for a Kerr black hole, $\Mdot = O(M^5/{\cal R}^5)$
and $\Jdot = O(M^5/{\cal R}^4)$. The scalings are very different, and
the condition $M/{\cal R} \ll \chi$ implies that the Schwarzschild
results cannot straightforwardly be obtained from the Kerr results in
a limit $\chi \to 0$. 

The origin of the difference in scalings can easily be understood in
the special case of rigid rotation, for which $\Mdot$ and $\Jdot$ are 
given by Eqs.~(\ref{4.29})--(\ref{4.31}), 
\[
\Mdot = \Omega(\Omega-\Omega_{\rm H}) {\cal K}, \qquad 
\Jdot = (\Omega-\Omega_{\rm H}) {\cal K},  
\]
where ${\cal K}$ is defined by Eq.~(\ref{4.32}) and $\Omega 
= O({\cal R}^{-1})$ is the hole's angular velocity in the external 
spacetime. This argument was first presented to me by Kip Thorne
(personal communication), and it was then elaborated on by Alvi
\cite{alvi:01}.   

Suppose, as we shall show below, that ${\cal K} = 
O(M^6/{\cal R}^4)$. In the case of nonrotating black hole we have
$\Omega_{\rm H} = 0$, and it follows that $\Mdot = \Omega^2 {\cal K} =
O(M^6/{\cal R}^6)$ and $\Jdot = \Omega {\cal K} = O(M^6/{\cal R}^5)$;
those are precisely the scalings obtained previously for a
Schwarzschild black hole. The situation is different for a rotating
black hole. In this case the condition $M/{\cal R} \ll \chi$ implies
that $\Omega \ll \Omega_{\rm H}$, and we have instead $\Mdot = -\Omega
\Omega_{\rm H} {\cal K} = O(M^5/{\cal R}^5)$ and $\Jdot = 
-\Omega_{\rm H} {\cal K} = O(M^5/{\cal R}^4)$; those are precisely the
scalings obtained previously for a Kerr black hole. Notice that in
the case of Kerr, $\Mdot$ and $\Jdot$ are both proportional to
$\Omega_{\rm H}$ and therefore to $\chi$; this observation is
confirmed by Eq.~(\ref{9.39}). 

The different scalings reflect the different technical meanings
assigned to the phrase ``small-hole/slow-motion approximation:'' For a
Schwarzschild black hole we impose $M/{\cal R} \ll 1$ and we naturally
have $\Omega \gg \Omega_{\rm H}$; for a Kerr black hole we impose
instead $M/{\cal R} \ll \chi$ and we consequently have $\Omega \ll 
\Omega_{\rm H}$. In generic situations (that is, in the absence of
rigid rotation) the scaling argument given previously continues to
apply, but $\Omega \sim {\cal R}^{-1}$ is now interpreted as an
inverse time scale associated with changes in the Weyl tensor of the
external spacetime.  

An expression for ${\cal K}$ can be obtained from the approximate
relation $\Jdot = -\Omega_{\rm H} {\cal K}$ which holds for a rotating
black hole. We obtain 
\begin{eqnarray}
{\cal K} &=& \frac{4}{45} M^6 \bigl(1 + \sqrt{1-\chi^2} \bigr) 
\Bigl[ 8(1 + 3\chi^2) \langle E_1 + B_1 \rangle 
\nonumber \\ & & \mbox{} 
- 3(4 + 17\chi^2) \langle E_2 + B_2 \rangle 
\nonumber \\ & & \mbox{} 
+ 15\chi^2 \langle E_3 + B_3 \rangle \Bigr], 
\label{9.46} 
\end{eqnarray}
and we confirm that indeed, ${\cal K} = O(M^6/{\cal R}^4)$. Notice
that there is no obstacle to taking the limit $\chi \to 0$ of this 
expression.      

\subsection{Black hole in a circular binary: Slow-motion
approximation} 

We now specialize the results of Sec.~IX D to a Kerr black hole placed
on a circular orbit in the weak gravitational field of an external
body of mass $M_{\rm ext}$. This is the slow-motion approximation, and
we shall repeat here most of the steps described in Sec.~VIII H.  

As before the tidal gravitational fields of the external universe are
approximated by ${\cal E}_{ab} \simeq \Phi_{,ab}$ and ${\cal B}_{ab}
\simeq 0$, where $\Phi = -M_{\rm ext}/|\bm{x}-\bm{r}_{\rm ext}|$ is
the Newtonian potential associated with the external body. As before
the black hole is moving on a circular orbit, and we assume that the
orbital angular momentum vector is either aligned or anti-aligned with
the hole's spin vector: $\bm{\hat{L}} \cdot \bm{s} \equiv \epsilon =
\pm 1$. The hole's orbital angular velocity is then  
\begin{equation}
\Omega = \epsilon \sqrt{\frac{M+M_{\rm ext}}{b^3}},   
\label{9.47} 
\end{equation}
where $b$ is the orbital radius; the angular velocity is positive when
the orbital and spin angular momenta are aligned, and it is negative
when they are anti-aligned.    

A simple calculation, along the lines of what was presented in
Sec.~VIII H, yields 
\begin{equation} 
\Mdot = -\epsilon \frac{8}{5} \eta^2 \biggl( \frac{M}{M+M_{\rm ext}}
\biggr)^3 \chi (1 + 3 \chi^2) V^{15}
\label{9.48}
\end{equation}
and 
\begin{equation} 
\Jdot = -\frac{8}{5} \eta^2 \biggl( \frac{M}{M+M_{\rm ext}}
\biggr)^3 (M + M_{\rm ext}) \chi (1 + 3 \chi^2) V^{12}, 
\label{9.49}
\end{equation}
where $\eta = M M_{\rm ext}/(M+M_{\rm ext})^2$ is a dimensionless
reduced-mass parameter and 
\begin{equation} 
V = \sqrt{ \frac{M + M_{\rm ext}}{b} } \ll 1 
\label{9.50}
\end{equation}
is the relative orbital velocity. Notice that while we could not
calculate $\Mdot$ in the general case described in Sec.~IX D, here it
is simply given by $\Omega \Jdot$ because the black hole is in rigid
rotation around $M_{\rm ext}$. The results of Eq.~(\ref{9.48}) and
(\ref{9.49}) agree with earlier expressions obtained by Alvi 
\cite{alvi:01}. 
In the regime $M_{\rm ext} \gg M$ they also agree with earlier 
results derived by Tagoshi, Mano, and Takasugi
\cite{tagoshi-etal:97}.    

\subsection{Black hole in a circular binary: Small-hole
approximation}

We now allow the Kerr black hole to move rapidly in the strong 
gravitational field of a Schwarzschild hole of mass 
$M_{\rm ext}$. We no longer restrict the size of $V$ but we now impose
$M \ll M_{\rm ext}$; this is the small-hole approximation, and we 
shall repeat here most of the steps described in Sec.~VIII I. 

Once more we take the orbit to be circular. In the standard
Schwarzschild coordinates $(t,r,\theta,\phi)$ used in the background
spacetime of the large black hole, the orbital radius is $b$ and the 
four-velocity of the small hole is $u^\alpha = \gamma(1,0,0,\Omega)$,
where $\gamma = (1-3M_{\rm ext}/b)^{-1/2}$ is a normalization factor
and 
\begin{equation}
\Omega = \epsilon \sqrt{ \frac{M_{\rm ext}}{b^3} } 
\label{9.51}
\end{equation}
is the angular velocity; as in the preceding subsection $\epsilon =
\pm 1$ gives the orientation of the orbital angular momentum vector
relative to the hole's spin vector, $s^\alpha =
(0,0,-1/b,0)$. Calculation of ${\cal E}_{\alpha\beta}$,  
${\cal B}_{\alpha\beta}$ using Eqs.~(\ref{8.4}), (\ref{8.5}), and
substitution into Eqs.~(\ref{9.33})--(\ref{9.39}) gives 
\begin{eqnarray} 
\Mdot &=& -\epsilon \frac{8}{5} \biggl( \frac{M}{M_{\rm ext}}
\biggr)^5 \chi (1 + 3\chi^2) V^{15} 
\nonumber \\ & & \mbox{} \times 
\frac{ (1-2V^2) \bigl(1- \frac{4 + 27\chi^2}{4 + 12\chi^2}V^2 \bigr) 
}{ (1-3V^2)^2 }  
\label{9.52}
\end{eqnarray}
and 
\begin{eqnarray} 
\Jdot &=& -\frac{8}{5} \biggl( \frac{M}{M_{\rm ext}} \biggr)^5 
M_{\rm ext} \chi (1 + 3\chi^2) V^{12} 
\nonumber \\ & & \mbox{} \times 
\frac{ (1-2V^2) \bigl(1- \frac{4 + 27\chi^2}{4 + 12\chi^2}V^2 \bigr)
}{ (1-3V^2)^2 },   
\label{9.53}
\end{eqnarray}
where $V = \sqrt{M_{\rm ext}/b} \leq 6^{-1/2}$ is a measure of the 
hole's orbital velocity. These results agree with those of the
preceding subsection in a common domain of validity defined by
$M \ll M_{\rm ext}$ and $V \ll 1$. To the best of my knowledge, the 
results of Eqs.~(\ref{9.47}) and (\ref{9.48}), complete with all-order 
relativistic corrections, have never appeared before in the
literature.  

\subsection{Black hole in a static tidal field}  

For completeness we explore another special case of Eq.~(\ref{9.39}),
in which the rotating black hole is at rest in a static tidal
gravitational field. We wish to calculate the rate at which this black 
hole loses its angular momentum. This calculation was presented many
times before, most notably by Hartle 
\cite{hartle:73, hartle:74}, 
Teukolsky \cite{teukolsky:phd}, 
Chrzanowski \cite{chrzanowski:75b}, 
Thorne, Price, and Macdonald \cite{thorne-etal:86}, 
and Alvi \cite{alvi:01}.  
The point of this subsection is to illustrate how easily the classic
spin-down result of Eq.~(\ref{9.55}) follows from Eq.~(\ref{9.39}).   

We assume that the tidal gravitational field is purely electric in the
local asymptotic rest frame of the black hole, and that it is axially
symmetric in the arbitrary direction of the unit vector $n^a$. With
these specifications we have 
\begin{equation}
{\cal E}_{ab} = -\frac{1}{2} {\cal E} (\delta_{ab} - 3 n_a n_b), 
\qquad
{\cal B}_{ab} = 0, 
\label{9.54}
\end{equation}
where ${\cal E} \equiv {\cal E}_{ab} n^a n^b$. If, for example, the
tidal field is produced by a body of mass $M_{\rm ext}$ maintained at
a fixed position $\bm{r} = b \bm{n}$ relative to the black hole, then  
${\cal E} = -2M_{\rm ext}/b^3$. We assume that the black hole's
angular momentum makes an angle $\alpha$ with respect to the
direction of $n^a$, so that $s_a n^a = \cos\alpha$. We then have
${\cal E}_{ab} s^b = \frac{1}{2} {\cal E} (3 \cos\alpha\, n_a - s_a)$,   
${\cal E}_{ab} s^a s^b = \frac{1}{2} {\cal E} (3 \cos^2\alpha 
- 1)$, and the invariants of Eqs.~(\ref{9.33})--(\ref{9.35}) are
easily computed. After simplification we find that Eq.~(\ref{9.39})
reduces to 
\begin{equation} 
\dot{J} = -\frac{2}{5} {\cal E}^2 M^5 \chi \sin^2\alpha \biggl[ 1 
- \frac{3}{4} \Bigl( 1 - 5 \sin^2\alpha \Bigr) \chi^2 \Biggr], 
\label{9.55}
\end{equation}
which is the classic spin-down formula.  

\acknowledgments 

This work was supported by the Natural Sciences and Engineering
Research Council of Canada. I am grateful to Eanna Flanagan, David
Garfinkle, Karl Martel, and Clifford Will for helpful discussions.   

\section*{Note added} 

Conversations with Kip Thorne and John Friedman (whom I thank) made me 
understand that the discussion of induced quadrupole moments inserted
in Secs.~VIII G and IX E is incomplete. I should have realized that
the tidal-heating formulae of Eqs.~(\ref{8.42}) and (\ref{8.43}) allow
the determination of $M_{ab}$ only up to a term proportional to 
${\cal E}_{ab}$, and the determination of $J_{ab}$ up to a term
proportional to ${\cal B}_{ab}$. Such terms do not participate in the 
tidal heating and leave $\langle \dot{M} \rangle$ and $\langle
\dot{J}^a \rangle$ unchanged. It is therefore possible for a tidally
distorted Schwarzschild black hole to have quadrupole moments given by 
$M_{ab} = a M^5 {\cal E}_{ab} + (32/45) M^6 \dot{\cal E}_{ab}$ and 
$J_{ab} = a' M^5 {\cal B}_{ab} + (8/15) M^6 \dot{\cal B}_{ab}$, where 
$a$ and $a'$ are undetermined dimensionless constants. Such moments
would scale as $M^5/{\cal R}^2$, which is the expected scaling.   
  
\appendix*
\section{Perturbations of a Schwarzschild black hole} 

In this Appendix I collect a few key results from the theory of
gravitational perturbations of a Schwarzschild black hole
\cite{regge-wheeler:57, zerilli:70, moncrief:74, chandrasekhar:83}. I
employ a covariant/gauge-invariant formalism that was inspired by the
work of Gerlach \& Sengupta 
\cite{gerlach-sengupta:79, gerlach-sengupta:80}  
and Sarbach \& Tiglio
\cite{sarbach-tiglio:01}. 
These results are presented without derivation; details can be found
in Martel's PhD dissertation \cite{martel:phd}.    

\subsection{Background metric}

The Schwarzschild metric is expressed as
\begin{equation}  
ds^2 = g_{ij}\, dx^i dx^j + r^2 \Omega_{AB}\, d\theta^A
d\theta^B, 
\label{A.1} 
\end{equation} 
in a form that is covariant under two-dimensional coordinate
transformations $x^{i} \to x^{i'}$. The indices $i, j, k, \ldots$ run
over the values 0 and 1, and the indices $A, B, C, \ldots$ run over
the values 2 and 3. The traditional Schwarzschild coordinates are $x^i
= (t,r)$, and in the text we use the ingoing Eddington-Finkelstein
coordinates $x^i = (v,r)$, where $v = t + r + 2M\ln(r/2M - 1)$, with
$M$ denoting the black-hole mass. In the metric of Eq.~(\ref{A.1}),
$r$ is viewed as a scalar function of the arbitrary coordinates $x^i$,
and $\Omega_{AB} = \mbox{diag}(1,\sin^2\theta)$ is the metric on the
unit two-sphere.  

We use $g_{ij}$ and its inverse to lower and raise all lower-case 
Latin indices. And in this Appendix, contrary to previous usage in the 
body of the paper, we use $\Omega_{AB}$ and its inverse to lower and
raise all upper-case Latin indices. We indicate covariant
differentiation with respect to a connection compatible with $g_{ij}$
with a dot: $g_{ij.k} = 0$. And we indicate covariant differentiation
with respect to a connection compatible with $\Omega_{AB}$ with a
colon: $\Omega_{AB:C} = 0$.  
 
\subsection{Spherical harmonics} 

The tensorial nature of the spherical harmonics refers to the unit
two-sphere, whose metric is $\Omega_{AB}$. The definitions adopted
below agree with those of Regge and Wheeler 
\cite{regge-wheeler:57}. 
The Levi-Civita tensor on the unit two-sphere is denoted
$\varepsilon_{AB}$, with $\varepsilon_{\theta\phi} = \sin\theta$.  

The scalar harmonics are the usual spherical-harmonic functions
$Y^{lm}(\theta^A)$, which satisfy the eigenvalue equation
$\Omega^{AB} Y^{lm}_{:AB} + l(l+1) Y^{lm} = 0$.  

Vectorial spherical harmonics come in two types. The {\it even-parity} 
harmonics are 
\begin{equation} 
Y^{lm}_{\ :A} \qquad \mbox{(even parity)}, 
\label{A.2}
\end{equation}  
while the {\it odd-parity} harmonics are 
\begin{equation} 
X_{A}^{lm} = - \varepsilon_A^{\ B} Y^{lm}_{\ :B}
\qquad \mbox{(odd parity)}. 
\label{A.3}
\end{equation} 
The vectorial harmonics satisfy the following orthogonality relations:     
\begin{equation} 
\int \bar{Y}^{:A}_{lm} Y^{l'm'}_{:A}\, d\Omega 
= \int \bar{X}^{A}_{lm} X^{l'm'}_A\, d\Omega 
= l(l+1)\, \delta_{ll'}\delta_{mm'}
\label{A.4}
\end{equation} 
and 
\begin{equation} 
\int \bar{Y}^{:A}_{lm} X^{l'm'}_A\, d\Omega = 0, 
\label{A.5}
\end{equation} 
where an overbar indicates complex conjugation and $d\Omega =
\sin\theta\, d\theta d\phi$.   

Tensorial spherical harmonics come in the same two types. The 
{\it even-parity} harmonics are 
\begin{equation} 
Y^{lm} \Omega_{AB}, \qquad Y^{lm}_{\ :AB}
\qquad \mbox{(even parity)}. 
\label{A.6}
\end{equation} 
It is useful to define also 
\begin{equation} 
Z_{AB}^{lm} = Y^{lm}_{\ :AB} + \frac{1}{2} l(l+1) Y^{lm} \Omega_{AB}.  
\label{A.7}
\end{equation} 
By virtue of the eigenvalue equation for the scalar harmonics,
$\Omega^{AB} Z_{AB}^{lm} = 0$; these harmonics are therefore
tracefree. The {\it odd-parity} harmonics are 
\begin{equation} 
X_{AB}^{lm} = - X^{lm}_{(A:B)} \qquad \mbox{(odd parity)}; 
\label{A.8}
\end{equation} 
these are also tracefree: $\Omega^{AB} X_{AB}^{lm} = 0$. We
record the following relations between the tensorial harmonics and 
the spherical harmonics of spin-weight $s=\pm 2$
\cite{goldberg-etal:67}:  
\begin{eqnarray}
Z_{AB}^{lm} &=& \frac{1}{2} \sqrt{(l-1)l(l+1)(l+2)}  
\nonumber \\ & & \mbox{} \times 
\bigl( \mbox{}_{-2}Y^{lm} \epsilon_A \epsilon_B 
+ \mbox{}_{2} Y^{lm} \bar{\epsilon}_A \bar{\epsilon}_B \bigr) 
\label{A.9}
\end{eqnarray} 
and 
\begin{eqnarray} 
X_{AB}^{lm} &=& \frac{i}{2} \sqrt{(l-1)l(l+1)(l+2)} 
\nonumber \\ & & \mbox{} \times 
\bigl( \mbox{}_{-2}Y^{lm} \epsilon_A \epsilon_B 
- \mbox{}_{2} Y^{lm} \bar{\epsilon}_A \bar{\epsilon}_B \bigr),    
\label{A.10}
\end{eqnarray}
where the vectors $\epsilon_A \equiv (1,i\sin\theta)/\sqrt{2}$ satisfy 
$\Omega^{AB} \epsilon_A \epsilon_B = \Omega^{AB} \bar{\epsilon}_A
\bar{\epsilon}_B = 0$ and $\Omega^{AB} \epsilon_A \bar{\epsilon}_B 
= 1$. The tensorial harmonics satisfy the following orthogonality
relations:    
\begin{eqnarray} 
\int \bar{Z}^{AB}_{lm} Z^{l'm'}_{AB}\, d\Omega &=& 
\int \bar{X}^{AB}_{lm} X^{l'm'}_{AB}\, d\Omega 
\nonumber \\ 
&=& \frac{1}{2} (l-1)l(l+1)(l+2)\, \delta_{ll'}\delta_{mm'} 
\nonumber \\ & & \mbox{} \label{A.11}
\end{eqnarray}
and 
\begin{equation} 
\int \bar{Z}^{AB}_{lm} X^{l'm'}_{AB}\, d\Omega = 0. 
\label{A.12}
\end{equation} 

The spherical-harmonic functions constructed thus far are complex, and
they are all proportional to $e^{i m \phi}$. We shall also need a set
of real spherical harmonics of degree $l = 2$, which we denote
$Y^{\sf m}$, $Y^{\sf m}_{:A}$, $X^{\sf m}_A$, $Y^{\sf m}_{:AB}$,
$Z^{\sf m}_{AB}$, and $X^{\sf m}_{AB}$. The label $\sf m$ runs over
the set $\{0,1c,1s,2c,2s\}$; the numerical part of this label refers
to the azimuthal index $m$, and the letter indicates whether the
corresponding scalar-harmonic function is proportional to
$\cos(m\phi)$ or $\sin(m\phi)$. Explicitly, 
\begin{eqnarray} 
Y^0 &=& \frac{1}{2} (3\cos^2\theta - 1), \nonumber \\ 
Y^{1c} &=& \sin\theta \cos\theta \cos\phi, \nonumber \\ 
Y^{1s} &=& \sin\theta \cos\theta \sin\phi, \nonumber \\ 
Y^{2c} &=& \sin^2\theta \cos 2\phi, \nonumber\\ 
Y^{2s} &=& \sin^2\theta \sin 2\phi. 
\label{A.13}
\end{eqnarray}  
The vectorial and tensorial harmonics are generated by acting on
$Y^{\sf m}$ with the same differential operators as those involved in 
Eqs.~(\ref{A.2}), (\ref{A.3}), (\ref{A.6})--(\ref{A.8}).    

\subsection{Odd-parity perturbations}  

The odd-parity perturbations of the Schwarzschild metric are those 
which are expanded in terms of odd-parity spherical harmonics. This
sector of the metric perturbation is given by 
\begin{eqnarray} 
\delta g_{iA}(x^i,\theta^A) &=& h_{i}(x^i) X^{lm}_{A}(\theta^A),  
\label{A.14} \\
\delta g_{AB}(x^i,\theta^A) &=& h_2(x^i) X^{lm}_{AB}(\theta^A).
\label{A.15}
\end{eqnarray}  
We suppress usage of the $lm$ label on the fields $h_{i}$ and $h_2$, 
and it is understood that the right-hand sides are summed over 
$l$ and $m$. It can be shown that the combinations   
\begin{equation} 
\tilde{h}_i = h_{i} + \frac{1}{2}\, h_{2,i} - \frac{1}{r}\, r_{,i} h_2    
\label{A.16}
\end{equation} 
are invariant under odd-parity gauge transformations. The linearized
Einstein field equations are then naturally expressed in terms of
$\tilde{h}_i$ and its covariant derivatives. One of these equations is
required in the text: In the absence of sources it can be shown that 
\begin{equation} 
\tilde{h}^i_{\ .i} = 0. 
\label{A.17}
\end{equation} 
The remaining field equations can be manipulated to form a
one-dimensional wave equation for the master variable 
\begin{equation} 
\Psi_{\rm RW} \equiv \frac{1}{r} r^{,i} \tilde{h}_i, 
\label{A.18}
\end{equation} 
which is evidently gauge invariant. The function $\Psi_{\rm RW}(x^i)$
is known as the Regge-Wheeler function
\cite{regge-wheeler:57}, 
and in the absence of sources it satisfies the differential equation  
\begin{equation} 
\Box \Psi_{\rm RW} - \biggl[ \frac{l(l+1)}{r^2} - \frac{6M}{r^3}
\biggr] \Psi_{\rm RW} = 0, 
\label{A.19}
\end{equation}
where $\Box \Psi \equiv g^{ij} \Psi_{.ij}$ is the one-dimensional
wave operator acting on the scalar function $\Psi(x^i)$. It is well
understood that in a specified gauge, all components of the odd-parity  
metric perturbation can be reconstructed from the Regge-Wheeler
function.   

\subsection{Even-parity perturbations}

The even-parity perturbations are expanded in terms of even-parity
spherical harmonics. This sector of the metric perturbation is given
by 
\begin{eqnarray} 
\delta g_{ij}(x^i,\theta^A) &=& h_{ij}(x^i) Y^{lm}(\theta^A), 
\label{A.20} \\
\delta g_{iA}(x^i,\theta^A) &=& j_{i}(x^i) Y^{lm}_{:A}(\theta^A), 
\label{A.21} \\
\delta g_{AB}(x^i,\theta^A) &=& r^2 \bigl[ K(x^i) Y^{lm}(\theta^A)
\Omega_{AB} 
\nonumber \\ & & \mbox{} 
+ G(x^i) Y^{lm}_{:AB}(\theta^A) \bigr].
\label{A.22}
\end{eqnarray} 
Once more we suppress usage of the $lm$ label on the fields $h_{ij}$,
$j_i$, $K$, and $G$, and it is understood that the right-hand sides
are summed over $l$ and $m$. The combinations 
\begin{equation} 
\tilde{h}_{ij} = h_{ij} - 2\varepsilon_{(i.j)}, \qquad  
\tilde{K} = K - \frac{2}{r}\, r^{,i} \varepsilon_{i}, 
\label{A.23}
\end{equation} 
where $\varepsilon_i = j_i - \frac{1}{2} r^2 G_{,i}$, are invariant
under even-parity gauge transformations. The linearized Einstein field
equations are then naturally expressed in terms of these fields and
their covariant derivatives. They can be manipulated to form a
one-dimensional wave equation for the master variable 
\begin{equation} 
\Psi_{\rm ZM} \equiv \frac{2r}{l(l+1)} \biggl[ \tilde{K} +
\frac{2}{\Lambda} \bigl( r^{,i} r^{,j} \tilde{h}_{ij} - r r^{,i}
\tilde{K}_{,i} \bigr) \biggr], 
\label{A.24}
\end{equation} 
where $\Lambda \equiv (l-1)(l+2) + 6M/r$. The function 
$\Psi_{\rm ZM}(x^i)$ is known as the Zerilli-Moncrief function
\cite{zerilli:70, moncrief:74}, 
and it is evidently gauge-invariant; it satisfies a differential
equation similar to Eq.~(\ref{A.19}), but with a more complicated
potential. The normalization of the Zerilli-Moncrief function is
chosen so as to agree with the definition proposed by Lousto and Price
\cite{lousto-price:97}.  

\subsection{Waveforms and energy radiated at infinity} 

When examined near future null infinity, the gravitational
perturbations of Eqs.~(\ref{A.14}), (\ref{A.15}),
(\ref{A.20})--(\ref{A.22}) can be presented in an outgoing-radiation
gauge that permits an easy identification of the radiative field.   
It can be shown that the two fundamental polarizations of the
gravitational waves are given by 
\begin{eqnarray} 
h_+ - i h_\times &=& \frac{1}{2r} \sum_{lm} \sqrt{(l-1)l(l+1)(l+2)}\,     
\biggl[ \Psi^{lm}_{\rm ZM}(u) 
\nonumber \\ & & \mbox{}
- 2 i \int^u \Psi^{lm}_{\rm RW}(u')\,
du' \biggr]\, \mbox{}_{-2}Y^{lm}(\theta^A), \qquad\quad 
\label{A.25}
\end{eqnarray}
where $u = t - r - 2M\ln(r/2M - 1)$ is retarded time, and
$\mbox{}_{-2}Y^{lm}(\theta^A)$ are spherical harmonics of spin-weight
$s=-2$. The fact that the waveforms are expressed in terms of an 
{\it integral} of the Regge-Wheeler function means that this master
variable is rather ill-suited to describe the radiative aspects of the
metric perturbation. An alternative choice of master variable, which
is free of this blemish, was proposed by Cunningham, Price, and
Moncrief
\cite{cunningham-etal:78, cunningham-etal:79}; 
it was recently revived by Jhingan and Tanaka
\cite{jhingan-tanaka:03}.   

The energy and angular momentum radiated to infinity are given by 
\begin{eqnarray} 
\langle \dot{E} \rangle &=& \frac{1}{64\pi} \sum_{l m} (l-1) l (l+1)
(l+2) 
\nonumber \\ & & \mbox{} \times 
\Bigl\langle 4 \bigl| \Psi^{lm}_{\rm RW}(u) \bigr|^2 
+ \bigl| \dot{\Psi}^{lm}_{\rm ZM}(u) \bigr|^2 \Bigr\rangle, 
\label{A.26} \\
\Jdot &=& \frac{1}{64\pi} \sum_{l m} (l-1) l (l+1) (l+2) (im) 
\nonumber \\ & & \mbox{} \times 
\Bigl\langle 4 \Psi^{lm}_{\rm RW}(u) \int^u 
\bar{\Psi}^{lm}_{\rm RW}(u')\, du' 
\nonumber \\ & & \mbox{} 
+ \dot{\Psi}^{lm}_{\rm ZM}(u)
\bar{\Psi}^{lm}_{\rm ZM}(u) \Bigr\rangle.
\label{A.27}
\end{eqnarray} 
The expressions are very similar to the horizon-flux formulae of
Eqs.~(\ref{7.8}) and (\ref{7.9}). 

\newpage
\bibliography{flow}
\end{document}